\documentclass[12pt]{article}
\usepackage[margin=2 cm]{geometry}
\usepackage{comment}
\usepackage{graphicx}
\usepackage{tikz}
\usetikzlibrary{arrows}
\usepackage{mathtools}
\usepackage[font={small,it}]{caption}
\usepackage{subcaption}
\usepackage{mwe}
\usepackage[nottoc]{tocbibind}
\usepackage{amsmath,amssymb,extarrows,mathtools,graphicx,setspace}
\usepackage{cite}
\usepackage{braket}
\usepackage{epsfig}
\usepackage[section]{placeins}
\usepackage{slashed}
\usepackage{color}
\usepackage{xcolor}
\usepackage{amsmath}
\usepackage[implicit]{hyperref}
\makeatother

\newcommand{\be}{\begin{equation}}
\newcommand{\bea}{\begin{eqnarray}}
\newcommand{\eea}{\end{eqnarray}}
\newcommand{\ba}{\begin{array}}
\newcommand{\ea}{\end{array}}
\newcommand{\ee}{\end{equation}}
\newcommand{\bes}{\begin{equation*}}
\newcommand{\beas}{\begin{eqnarray*}}
\newcommand{\eeas}{\end{eqnarray*}}
\newcommand{\bas}{\begin{array*}}
\newcommand{\eas}{\end{array*}}
\newcommand{\ees}{\end{equation*}}

\numberwithin{equation}{section}

\begin{document}

\onehalfspacing
\vfill
\begin{titlepage}
\vspace{10mm}

\begin{center}

\vspace*{10mm}
\vspace*{1mm}
{\Large  \textbf{Complexity growth rate during phase transitions}} 
 \vspace*{1cm}
 
{$\text{Mahdis Ghodrati}$}

\vspace*{8mm}
{ \textsl{
$ $Center for Gravitation and Cosmology, College of Physical Science and Technology,\\
Yangzhou University, Yangzhou 225009, China}} 
 \vspace*{1cm}

\textsl{E-mail: {\href{mailto: mahdisg@yzu.edu.cn }{mahdisg@yzu.edu.cn}}}
 \vspace*{2mm}

\vspace*{1.7cm}

\end{center}

\begin{abstract}
We present evidences for the connection between the potential of different fields and complexity growth rates both in conformal and confining cases. By studying different models, we also establish a strong connection between phase transitions and the discontinuities  in the complexity growth rates. In the first example, for the dyonic black holes which are dual to van der Waals fluids, we find a similar first order phase transition in the behavior of complexity growth rate. We then compare the Schwinger effect and also the behavior of complexity in the AdS and AdS soliton backgrounds and comment on the connection between them. Finally, in a general Gubser model of QCD, we present the connections between the potentials, entropies, speed of sounds and complexity growth rates during crossover, first and second order phase transitions and also the behavior of quasinormal modes. 
\end{abstract}

\end{titlepage}

\tableofcontents


\section{Introduction}

In addition to Ryu-Takayanagi formula \cite{Ryu:2006bv} which built a connection between the entanglement entropy as a quantum information quantity in the boundary, and the area of a codimension-two hypersurface, as a geometric quantity in the bulk, recently holographic quantum complexity \cite{Susskind:2014rva, Brown:2015lvg, Alishahiha:2015rta } has been proposed which builds another connection between quantum information and geometry.  

In the quantum information side, complexity would be the number of gates needed to go from one specific quantum state to another one and therefore, it quantifies how difficult a computational task would be. In the geometrical side, it could be calculated from the volume of a codimension-one surface which extends between two boundaries (complexity$=$volume, CV conjecture) or the action on the Wheeler-DeWitt patch (complexity$=$action, CA conjecture) \cite{Brown:2015bva,Brown:2015lvg}. These conjectures have been studied extensively recently, see \cite{Cai:2017sjv, Bhattacharyya:2018wym,Carmi:2017jqz, Ghodrati:2017roz, Auzzi:2018zdu, Auzzi:2018pbc,Abt:2018ywl, Cottrell:2017ayj, Agon:2018zso, Swingle:2017zcd,Hashimoto:2018bmb, Barbon:2018mxk, Ageev:2018nye, Du:2018uua ,Chen:2018mcc, Ovgun:2018jbm, Fu:2018kcp,Flory:2018akz,Caputa:2018kdj,Caputa:2017yrh,Abt:2017pmf} for some examples. 

In order to make this new connection more precise and use it in the holographic context, a more exact definition of complexity for the field theories are needed. The initial attempts were for the free scalar fields \cite{Jefferson:2017sdb,Chapman:2017rqy}, simple fermionic fields \cite{Khan:2018rzm,Hackl:2018ptj,Reynolds:2017jfs}, and recently coherent \cite{Guo:2018kzl} and simple interacting QFTs \cite{Bhattacharyya:2018bbv}. Generally there would be some ambiguities in defining the specific gates needed for each case.

In this work, first, we examine if there exist a relationship between quantum fluctuations of the system in different models, for instance between particle pair creations and annihilation rates, and the rates of growth of complexity.  There are actually several reasons one might conjecture that such a connection between them exists.  One reason is that the only physical process that still would occur even after the thermal equilibrium is just the quantum fluctuations which could be the main source of increasing the complexity. Also, it has been found that the complexity growth rate of black holes and also the worldsheet of a string dual to entangled particle-antiparticle would both saturate the Lyapunov bound \cite{deBoer:2017xdk}. Also, generally increasing the energy of the system would increase complexity growth rate. 

To further examine this idea, in section \ref{sec:one}, we study the potential wells and barriers of different fields in different models and also the complexity growth rates in them and in fact we observe such a relationship.

To learn more about the nature of holographic complexity, one could study its behavior in more exotic cases \cite{Ghodrati:2014spa, Zhang:2017nth, Ghodrati:2015rta, Zangeneh:2017tub, Ghodrati:2016ggy,Camargo:2018eof, Ghodrati:2016tdy, Ghodrati:2016vvf,Alishahiha:2018tep} such as various phase transitions and then try to establish the connections with other physical quantities of the system. This quest is the main purpose of this work. 

In \cite{Zhang:2017nth},  the complexity in an interesting QCD model has been studied and the similarities between the first, second and crossover phase transitions both in entropy and complexity growth rate have been observed. Therefore, in that work, it has been shown that complexity could in fact act as a probe of different phase transitions and specially could probe the confinement. Also, in \cite{Camargo:2018eof}, using models of quantum harmonic oscillators, it has been shown that complexity can act as a probe of quantum quenches and can even capture features that entanglement entropy is not capable of probing.  Along the works of those papers, we would like to study the complexity growth rates in various phase transition setups. 

Therefore, first, in section \ref{sec:two}, we study charged dyonic black hole model which are the system which could mimic the properties of van der Waals fluid, and already using the Gibbs free energy, the first and second order phase transitions have been observed in them. Using the full time dependent complexity growth rate, we also observe phase transition in the complexity growth rate of this model. 

Then in section \ref{sec:soliton}, we consider AdS and AdS soliton backgrounds, study their potentials, the complexity growth rate in each background and examine the connection between the phase transitions such as tachyon condensation and complexity growth rate behaviors specifically in the IR and UV regions and again illustrate the connection. We also compare the Schwinger effect phase diagrams in these two cases and comment on the relationship with the complexity.

Finally, in section \ref{sec:three}, we study the phase transitions in the Gubser model of QCD. In this model, by just fine tuning the parameters of the confining potential, in the diagram of entropy versus temperature, one could obtain crossover, first and second order first transitions for the $V_{\text{QCD}}$, $V_1$ and $V_2$ respectively. Also, we study another model of improved holographic QCD, $V_{\text{IHQCD}}$, which could present a first order phase transition while give a more effective model for the dynamical properties of QCD such as asymptotic freedom and a more realistic bulk viscosity. This model shows  substantial differences in the behavior of quasinormal modes and also complexity, which we will discuss in details.

Therefore, we study the behavior of entropy, complexity growth rate, speed of sound and potential for all these four models and compare the behavior of them at lower temperatures and specifically near the phase transitions. We also outline the relationship between the cross over from hydrodynamic to non-hydrodynamic modes of each model, and the resulting topological jump in the complexity growth rate around the phase transition point.

Finally, we conclude with a discussion in section \ref{sec:discuss}.

\section{The relationship between physical parameters and complexity growth rates}\label{sec:one}

By comparing the results for the complexity growth rates in different black hole solutions with different fields, in this section, we study the connections between action growth rates and mass, charges with different physical properties and its coupling to gravity. Specifically we study solutions with dilaton, Maxwell, and phantom fields which has ghost and compare the behaviors. 

 So in this section we first review the already calculated results for different backgrounds which have fields with different natures, and then we discuss the connection between the complexity growth rate and the parameters of the solution and the physical interpretations.

\subsection{Charged dilaton black hole in AdS space}

First, we consider the Einstein-Maxwell-dilaton model \cite{Cai:2017sjv}
\begin{gather}
S=\frac{1}{16\pi} \int d^4 x \sqrt{-g} ( R-2 (\partial \phi)^2 -V(\phi) -e^{-2\phi} F^2),
\end{gather}
where the potential $V(\phi)$ for the dilaton field is
\begin{gather}\label{eq:pot1}
V(\phi)= -\frac{4}{l^2}-\frac{1}{l^2} [ e^{2(\phi-\phi_0)}+e^{-2(\phi-\phi_0)}].
\end{gather}

Here $\phi_0$ is a constant and $l$ is the AdS radius.  So, this potential is the combination of a constant value and two Liouville-type potentials. 
 \begin{figure}[ht!]
 \centering
  \includegraphics[width=5.5cm] {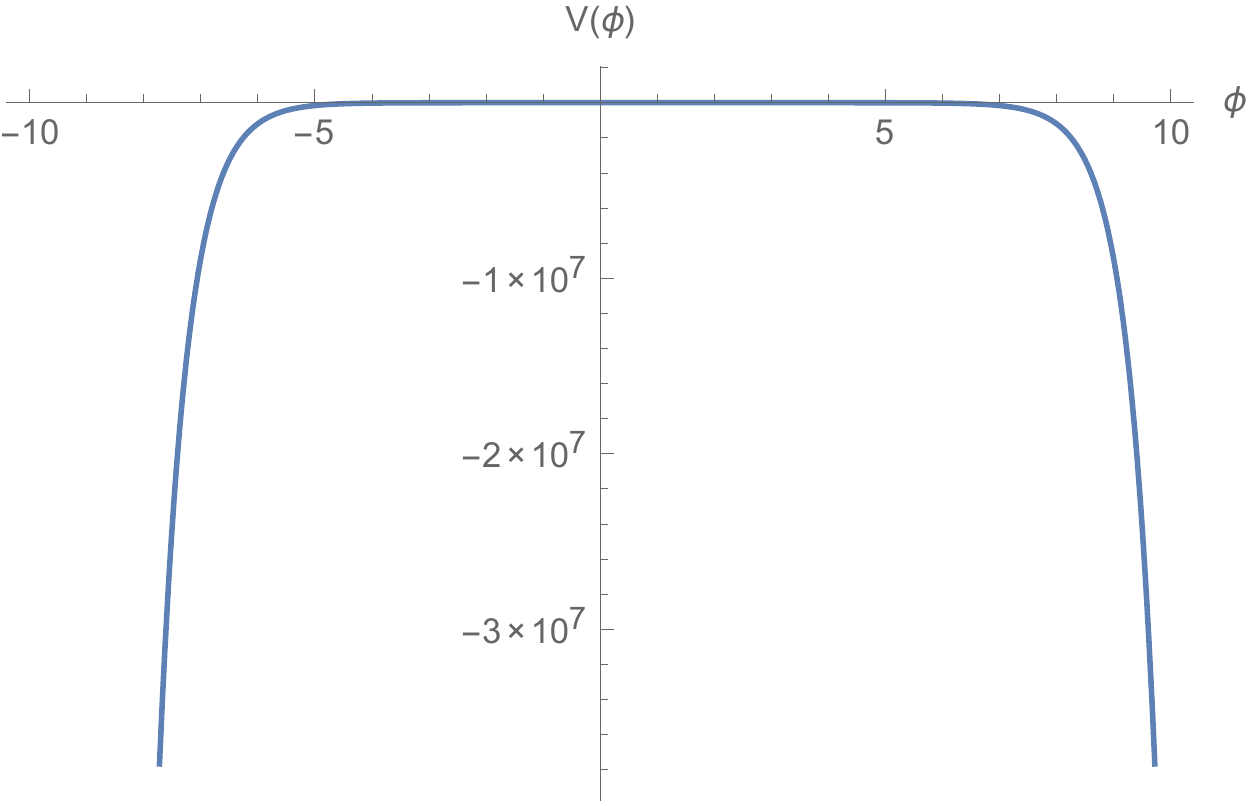}
  \caption{Two Liouville-type potential \ref{eq:pot1} , for $l=\phi_0=1$. }
 \label{fig:dcdt}
\end{figure}
The spherically symmetric charged dilaton black hole solution of this action would be
\begin{gather}
ds^2=-f( r  ) dt^2+f^{-1}( r  )dr^2+ U^2( r  ) d\Omega^2, \nonumber\\
F_{rt}= \frac{Q e^ {2\phi} }{U^2}, \ \ \ \ \ \ \ \  e^{2\phi}=e^{2\phi_0} \left(1-\frac{2D}{r} \right),
\end{gather}
where
\begin{gather}
f (  r ) =1-\frac{2M}{r}+\frac{r (r-2D)}{l^2}, \ \ \ \ \ \ \ 
U^2 ( r  )= r(r-2D), \ \ \ \ \ \ \ D=\frac{Q^2 e^{2\phi_0}} {2M},
\end{gather}
and $D$ is the dilaton charge and $\phi_0$ is the asymptotic constant value of dilaton \cite{Gao:2004tu}. Note that when $\phi=\phi_0= 0$, this solution reduces to the Reissner-Nordstrom black hole.

In \cite{Cai:2017sjv}, for the charged dilaton AdS black hole, the complexity growth rate at late times has been found as
\begin{gather}
\text{charged dilation AdS BH : } \ \ \ \ \ \dot{C} \sim \frac{dS}{dt}=2M-Q^2 e^{2\phi_0} \left( \frac{1}{2M}+\frac{1}{r_+} \right),
\end{gather}
where as one expects, for the case of $Q=0$, the result reduces to the one for AdS-Schwarzschild solution. Also, note that this black hole has only one horizon $r_+$, which is a function of $M$, $Q$ and $l$. It could simply be found by $f(r) \big |_{r=r_+}=0$, but as it is a long relation, we don't bring it here.

We want to study the effect of each parameter on the complexity growth rate while other parameters are fixed and try to find a physical explanation for the specific behavior.

\begin{figure*}[ht!]
        \centering
        \begin{subfigure}[b]{0.3\textwidth}
            \centering
            \includegraphics[width=5.5cm] {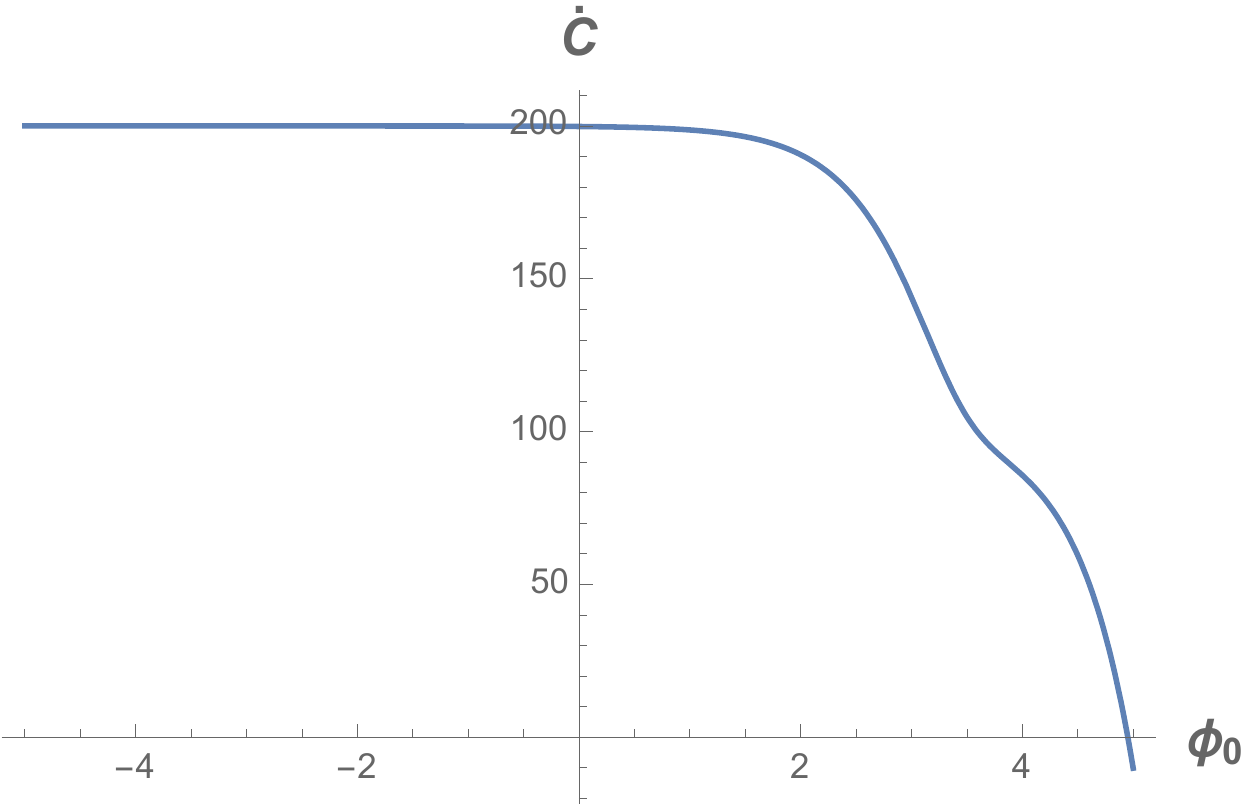}
            \caption[Network2]%
            {$M=100$, $Q=1$ }    
            \label{cdotphi}
        \end{subfigure}
          \quad
         \centering
        \begin{subfigure}[b]{0.3\textwidth}   
            \centering 
            \includegraphics[width=5.5cm] {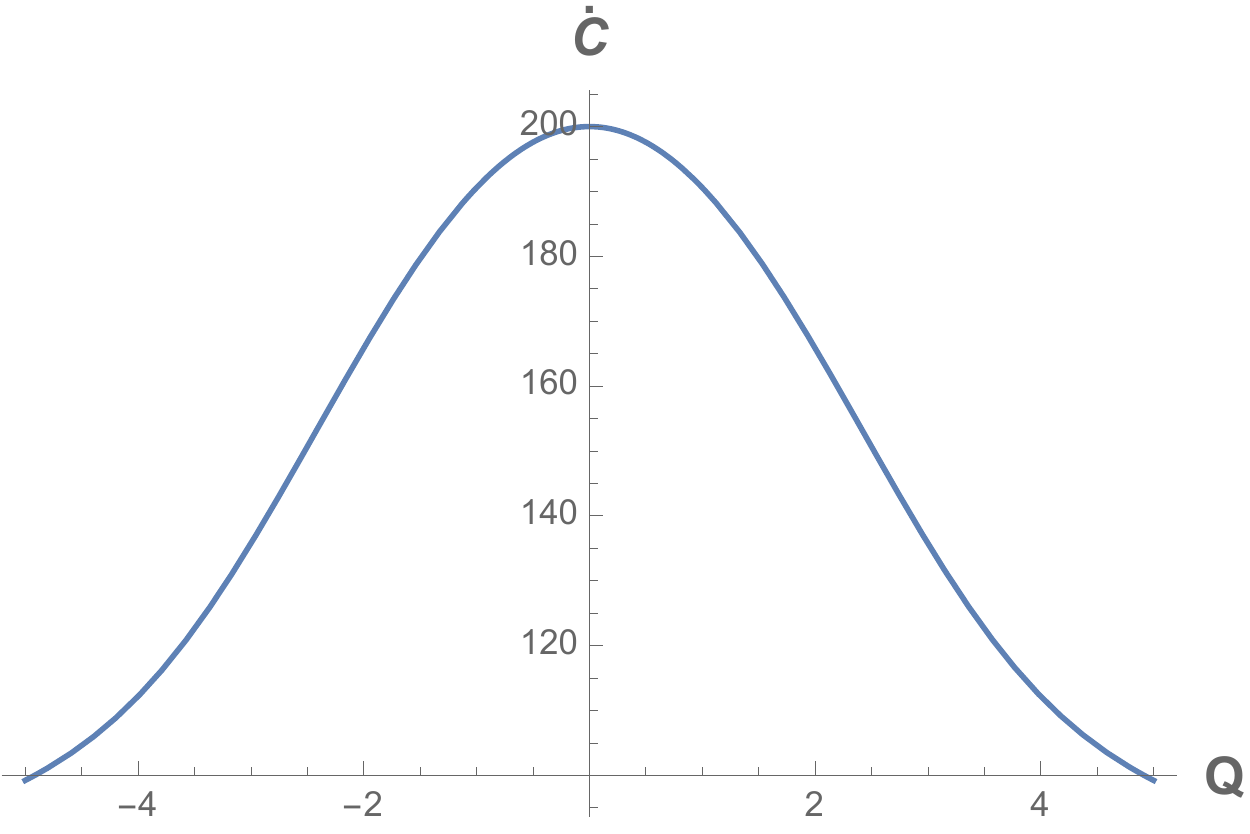}
            \caption[]%
            { $M=100$, $\phi_0=2$}    
            \label{cdotcharge}
        \end{subfigure}
           \begin{subfigure}[b]{0.3\textwidth}   
            \centering 
            \includegraphics[width=5.5cm] {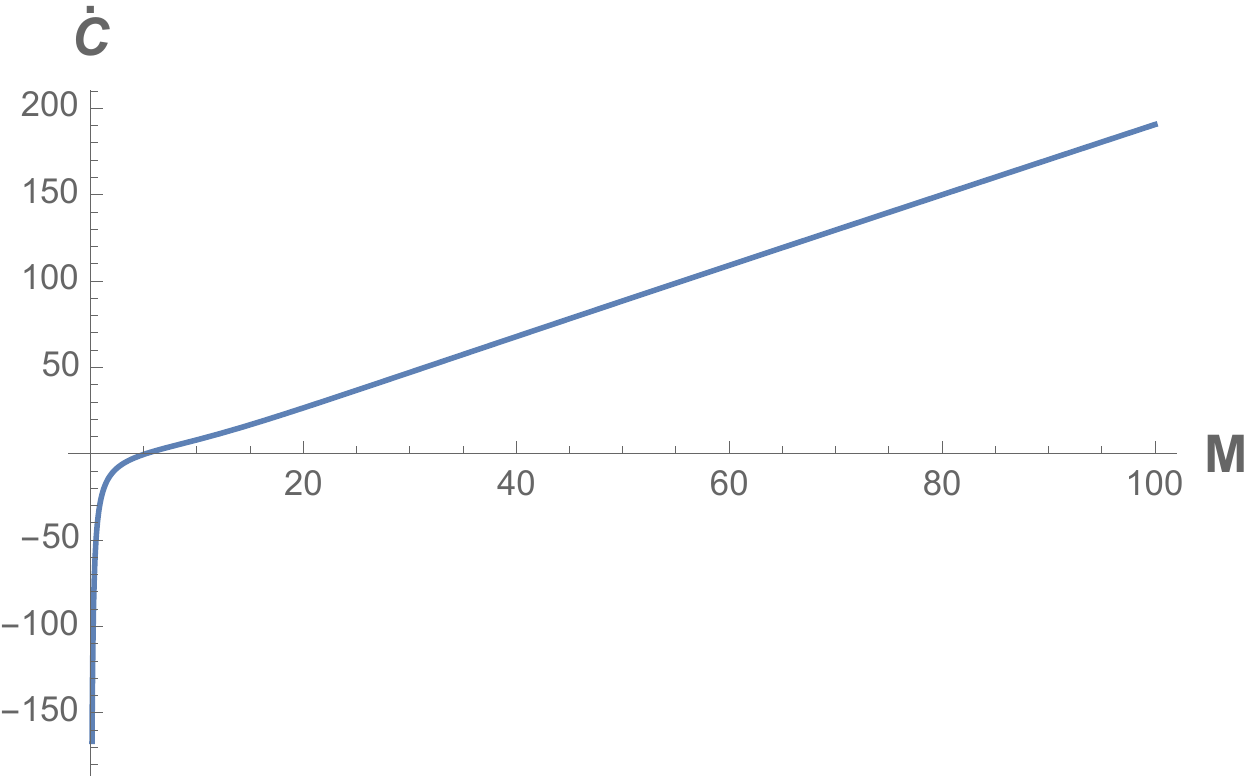}
            \caption[]%
            {$Q=1$, $\phi_0=2$}    
            \label{cdotmass}
        \end{subfigure}
        \caption{The dependence of complexity growth rate on each parameter of charged dilaton black hole. }
        \label{fig:ChargedDilaton}
    \end{figure*}

First, we keep $M$ and $Q$ constant and only study the effect of boundary dilaton field $\phi_0$ on the complexity growth rate.  The plot is shown in Fig. \ref{cdotphi}. (In all cases we take $l=1$.)

From Figure \ref{cdotphi}, one can notice that for the all negative and small positive $\phi_0$, the complexity is constant. Then at a specific positive value of $\phi_0$, it decreases exponentially. Therefore, as a first observation, one could say that generally by increasing the dilaton field, the complexity growth rate would decrease. The specific value of $\phi_0$ which this sudden decrease would happen, would decrease by increasing $Q$, but it does not change much with varying $M$. Though, increasing $M$ would increase greatly the initial value of $\dot{C}$ at negative dilaton fields.
 
Recently in \cite{An:2018xhv}, the effect of dilaton field on the complexity growth rate has been studied further where the \textit{full time} behavior of the action growth rate for the AdS dilaton black hole and also asymptotically Lifshitz black holes have been studied. Generally as it was shown in \cite{An:2018xhv} with another method, the dilaton field would actually slow down the rate of growth of complexity which we observed here as well.  In section \ref{sec:three}, the same result would be seen in models of confining potentials, so we conjecture it would be a universal effect.  

The physical reason behind this could be explained by noting that in string theory, the dilaton field actually describes how strongly open strings couple to one another. In fact, in perturbation theory, the coupling constant of open strings would be proportional to the exponential value of the dilaton's expectation value, $g_s =e^{<\phi>}$. So the more the strings are coupled to each other, the more difficult it would be to go from one state to another and the rate of computation and complexification would be lower. 

In Fig. \ref{cdotmass}, the relationship between complexity growth rate and mass is shown.  The explanation for the relation to mass would be easier. For bigger masses, there would be more degrees of freedom and correspondingly higher number of gates in the dual boundary and therefore the rate of complexity growth would be higher. In fact, the growth is linear which intuitively makes sense as mass and number of gates would linearly be connected to each other. 

The behavior of complexity growth rate with respect to charge, Fig. \ref{cdotcharge}, could also be explained by considering the relation between complexity growth rate and the height of potential barriers. Note that for the case of Schwarzschild black hole, which has no charge, the height of potential barrier is just equal to the temperature, while for the case of charged black holes, such as Reissner-Nordstrom (RN) black hole this barrier, due to its charge, is much higher, \cite{Brown:2018kvn}, which would cause the rate of complexity growth would be lower. 

In \cite{Brown:2018kvn}, the scrambling behavior in the field of a ``near-extremal charged black hole" has been studied. There it has been noted that due to the buildup of momentum of accelerating particle through the long throat of Reissner-Nordstrom geometry, some degrees of freedom would be decoupled and cannot enter in the process of computation and complexification. The more the charge of the black hole, the longer the throat of RN black hole would be, which causes more degrees of freedom decouple and therefore the complexity growth rate would fall down.

Also, one could note that a charge in the bulk would be dual to a current in the boundary field theory side. Higher amount of charge of the black hole would be dual to higher rate of current with higher momentum. The bigger momentum of current would actually decrease the complexification rate.

\begin{figure*}[ht!]
        \centering
        \begin{subfigure}[b]{0.3\textwidth}
            \centering
            \includegraphics[width=5.5cm] {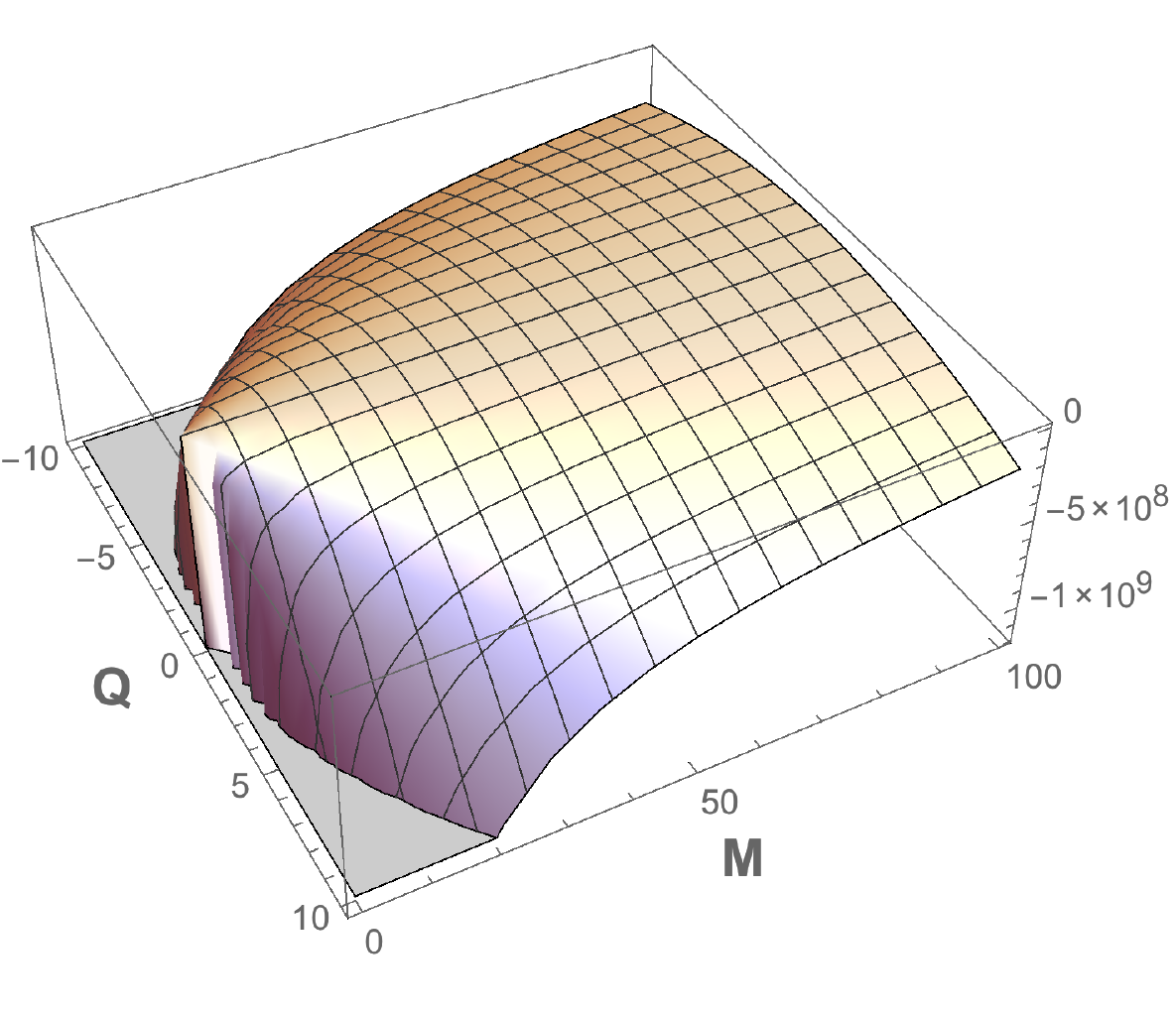}
            \caption[Network2]%
            {$\phi_0=10$ }    
            \label{QMi}
        \end{subfigure}
          \quad
         \centering
        \begin{subfigure}[b]{0.3\textwidth}   
            \centering 
            \includegraphics[width=5.5cm] {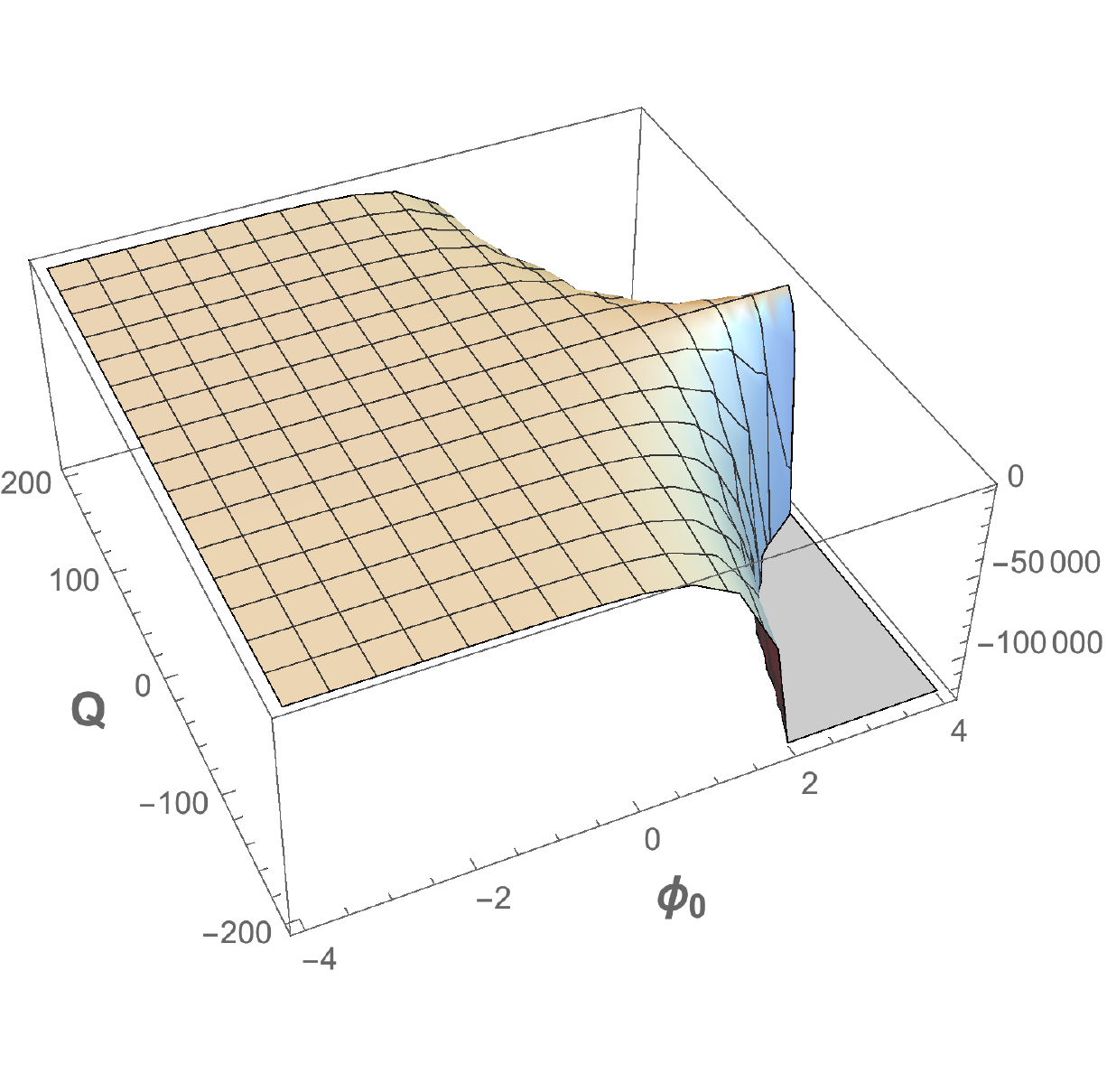}
            \caption[]%
            {$M=10$}    
            \label{Qphi}
        \end{subfigure}
           \begin{subfigure}[b]{0.3\textwidth}   
            \centering 
            \includegraphics[width=5.5cm] {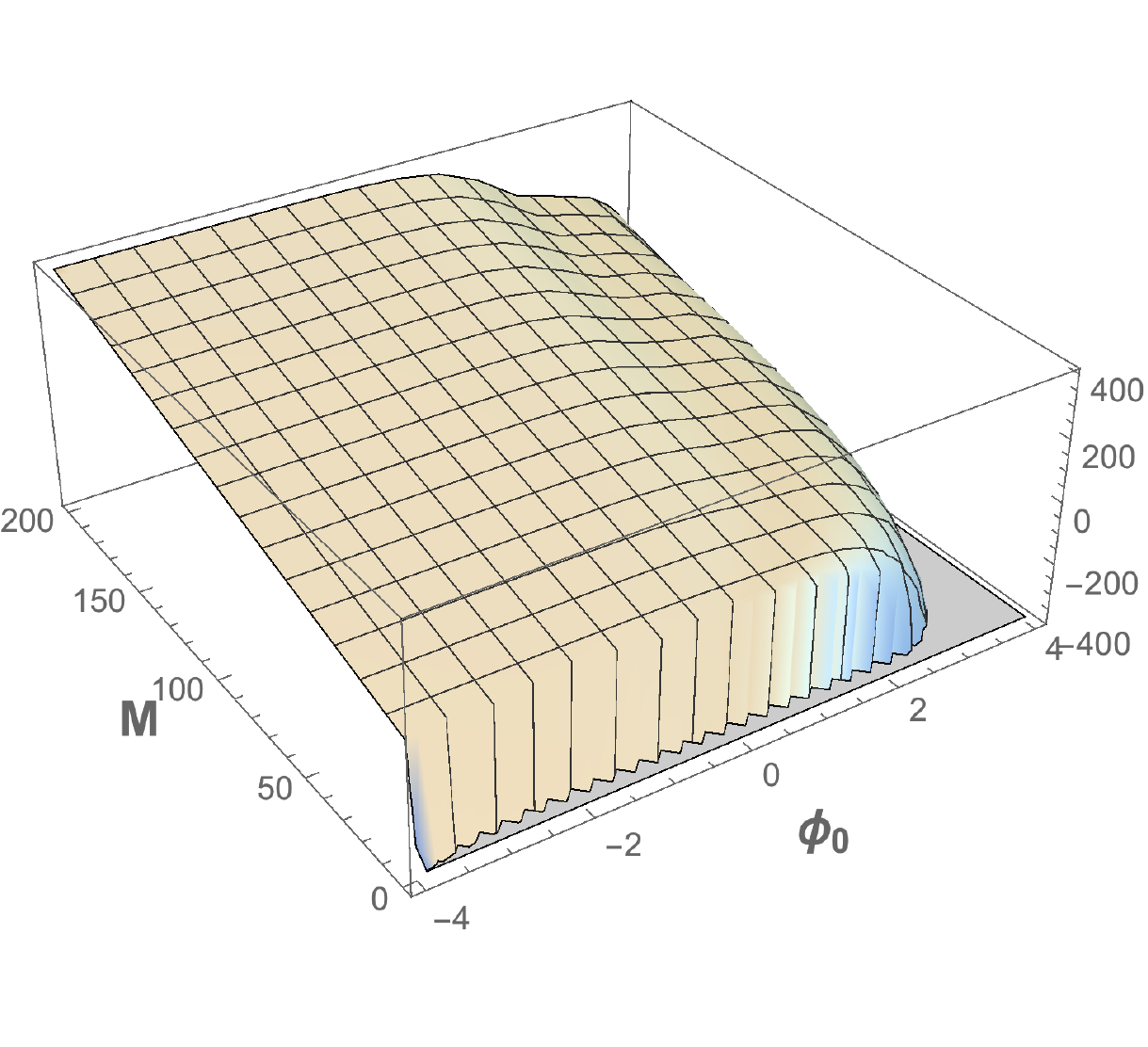}
            \caption[]%
            {$Q=5$}    
            \label{Mphi}
        \end{subfigure}
        \caption{The dependence of complexity growth rate on two parameters of charged dilaton black hole. }
        \label{fig:ChargedDilaton2}
    \end{figure*}

 The relationship of complexity with two parameters of black hole keeping the third one constant, is shown in Fig \ref{fig:ChargedDilaton2}.  From figure \ref{QMi}, one could notice that for higher masses, changing charge would not change complexity growth rate much while for lower masses, increasing charge would have a dramatic effect on decreasing the rate. 

From figure \ref{Qphi}, one can notice that when the dilaton field is negative or very small and close to zero, even for higher charges, the complexity growth rate would not change. This is because for small and negative $\phi_0$, the string couplings to each other would be very weak and therefore even for high charges the rate of growth of complexity could not change. While, increasing dilaton would increase the couplings between string modes, which subsequently makes the dependence of growth rate to change of charge very strong.

From figure \ref{Mphi}, one can again see that for negative dilaton field, the rate of complexity growth is constant and even increasing mass $M$ cannot change this flat behavior. However, increasing mass at each $\phi_0$ would linearly increase the rate of growth of complexity. Also, note that in bigger masses, increasing dilaton field would cause a milder fall down of the rate of complexity growth.

\subsection{Born-Infeld black hole in AdS Space}
Another example of charged black hole in AdS is Born-Infeld, with the following action
\begin{gather}
S=\frac{1}{16\pi} \int d^4x \sqrt{-g} \left[R+\frac{6}{l^2}+4\beta^2 \left(1-\sqrt{1+\frac{F_{\mu \nu} F^{\mu \nu} }{2 \beta^2}} \right) \right],
\end{gather}
where one can get a static symmetrically symmetric black hole solution as follows
\begin{gather}
ds^2=-f (  r) dt^2+f^{-1} (  r) dr^2+r^2 d\Omega^2, \  \ \ \ \ \ \
F_{rt}=\frac{Q}{\sqrt{r^4+Q^2/\beta^2}},
\end{gather}
where
\begin{gather}
f( r )=1-\frac{2M}{r}+\frac{r^2}{l^2}+\frac{2\beta^2}{r} \int_r^\infty dx \left(\sqrt{x^4+Q^2/\beta^2}-x^2\right)   \nonumber\\ =1-\frac{2M}{r}+\frac{r^2}{l^2}+\frac{2\beta^2}{r} \left [I( r=\infty; a^2=\frac{Q^2}{\beta^2}-I(r; a^2=\frac{Q^2}{\beta^2} )   \right],
\end{gather}
and $I(r;a^2) \equiv \int^r dx (\sqrt{x^4+a^2}-x^2)$. 

Here $M$ and $Q$, are the mass and charge of the black hole. Also note when $\beta \to \infty$, the Born-Infeld theory reduces to the Maxwell theory with $\mathcal{L} (F)=-F^2$. The chemical potential of this black hole could also be obtained as
\begin{gather}
\mu (r ) Q = \frac{Q^2}{r } {}_2F_1[\frac{1}{4},\frac{1}{2}, \frac{5}{4}, -\frac{Q^2}{\beta^2 r^4}].
\end{gather}

As found in \cite{Cai:2017sjv} when $\beta^2 Q^2 \ge 1/4$ the black hole solution has two horizons and when $\beta^2 Q^2 < 1/4$, it has only one horizon where the inner one is absent. So changing the coupling $\beta$ or the charge $Q$ could greatly impact the complexity as it significantly changes the geometry. 

For the case with two horizons, \cite{Cai:2017sjv} found the action growth as
\begin{gather}
\frac{dS}{dt}=\left(\frac{Q^2}{r} {}_2F_1[\frac{1}{4},\frac{1}{2}, \frac{5}{4}, -\frac{Q^2}{\beta^2 r^4}]  \right) \bigg |_{r_+}^{r_-}= \mu_- Q- \mu_+ Q,
\end{gather}
and for the case with one horizon, it would be
\begin{gather}
\frac{dS}{dt} =2M-\frac{Q^2}{r_+}  {}_2F_1[\frac{1}{4},\frac{1}{2}, \frac{5}{4}, -\frac{Q^2}{\beta^2 r_+^4}] -\beta^{\frac{1}{2}} Q^{\frac{3}{2}} \frac{\Gamma(\frac{1}{4}) \Gamma(\frac{5}{4})}{ 3\Gamma(\frac{1}{2})},
\end{gather}
where
\begin{gather}
r_+= \frac{l^2}{12 \beta ^2 l^2+9}\left( -3-2 \beta ^2 l^2+2 \sqrt{\beta ^4 l^4+3 \beta ^2 Q^2 \left(4 \beta ^2 l^2+3\right)}\right)
\end{gather}

\begin{figure}[ht!]
 \centering
  \includegraphics[width=6cm] {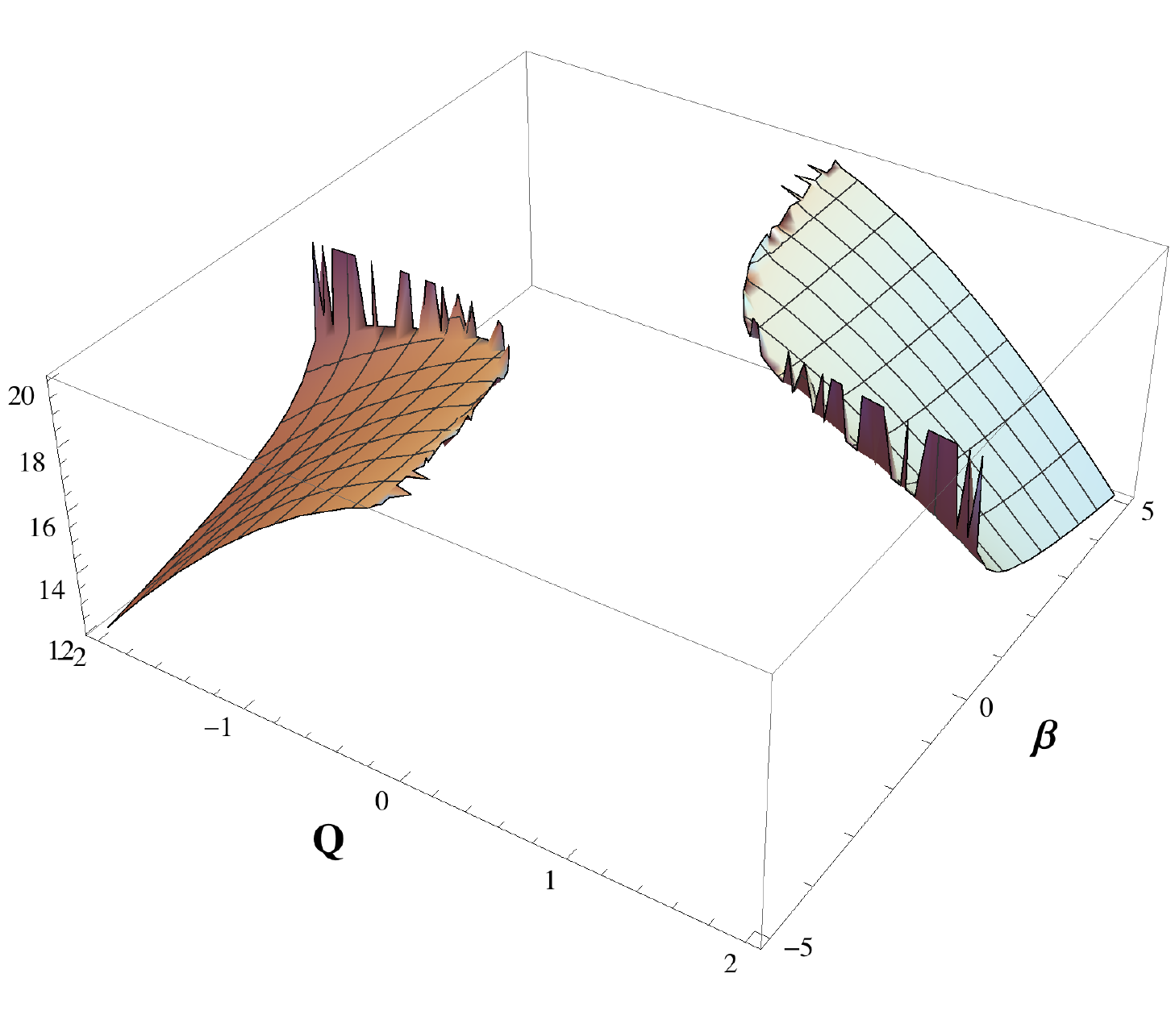}
  \caption{Complexity growth rate for the case of black hole with one horizon in Born-Infeld theory, $M=10$, $l=1$. }
 \label{fig:oneHorizon}
\end{figure}

Again, one can see the relationship with respect to mass $M$ is linearly increasing. From figure \ref{fig:oneHorizon}
one can see that in this case similar to previous case, increasing charge $Q$ would decrease the rate of growth of complexity, while increasing the coupling constant $\beta$, which couples the gauge field $F_{\mu\nu}$ to the metric $g$, would decrease the rate of growth of complexity.

Note that due to a strong singularity in this case, at low charges, low temperatures, or low $\beta$, the complexity growth would not be smooth which could be seen from the plot. 

\subsection{Charged black hole with phantom Maxwell field}
As another example, we consider the action of Einstein-phantom-Maxwell theory with a negative cosmological constant as 
\begin{gather}
S=\frac{1}{16\pi} \int d^4x \sqrt{-g} (R+\frac{6}{l^2}+F_{\mu \nu} F^{\mu \nu}),
\end{gather}
where its solution would be
\begin{gather}
ds^2=-f( r  ) dt^2 +f^{-1} ( r  ) dr^2+r^2 d\Omega^2, \ \ \ \ \ 
F_{tr}=\frac{Q}{r^2}, \ \ \ \ 
f(  r )=1+\frac{r^2}{l^2}-\frac{2M}{r}-\frac{Q^2}{r^2}.
\end{gather}
Note that in this case the maxwell term has a wrong sign. Its Penrose diagram is the same as AdS Schwarzschild black hole. The action growth rate has been found as $\frac{dS}{dt}=2M+\mu_+ Q$.

 \begin{figure}[ht!]
 \centering
  \includegraphics[width=6.5cm] {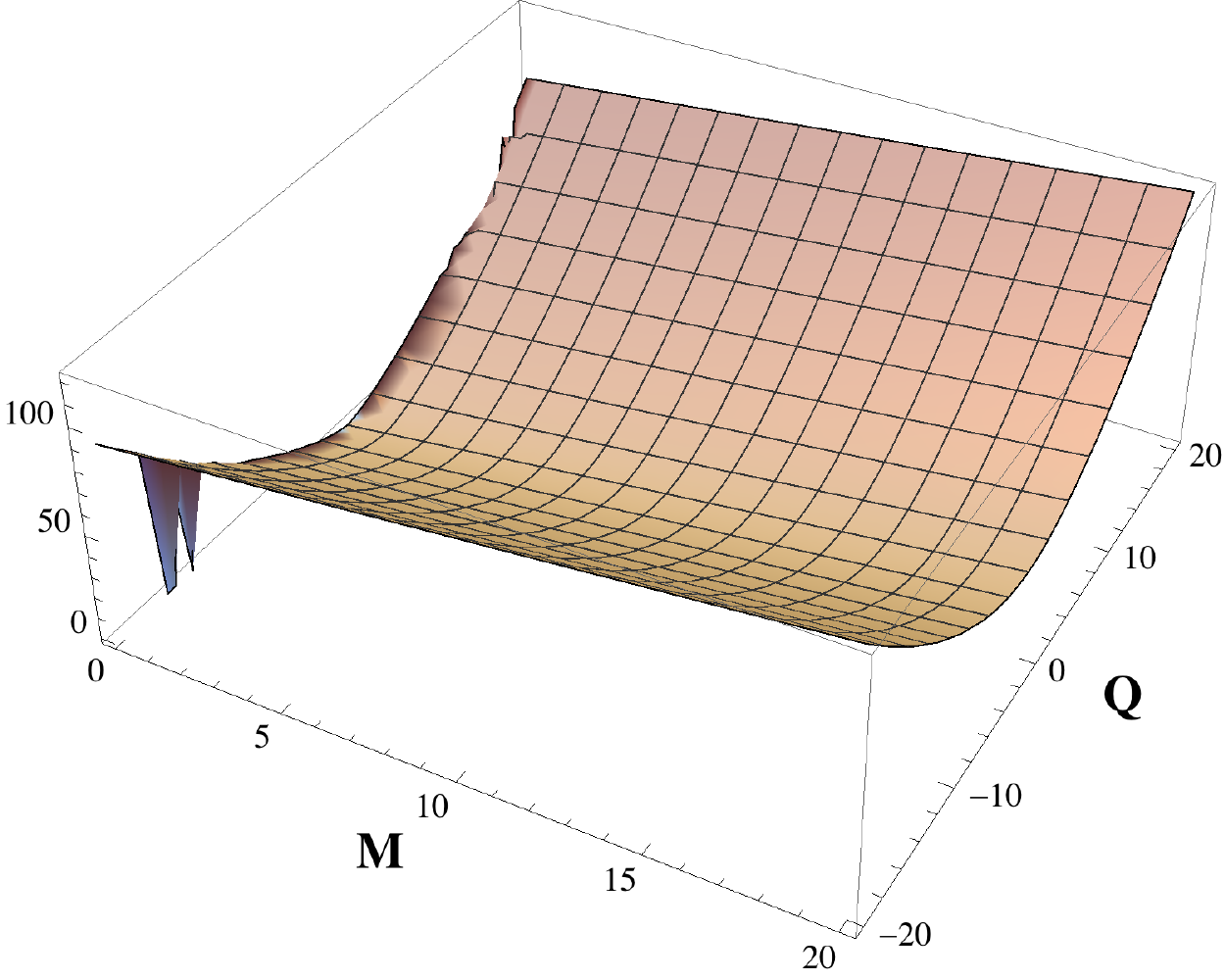}
  \caption{Complexity growth rate for the case of phantom Maxwell field, with ghost. }
 \label{fig:ghost}
\end{figure}

From figure \ref{fig:ghost}, one can see that increasing $M$ would increase the complexity growth rate linearly, however unlike previous cases, here increasing charge $Q$ would increase the rate of complexity growth.

 Note that the phantom Maxwell field is different from other cases in significant other aspects. As mentioned before, the Maxwell term has a wrong sign. Also, other cases satisfied the Lloyd bound at late times but this case violate it, even at late times and it is because the phantom field is actually a ``ghost field" which violates the strong energy condition. So generally one could say that the charges which violates the strong energy condition such as ghosts, would increase the rate of growth of complexity.
  \begin{figure}[ht!]
 \centering
  \includegraphics[width=6.5cm] {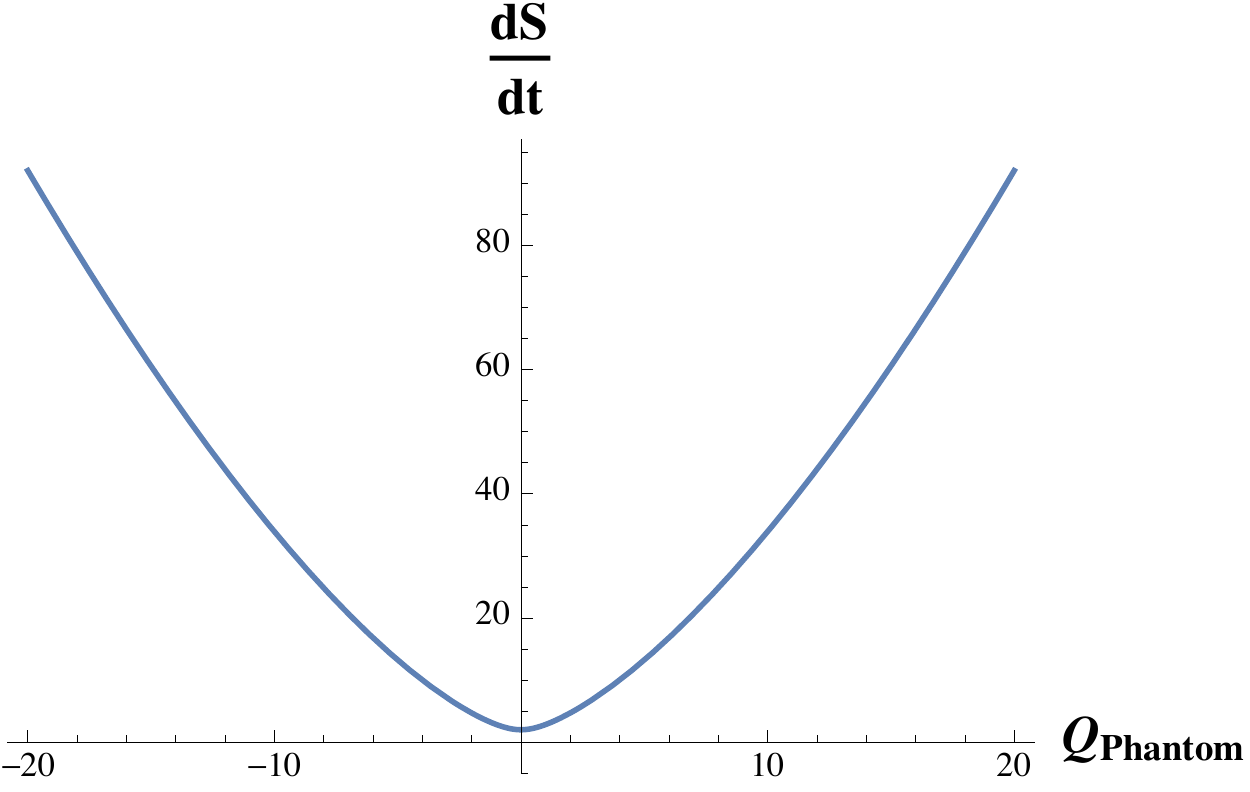}
  \caption{The behavior of complexity growth rate for the case of phantom Maxwell field (ghost) is opposite of other cases like \ref{cdotcharge} which has physical charges. }
 \label{fig:phantomfield}
\end{figure} 

\section{Phase transitions and complexity growth rate with charge or kink}

We now study the full time behavior of complexity growth rate for the dyonic and AdS soliton black holes and compare them with their different phase transitions.  We find that generally the complexity growth rate could probe any kind of phase transition and instability.  
 
\subsection{Complexity in dyonic black holes}\label{sec:two}

First, we consider the dyonic black hole in the holographic setup similar to \cite{Dutta:2013dca}. We chose this type of black hole as it has both electric and magnetic charges. It could be a solution to the Reissner-Nordstrom action with the negative cosmological constant,
\begin{gather}
I=\frac{1}{16\pi } \int d^d  \sqrt{g} \left [\mathcal{R}-2\Lambda-F^{\mu\nu}F_{\mu \nu} \right],
\end{gather}
where the static spherically symmetric solution would be
\begin{gather}
ds^2=-f ( r  ) dt^2 +\frac{dr^2}{ f (  r )} +r^2 d\Omega^2, \ \ \ \ \ \ f (  r)  = 1-\frac{\Lambda r^2}{3}-\frac{2m}{r}+\frac{q_E^2+q_M^2}{r^2}, \nonumber\\ 
A=\left( - \frac{q_E}{r}+\frac{q_E}{r_+} \right)dt+(q_M \cos \theta ) d \phi.
\end{gather}

The charge and mass of this solution are
\begin{gather}
M=\frac{2\Omega_2}{16\pi} \left( r_+ + r_- +\frac{1}{l^2} \frac{r_+^4-r_-^4}{r_+-r_-} \right), \ \ \ \ \ \ 
Q^2=\frac{2\Omega_2}{8\pi} r_+ r_- \left(1+\frac{1}{l^2} \frac{r_+^3-r_-^3}{r_+-r_-} \right),
\end{gather}
and the chemical potentials would be $\mu_-=Q/r_-$, $\mu_+=Q/r_+$ and also $Q^2 =q_M^2+ q_E^2$. 

The complexity growth rate for this solution \textit{at late times} has been found as \cite{Ovgun:2018jbm}
\begin{gather}
\frac{dS_{bk+bd}}{dt}=(M-\mu_+ Q)-(M-\mu_- Q).
\end{gather}

\begin{figure}[ht!]
 \centering
  \includegraphics[width=6.5cm] {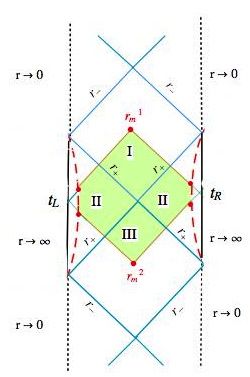}
  \centering
  \caption{Penrose diagram of the dyonic charged black hole. The Wheeler-DeWitt patch is shown in green. }
 \label{fig:potentials1}
\end{figure}

Now we find the full, time dependent, holographic complexity. First, we divide the bulk action as  
\begin{flalign*}
I_{I}^{\text{bulk}}=&\frac{\Omega_2}{8\pi G_N} \int_{r_+}^{r_{\text{max}}} dr \left[ -\frac{2r^3}{l^2}-\frac{2r_+ r_-}{r}\left( 1+\frac{1}{l^2} \frac{r_+^3-r_-^3}{r_+ - r_-}\right) \right]\left(\frac{t}{2}+r^*(0)-r^* (  r ) \right), \nonumber\\
I_{II}^{\text{bulk}} =& \frac{\Omega_2}{4\pi G_N} \int_{r_-}^{r_+} dr \left[ -\frac{2r^3}{l^2}-\frac{2 r_+ r_ -}{r} \left( 1+\frac{1}{l^2} \frac{r_+^3 -r_-^3}{r_+ -r_-} \right) \right] \left( r^* (0) -r^* (r  )  \right), \nonumber\\
I_{III}^{\text{bulk}}= &\frac{\Omega_2}{8\pi G_N} \int_{r_+}^{r_m} dr \left [ - \frac{2r^3}{l^2}- \frac{2 r_+ r_-}{r} \left ( 1+ \frac{1}{l^2} \frac{r_+^3 -r_-^3}{r_+ -r_-}  \right) \right ] \left(\frac{-t}{2} +r^*(0)-r^* ( r  ) \right ).
\end{flalign*}

The sum of these terms would be
\begin{flalign*}
I^{\text{bulk}}=&\frac{\Omega_2}{8\pi G_N} \int_{r_m}^{r_{\text{max}}} dr \left[ -\frac{2r^3}{l^2}-\frac{2r_+ r_-}{r}\left( 1+\frac{1}{l^2} \frac{r_+^3-r_-^3}{r_+ - r_-}\right) \right]\left(\frac{t}{2}-r^*(0)+r^* (  r ) \right), \nonumber\\+& \frac{\Omega_2}{4\pi G_N} \int_{r_-}^{r_{\text{max}}} dr \left[ -\frac{2r^3}{l^2}-\frac{2r_+ r_-}{r}\left( 1+\frac{1}{l^2} \frac{r_+^3-r_-^3}{r_+ - r_-}\right) \right]\Big(r^*(0)-r^* (  r ) \Big),
\end{flalign*}

By choosing the following normal vectors, 
\begin{gather}
k_1^a=\alpha \left( \frac{1}{f(  r )} (\partial_t)^a+(\partial_r)^a \right),\ \ \ \ \
k_2^a=\beta \left(-\frac{1}{f( r )  } (\partial_t)^a+(\partial_r)^a \right ),
\end{gather}
and the joint action
\begin{gather}
I^{\text{joint}} =\frac{1}{8\pi G_N} \int d^d x \sqrt{\gamma} \log \Big | \frac{k_1.k_2}{2} \Big|,
\end{gather}
the contribution from this term at $r=r_m$ would then be found as
\begin{gather}
I^{\text{joint}} =\frac{\Omega_2}{8\pi G_N} r_m^2 \log \Big | \frac{\alpha \beta}{f( r_m )  } \Big |.
\end{gather}

Also, for the counter term,
\begin{gather}
\frac{1}{8\pi G_N} \int d\lambda d^d x \sqrt{\gamma} \log \frac{\Theta}{d},
\end{gather}
we find
\begin{gather}
\Theta=\frac{1}{\sqrt{\gamma} } \frac{\partial \sqrt{\gamma} }{\partial \lambda}=\frac{2\alpha}{r}.
\end{gather}

Here, $\lambda$ is the affine parameter for the null surface. For the null vector $k_1$ it would be
\begin{gather}
\frac{\partial r}{\partial \lambda}=\alpha,	
\end{gather}

So finally we get
\begin{gather}
I^{ct}= \frac{\Omega_2}{8\pi G_N} r_m^2 \log \left | \alpha \beta \right | +\frac{\Omega_2}{8\pi G_N} \left( r_m^2-2 r_m^2 \log {r_m} \right),
\end{gather}
where the first term can eliminate the ambiguity coming from the normalization factors of the null vectors.
Summing all of these terms, taking the time derivative, removing some remaining constant terms, and using $\frac{dr_m}{dt}=-\frac{f(r_m)}{2}$, we finally get the relationship
 \begin{gather}
 \frac{d}{dt}\mathcal{C} \propto  -\frac{f(r_m) r_m^3}{16} \left( 1+ 4 \log r_m \right) + f(r_m) r_m \log (\frac{f(r_m)}{r_m^2} ) +\frac{r_m^2 f'(r_m)}{2}.
  \end{gather}

   \begin{figure*}[ht!]
        \centering
        \begin{subfigure}[b]{0.47\textwidth}
            \centering
            \includegraphics[width=\textwidth]{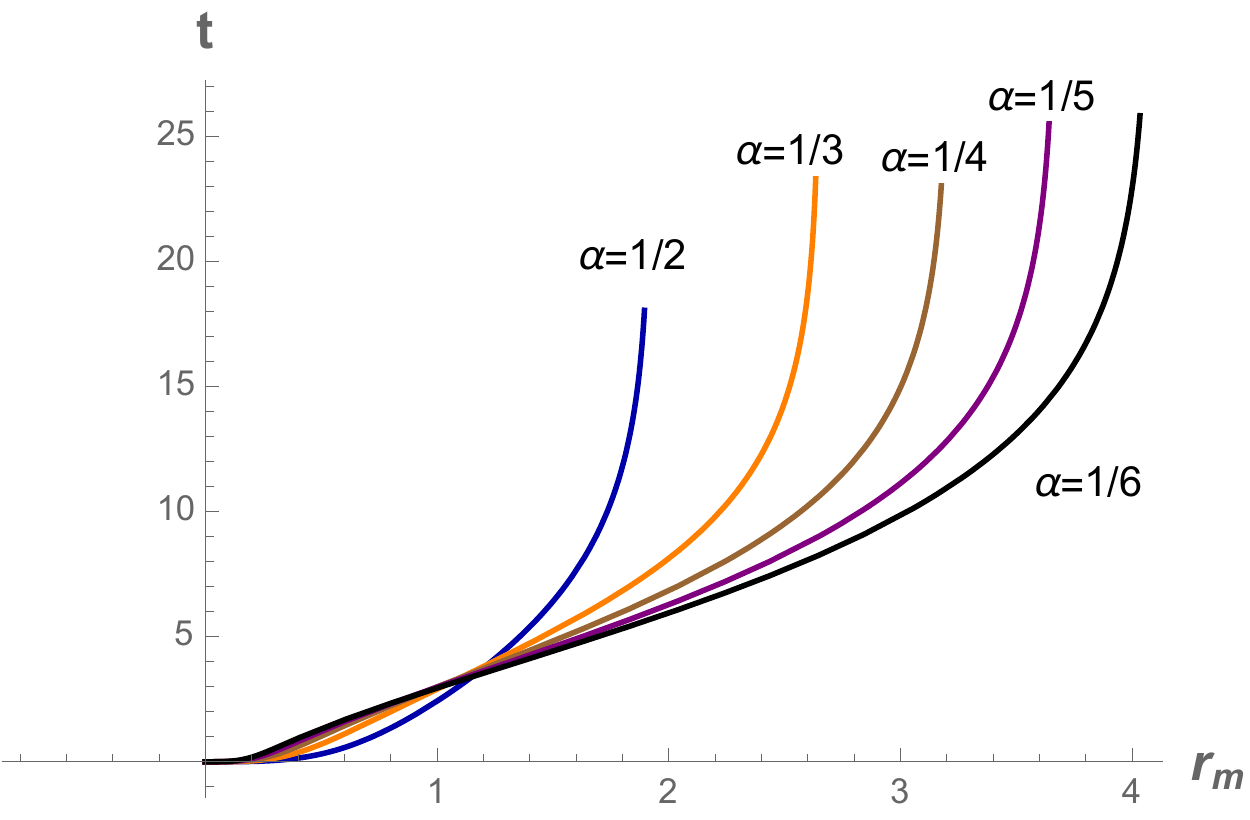}
            \caption[Network2]%
            {{\small time versus $r_m$}}    
            \label{tdyonic}
        \end{subfigure}
            \quad
         \centering
        \begin{subfigure}[b]{0.47\textwidth}   
            \centering 
            \includegraphics[width=\textwidth]{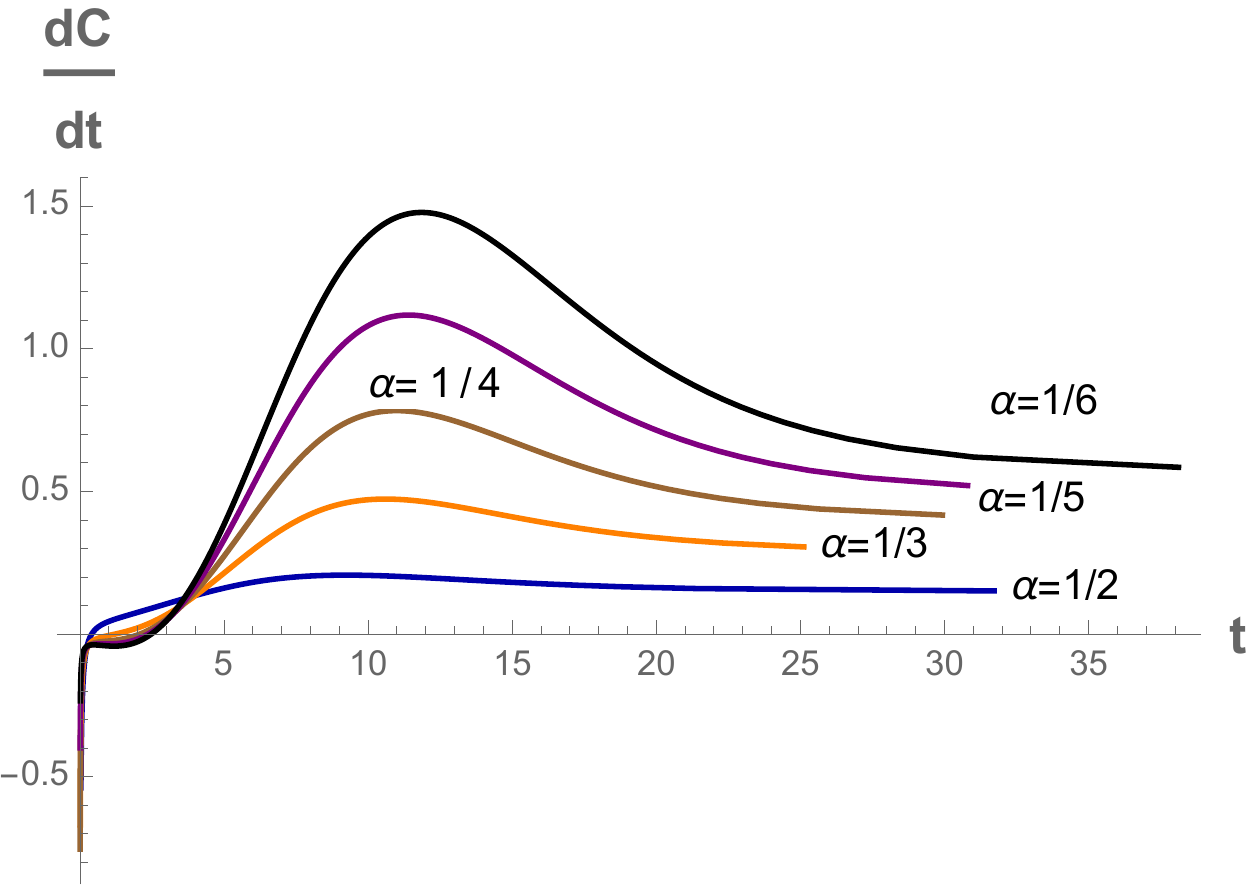}
            \caption[]%
            {{\small complexity growth rate versus time}}    
            \label{cdyonic2}
        \end{subfigure}
        \caption[  ]
        {\small The full time behavior of complexity growth rate of dyonic black holes.} 
        \label{fig:mean and std of nets}
    \end{figure*}

  The plots for the behavior of complexity of dyonic black holes are shown in Fig \ref{cdyonic2} where $\alpha$ is just a constant we chose which is defined as $ \Lambda=m=q_E=q_M= \alpha $.

 From these diagrams, one can see that the Lloyd's bound at early times would be violated and at later times it would be saturated from above similar to other studies in \cite{Chapman:2018dem,Chapman:2018lsv,Carmi:2017jqz, Alishahiha:2018tep}.

In addition, we could notice that by increasing the electric and magnetic charges, $q_E$, $q_M$,  the maximum of $\frac{dC}{dt}$ would increase and therefore the violation of Lloyd's bound at early times would become stronger. Also, the limit of Lloyd's bound at late times would increase as well. Note that this again could point to a relationship between quantum fluctuations like Schwinger mechanism and the complexity growth rate as we have conjectured before.  Of course to further study this relation, one could study the Schwinger effect for other charged black holes with varied setups of gauge fields. For instance, the electric and magnetic fields might have different directions with respect to each other and then one could study the complexity growth rate for those varied angle setups, and therefore further quantify this connection.
  
Additionally, from these diagrams, one could notice that, keeping the charges constant while increasing $m$ or $\Lambda$ would decrease the maximums in the plots of $\frac{dC}{dt}$ versus $t$, and also its final late time limit which is the Lloyd's bound. It would also decrease the time which is needed to reach to that specific Lloyd's bound.
  
\begin{figure*}[ht!]
        \centering
        \begin{subfigure}[b]{0.35\textwidth}
            \centering
            \includegraphics[width=\textwidth]{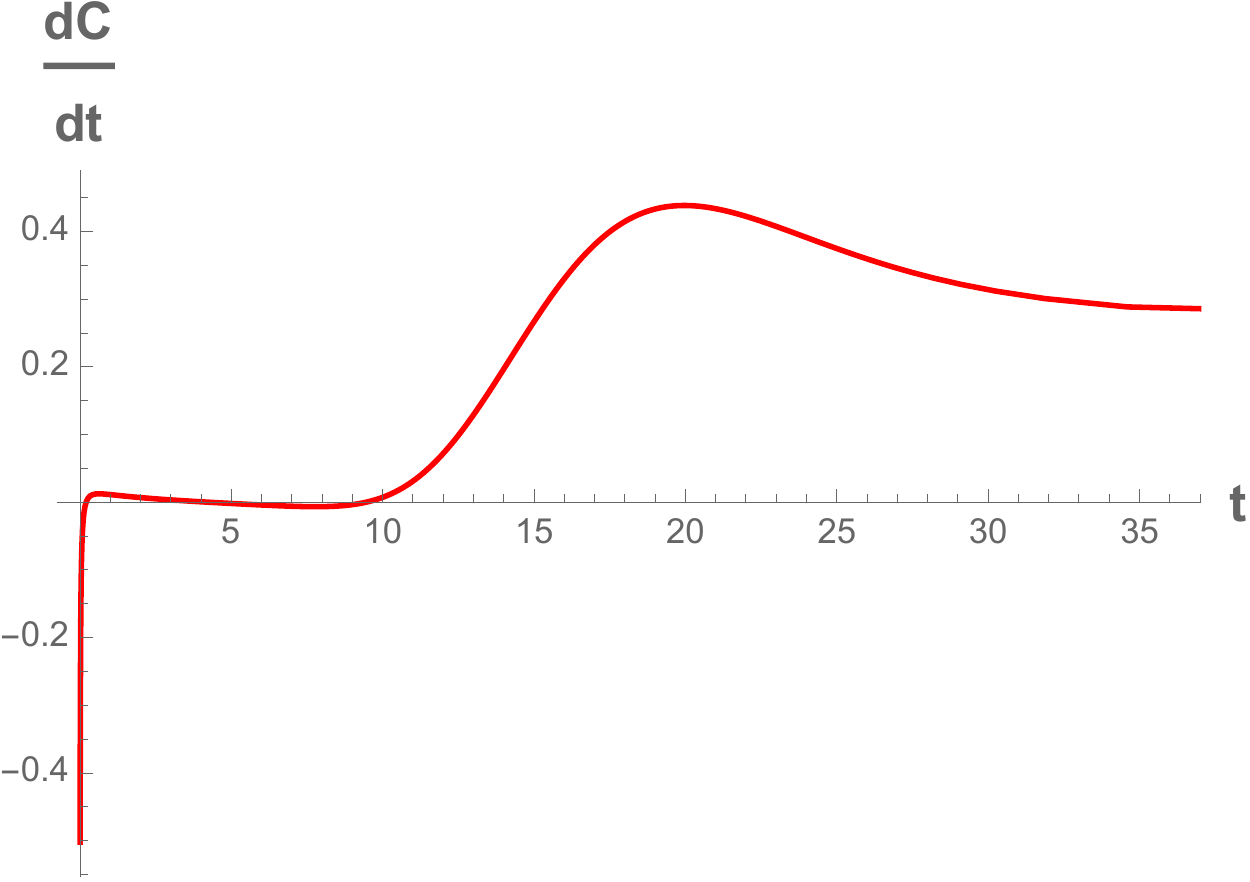}
            \caption[Network2]%
            {{\small  $q_E=q_M=\frac{1}{4}$}}    
            \label{lowCharge}
        \end{subfigure}
            \quad
         \centering
        \begin{subfigure}[b]{0.35\textwidth}   
            \centering 
            \includegraphics[width=\textwidth]{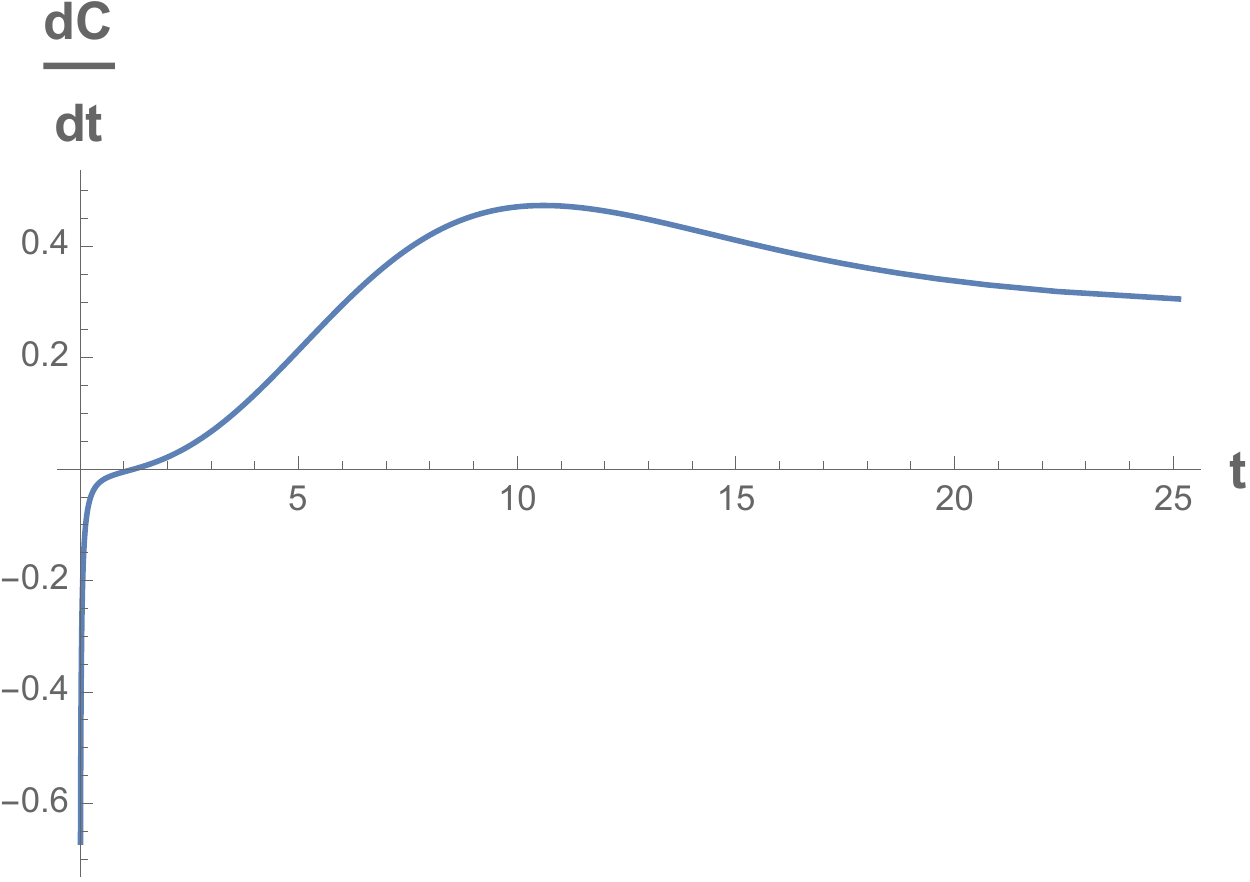}
            \caption[]%
            {{\small   $q_E=q_M=\frac{1}{3}$}}    
            \label{qhalf}
        \end{subfigure}
         \begin{subfigure}[b]{0.35\textwidth}   
            \centering 
            \includegraphics[width=\textwidth]{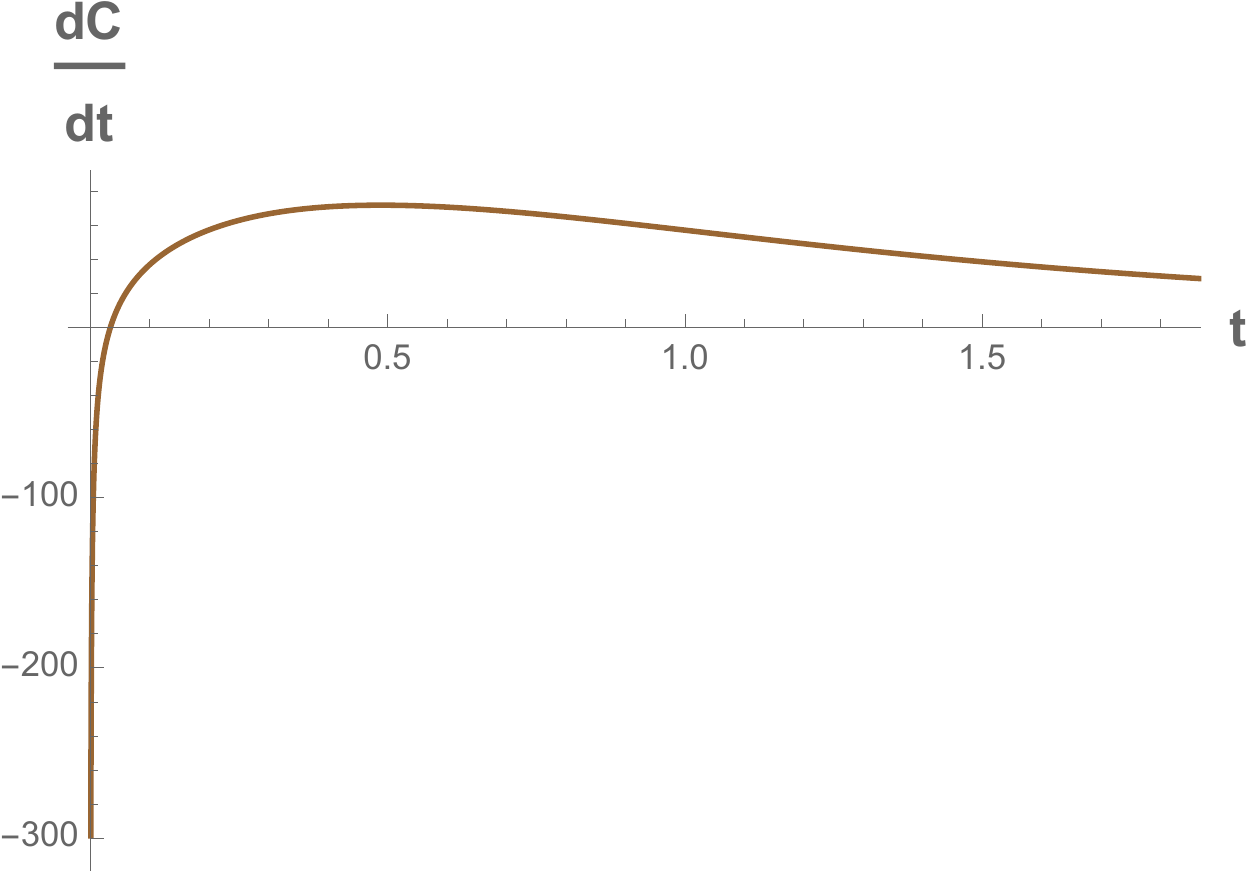}
            \caption[]%
            {{\small  $q_E=q_M=10$}}    
            \label{qten}
        \end{subfigure}
         \begin{subfigure}[b]{0.35\textwidth}   
            \centering 
            \includegraphics[width=\textwidth]{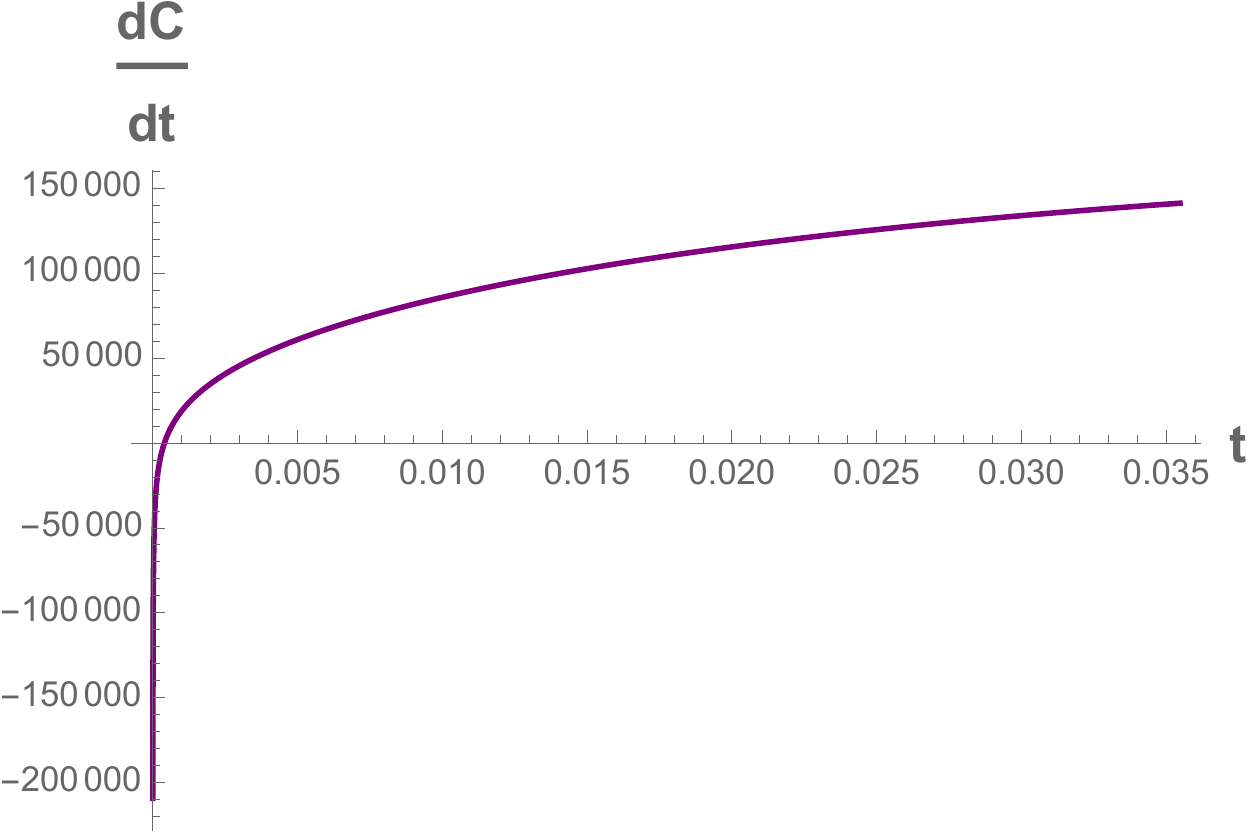}
            \caption[]%
            {{\small   $q_E=q_M=200$}}    
            \label{qtwohundred}
        \end{subfigure}
        \caption[  ]
        {\small The change of behavior in complexity growth rate by changing the charges of dyonic black holes.} 
        \label{dyonicphases}
    \end{figure*}

  By keeping $m$ and $\Lambda$ constant and changing $q_E$ and $q_M$, one can observe different phases which are shown in the plots of $\frac{dC}{dt}$ versus time in Fig. \ref{dyonicphases}. 
  
One could notice that for very small charges, the complexity growth rates at the beginning are very close to zero and only fluctuate a little bit at the early times, then suddenly increase and reach to their maximum and finally at the later times approach the corresponding Lloyd's bound, Fig. \ref{lowCharge}. 
 
 Increasing the charges would smooth out the fluctuations at the beginning and make the rate of complexity growth to reach to its maximum and then its Lloyd's bound much faster which could be noticed from the time axis of Figs \ref{qhalf}, \ref{qten}.  
 
For very large charges, the complexity growth rate would diverge. This could only mean that the dyonic black hole is unstable for those ranges of charge, see Fig. \ref{qtwohundred}.  So, in addition to the various phase transitions, complexity growth rate could even capture the instabilities in the solution of black holes.

One could also consider some additional counter terms for these dyonic black holes similar to \cite{Alishahiha:2018tep}, and consider their effects. This might solve the peculiar behavior at the beginning of our diagrams or it might even solve the issue of violation of Lloyd's bound at the early times.

Note that as the dyonic black holes are the dual holographic model for the van der Waals fluids \cite{Chamblin:1999tk,Dutta:2013dca,Xu:2017wvu}, we expect that the behavior of complexity growth rate for these systems would just be similar to the diagrams in \ref{cdyonic2} and this could  probably be tested in the lab.

\subsection{Complexity in AdS soliton background}\label{sec:soliton}

Now in this section we study the AdS soliton solution which is a confining geometry but still is very similar to the conformal AdS case as it could be transformed to it by a double Wick rotation.

The AdS soliton background is
\begin{gather}
ds^2=\frac{r^2}{\ell^2} \left[-dt^2+\left (1-\frac{r^d_+}{r^d} \right) d \chi^2+d\vec{x}^2 \right]+\left(1-\frac{r^d_+}{r^d}  \right)^{-1} \frac{\ell^2}{r^2} dr^2,
\end{gather}
where $\chi$ is the circle with smallest period which has antiperiodic boundary condition. The periodicity of $\chi$, $\Delta \chi$ and the parameter $r_+$ are related to each other by the relation $\Delta \chi = \frac{4\pi \ell^2}{d r_+} $, which comes from the smoothness at $r=r_+$. Note that as the boundary here is $d$-dimensional, there are $d-2$ coordinates $\vec{x}$ and the volume of these transverese $d-2$ $x^i$'s would be $V_x$.

This solution has the following negative boundary energy \cite{Reynolds:2017jfs},
\begin{gather}
E=-\frac{r_+^d \Delta \chi V_x}{\ell^{d+1}}=-\frac{V_x \ell^{d-1} (4\pi)^d } {d^d \Delta \chi ^{d-1}}.
\end{gather}

This ground state energy is the result of variance between the Casimir energies of bosons and fermions which exists due to the anti periodic boundary conditions for the fermions.  Increasing the anti periodicity would increase this negative Casimir energy and as a result the complexity.

Using the CV conjecture, the complexity in AdS soliton has been found as \cite{Reynolds:2017jfs}
\begin{gather}\label{cvolume}
\mathcal{C}_V=\frac{8 V_x \Delta \chi }{\pi} \frac{ r_{max}^{d-1}-r_+^{d-1} }{\ell^{d-1}}
\end{gather}
where $r=r_{max}$ is a UV cutoff.

One could notice that increasing periodicity $\Delta \chi$ and volume $V_x$ each would increase complexity linearly. This result could be confirmed by calculating complexity using ``complexity=action" (CA) conjecture too \cite{Reynolds:2017jfs}.

Note that for the case of pure AdS, the complexity would be $\mathcal{C}_V=\frac{8 \ell^{d-1} V_x }{\pi \epsilon^{d-1}}$ where $\epsilon$ is the cutoff in the radial direction. So from ``complexity=volume" (CV) conjecture one can find that the difference between these two complexities is negative and is in fact the second term of \ref{cvolume}.  So, from \ref{cvolume} one could see that the complexity of AdS soliton case is smaller than the AdS case. This could be explained by the fact that, relative to the AdS soliton, pure AdS is the excited state and therefore as its energy is higher, the complexity of AdS would also be higher than AdS soliton background. 

However, using the CA conjecture the authors of \cite{Reynolds:2017jfs}  found a different result for the relation between complexity and $r_+$. Using CA conjecture, they calculated the full IR/UV behavior of complexity growth rate and found that by increasing the IR scale, the action first increases until reaching a maximum and then it would decrease to zero at the UV cutoff. They justified this result, (that complexity of AdS soliton is higher than AdS case), by calculating complexity from Nielsen approach \cite{Nielsen:205eri}, and from field theory calculations for fermions on a rectangular lattice. However, for doing that calculation, they used Manhattan metric and took many approximations which using other metrics or assumptions could actually change the final result.

For instance, by considering gates with penalty factors which could take into account the entanglement between such gates and also take into account the non-locality, one could again calculate complexity. Doing that might change the result of \cite{Reynolds:2017jfs} from field theory and lattice calculations, so then their calculation of complexity from CV conjecture would match with the result of the field theory.  

\begin{figure}[ht!]
 \centering
  \includegraphics[width=7cm] {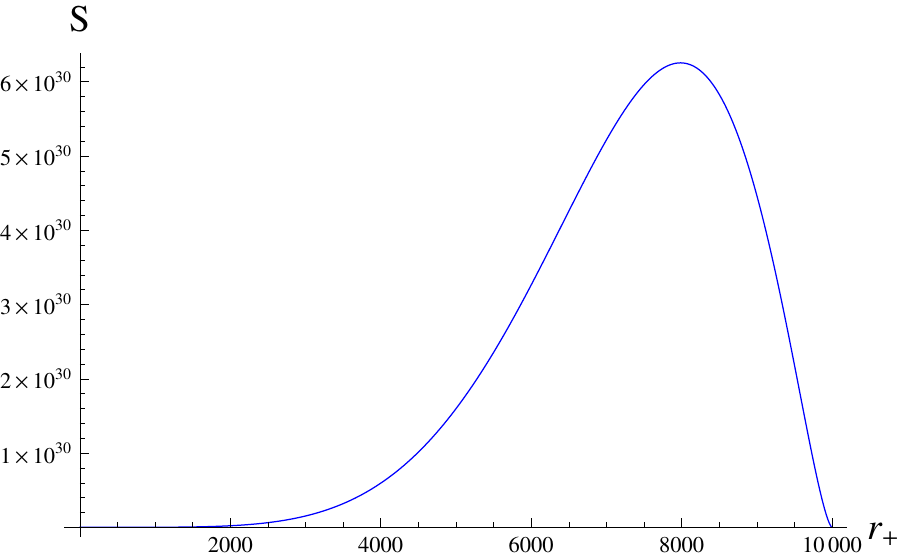}
  \caption{$S$ versus $r_+$ for $d=4$, $r_{max}=10000$ and $l=1$. }
 \label{fig:AdSsoliton}
\end{figure}

We reproduce their plot for the whole action versus the radial distance for the case of AdS soliton in Fig. \ref{fig:AdSsoliton} which actually comes from Eq. 38 of \cite{Reynolds:2017jfs}. Note that there is no time-dependence here as time is constant and the plot is the complexity, rather than its growth rate. The growth rate of complexity for the AdS soliton could be studied in future works.

\begin{figure}[ht!]
 \centering
 \begin{subfigure}{.45\textwidth}
  \centering
  \includegraphics[width=5.7cm] {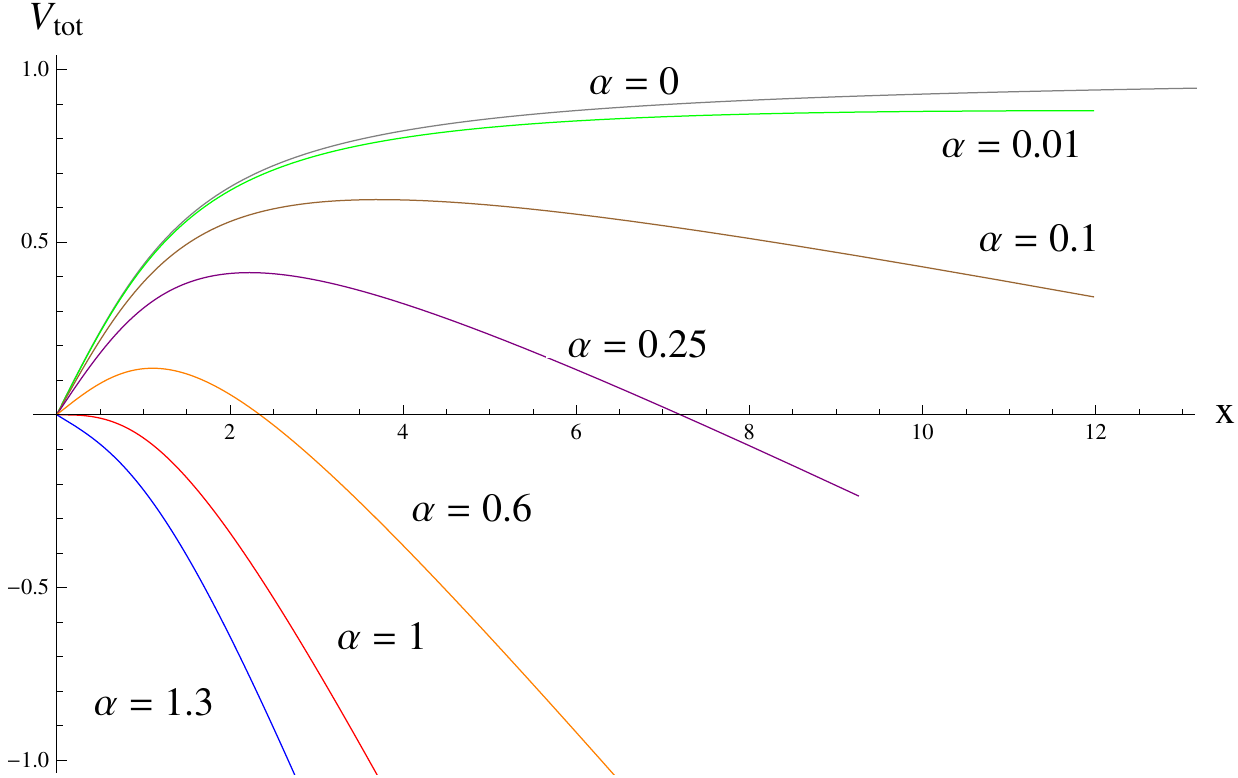}
    \captionof{figure}{ AdS}
  \end{subfigure}
   \begin{subfigure}{.45\textwidth}
    \centering
 \includegraphics[width=5.7cm] {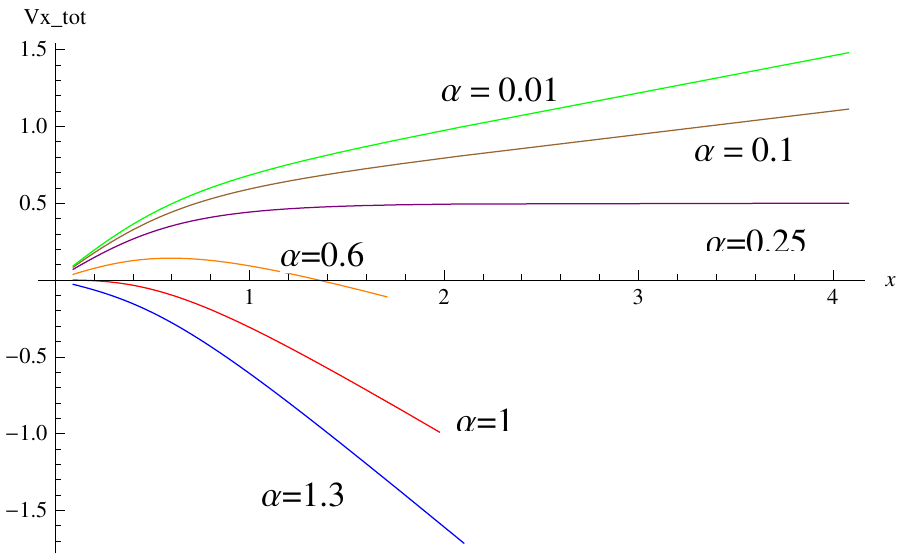}
   \captionof{figure}{ AdS soliton }
  \end{subfigure}%
    \caption{ Comparing Schwinger phase diagrams of AdS and AdS soliton case.  }
   \label{compare}
\end{figure}

Now one could examine which one of their results would be more compatible with other observations regarding the relationship between complexity and other physical quantities such as changes in potential.  Comparing the behavior of complexity, potentials, and also the phase diagrams of Schwinger pair creation in two backgrounds of AdS and AdS soliton, Fig. \ref{compare}, could give some hint about the correct answer for the complexity. 

In \cite{Nishioka:2006gr},  it has been shown that there is a connection between AdS bubbles (AdS solitons) and closed string tachyon condensations, and also it has been found in that work, that the degrees of freedom and therefore the entanglement entropy would decrease under this tachyon condensation, (which is a second order phase transition).  In addition, the energy density would decrease in this process \cite{Horowitz:2006mr, Horowitz:1998ha}. So as both energy and entanglement entropy of AdS soliton is lower than AdS case, and as AdS soliton is more stable than AdS background, one could propose that complexity of AdS soliton should be lower than AdS case, and therefore the calculation from CV in \cite{Reynolds:2017jfs} should be the correct answer. One could repeat the calculations for the ``twisted" AdS bubble which is a better gravity dual of the corresponding Yang-Mills theory on $S^1 \times R^3$ and compare the results. 

Moreover, recently in \cite{Kim:2017qrq} it has been found that between different holographic (CV and CA) and field theory methods (Fubini-Study metric (FS) and Finsler geometry (FG)) for calculating complexity, the holographic CV conjecture and field theoretic FG method are actually correlated to each other. So if actually one repeat the calculation of \cite{Reynolds:2017jfs} for AdS soliton using FG, a result more correlated with CV could be derived. We hope to come back to this problem in future works.

To get further information, one could compare the phase diagrams of Schwinger effect in both AdS and AdS soliton backgrounds while an electric field is present.

At each value for alpha which here is the dimensionless quantity for the strength of electric field, i.e, $\alpha=\frac{E}{E_c}$, the potential of AdS soliton would be higher, pointing to smaller fluctuations and therefore smaller complexity. There are also three phases in the AdS soliton background while there are two phases for the AdS case. This richer possibilities for the phases could also be seen from the behavior of complexity.

Also in \cite{Haehl:2012tw}, the quasinormal modes in the AdS soliton case has been plotted. One could notice that for higher radial distances, these modes exponentially damp which explains the reduction of complexity. 

This connection could be examined further in future works. For instance one could then turn on a magnetic field in addition to the electric field and change the angle between the directions of the two which could change the Schwinger pair creation rate greatly. As a universal result, \cite{Hashimoto:2014yya} found that when the electric field is parallel to the electric field the pair creation rate is higher relative to the setup where they are perpendicular to each other.
The complexity growth rates in these two setups would also be different.  The results could be compared with the pair creation rates so to find more evidences for the connection between quantum fluctuations, boundary periodicity and complexity.

\section{Complexity and phase transitions in QCD models}\label{sec:three}

To further study the relationship between potentials, quantum fluctuations, phase transitions and complexity growth, we study the Gubser model of QCD \cite{Gubser:2008yx, Gubser:2008ny,Janik:2015iry,Janik:2016btb, Janik:2017ykj }, as the main part of this work.

By tuning the parameters of the potential, in this model several kinds of phase transitions could be displayed. Based on the parameters of the dilaton potential, it could generate three types of phase transitions which are crossover, first and second order. We study the full time behavior of complexity growth rates around these phase transitions and the correlations with other thermodynamical quantities.

In \cite{Zhang:2017nth}, by using the CA conjecture, the late-time behavior of complexity of this holographic QCD model has been studied.  There, it has been shown that the growth rate of complexity could also be used as a parameter to detect the phase transitions. This is because the behavior of entropy and complexity growth rate versus temperature would match up for each type of phase transition. We then study the connection between the behavior of complexity growth rates, speed of sound, entropy and potential for four different models of confining potentials.

The action of this model is
\begin{gather}
S=\frac{1}{16 \pi G_5} \int d^5 x \sqrt{-g} \left [ R- \frac{1}{2} (\partial \phi )^2 -V(\phi) \right ] +\frac{1}{8 \pi G_5} \int_{\partial \mathcal{M}} K, 
\end{gather}
where $K$ is the trace of the extrinsic curvature. For simplicity we also take $16 \pi G_5=1$ and $\hbar=1$. 

 In this model the five-dimensional gravity is coupled to a single scalar field. This setup then contains the minimum freedom which is needed to match the equation of states and also reproduce the desired phase diagrams of QCD.

The general ansatz for the dilaton potential would be
\begin{gather}
V(\phi) = -12 (1+a \ \phi^2)^{1/4} \cosh(\gamma \phi)+b_2 \phi^2+b_4 \phi^4+b_6 \phi^6,
\end{gather}

where $(a, \gamma, b_2, b_4, b_6) $ are the parameters shown below.

\begin{center}\label{table}
 \begin{tabular}{||c| c c c c c c c ||} 
 \hline
 \text{potential} & $a$ & $\gamma$ & $b_2$ &  $b_4$ & $b_6$ & $\Delta$ & $T_c$  \\ [0.5ex] 
 \hline
$V_{\text{QCD}}$ & 0 & 0.606 & 1.4  & -0.1  & 0.0034 & 3.55 & 0.181033  \\ 
 $V_{\text{2nd}}$ & 0 & $1/\sqrt{2}$ & 1.958 & 0 & 0 & 3.38 & 0.243901\\
$V_{\text{1st}}$ & 0& $\sqrt{7/12}$ & 2.5 & 0 & 0 & 3.41 & 0.156841\\ 
$V_{\text{IHQCD}}$ & 1 & $\sqrt{2/3}$ & 6.25 & 0 & 0 & 3.58 & 0.295847 \\
\hline
\end{tabular}
\end{center}

   \begin{figure*}[ht!]
        \centering
        \begin{subfigure}[b]{0.35\textwidth}
            \centering
            \includegraphics[width=\textwidth]{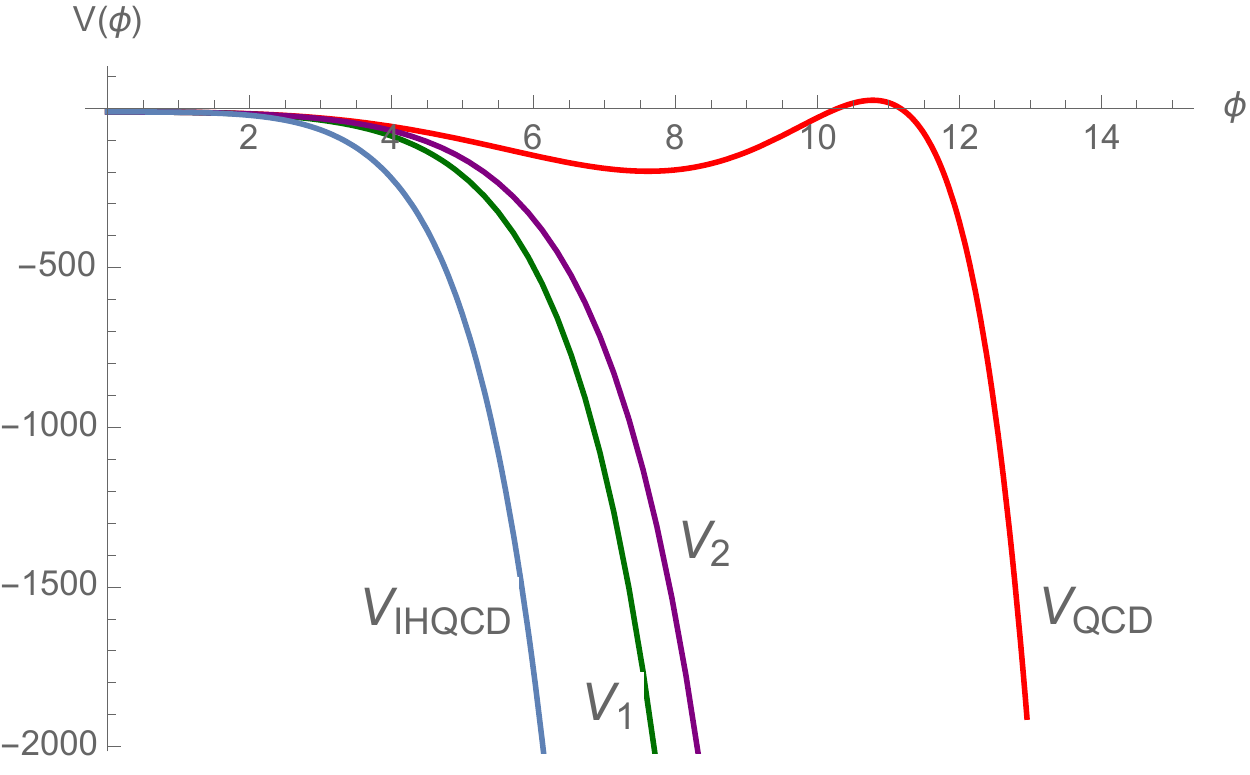}
            \caption[Network2]%
            {{\small $V(\phi)$ vs. $\phi$.}}    
            \label{VQCDcomplexity2}
        \end{subfigure}
          \quad
         \centering
        \begin{subfigure}[b]{0.35\textwidth}   
            \centering 
            \includegraphics[width=\textwidth]{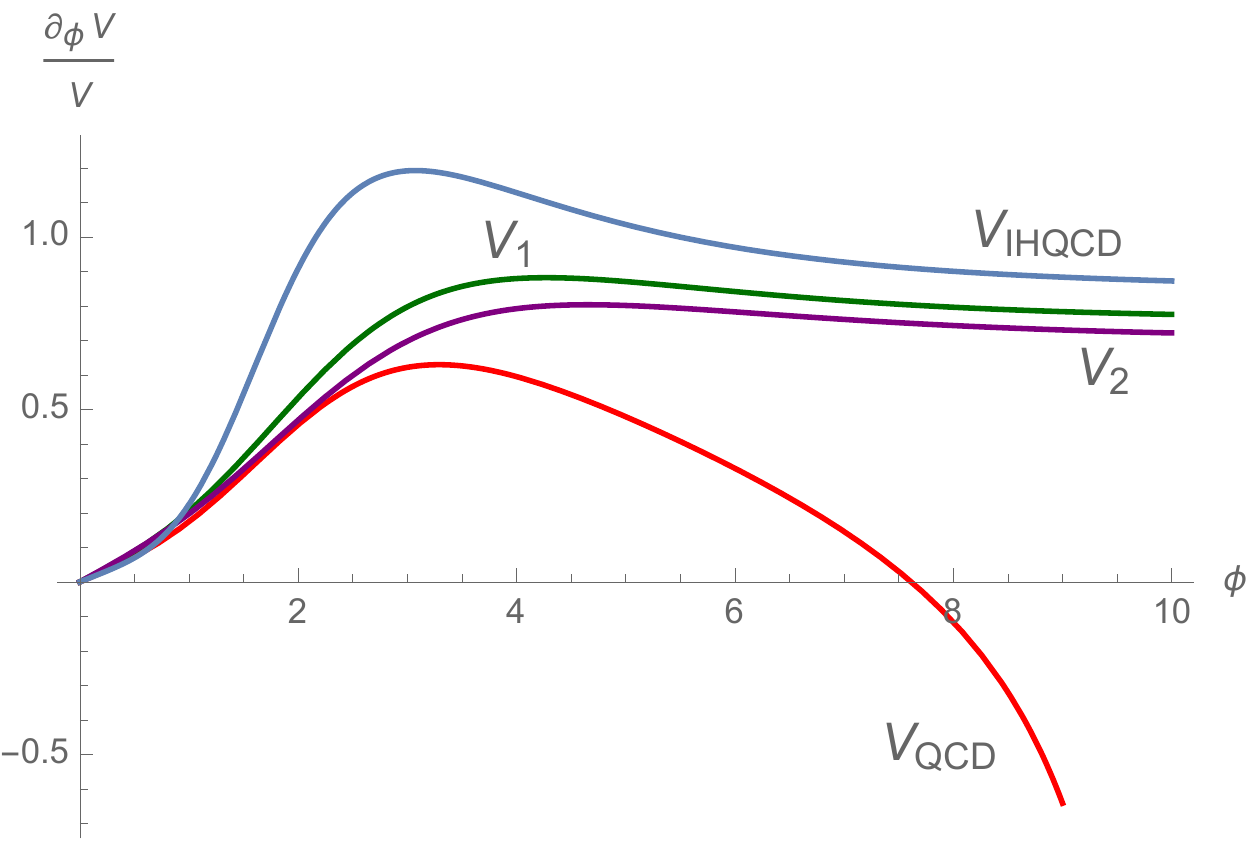}
            \caption[]%
            {{\small $\frac{\partial_\phi V}{V}$ vs. $\phi$.}}    
            \label{complexityVIHQ}
        \end{subfigure}
        \caption[ QCD potentials and their derivative vs. $\phi$. ]
        {\small QCD potentials and their derivative vs. $\phi$.} 
        \label{fig:potphases}
    \end{figure*}

To compare the behavior of potential, with their different parameters, which  then would lead to various phase transitions, the plots of $V(\phi)$ and $\frac{\partial_\phi V}{V}$ versus $\phi$ is shown in Fig. \ref{fig:potphases}. The parameters in this potential \cite{Gubser:2008ny, Gubser:2008yx, Zollner:2018uep} have been chosen in a way that the plots of $c_s^2$ versus $T/T_c$ would match with the phenomenological results of hadron gas and also the lattice models. We discuss the properties of each model at the later parts of this section. 

So first, we consider the following ansatz
\begin{equation} 
ds^2=e^{2A}(-h dt^2+d\vec{x}^2)+\frac{e^{2B}}{h} dr^2, \ \ \ \ \ \ \ 
\phi =r,
\end{equation}
where $A, B$ and $h$ are only the functions of $r$ (or $\phi$). Note that the asymptotic boundary is at $r \to 0$ and the singularity is where $ r \to \infty$. 

The equations of motion would be
\begin{equation}\label{equationsseries}
\begin{split}
A''-A' B' +\frac{1}{6} &= 0, \nonumber\\
h''+ (4 A' -B') h' &= 0 , \nonumber\\
6 A' h'+h(24 A'^2-1) + 2 e^{2B} V & =0, \nonumber\\
4A'-B'+\frac{h'}{h}-\frac{e^{2B}}{h} V' &=0,
\end{split}
\end{equation}

and the horizon is where $ h(\phi_H) = 0$.

In \cite{Zhang:2016rcm}, using the method of \cite{Gubser:2008yx, Gubser:2008ny}, the field equations have been solved as
\begin{equation} 
\begin{split}
A(\phi) &=A_H +\int_{\phi_H}^\phi d\tilde{\phi} G(\tilde{\phi}), \nonumber\\
B(\phi)&=B_H +\ln \left( \frac{G(\phi) }{ G (\phi_H)} \right) +\int_{\phi_H}^\phi \frac{d \tilde{\phi} }{6 G (\tilde {\phi} )} ,\nonumber\\
h(\phi)&=h_H+h_1\int_{\phi_H}^\phi d\tilde{\phi} e ^{-4 A(\tilde{\phi}) + B(\tilde{\phi} ) },
\end{split}
\end{equation}
where $G(\phi) \equiv A'(\phi) $ and then using the initial and boundary conditions, the constants have been found as
\begin{equation} 
\begin{split}
A_H &= \frac{\ln \phi_H}{\Delta-4} + \int_0^{\phi_H} d\phi \left[ G(\phi) - \frac{1}{ (\Delta-4) \phi} \right ],\nonumber\\
B_H & = \ln \left( - \frac{4V(\phi_H) }{ V(0) V'(\phi_H) } \right) +\int_0^{\phi_H} \frac{d\phi}{ 6 G(\phi)}, \nonumber\\
h_H & =0, \nonumber\\
h_1 & =  \frac{1}{ \int_{\phi_H} ^0  d\phi e^{-4 A(\phi) +B(\phi) } }.
\end{split}
\end{equation}

For finding the solution of $G(\phi)$, one actually needs to solve the following equation
\begin{gather}
\frac{G'}{G+V/3V'} =\frac{d}{d\phi} \ln \left( \frac{G'}{G}+\frac{1}{6G}-4G-\frac{G'}{G+V/3V'}\right),
\end{gather}
and from it, one could find the series expansion of $G(\phi)$ near the horizon $\phi=\phi_H$ as
\begin{gather}
G(\phi)= - \frac{V(\phi)}{3 V'(\phi) }+\frac{1}{6} \left ( \frac{V(\phi_H) V''(\phi_H) }{V'(\phi_H)^2} -1 \right) (\phi-\phi_H)+\mathcal{O} (\phi-\phi_H)^2.
\end{gather}
This result could be used as the boundary condition for solving $G(\phi)$.

Assuming $16\pi G_5=1$ and $ \hbar =1$, the temperature, entropy and the speed of sound      $c_s$, could be written as
\begin{gather}
T=\frac{ e^{A_H -B_H}}{4 \pi}|h'(\phi_H)|=\frac{ e^{A_H +B_H}}{4 \pi} | V'(\phi_H)  |, \ \ \ \ \ \ \  s=4\pi e^{3A_H}, \ \ \ \ \ \ \ c_s^2= \frac{d \log T}{ d \log s}.
\end{gather}

\begin{figure*}[ht!]
        \centering
        \begin{subfigure}[b]{0.35\textwidth}
            \centering
            \includegraphics[width=\textwidth]{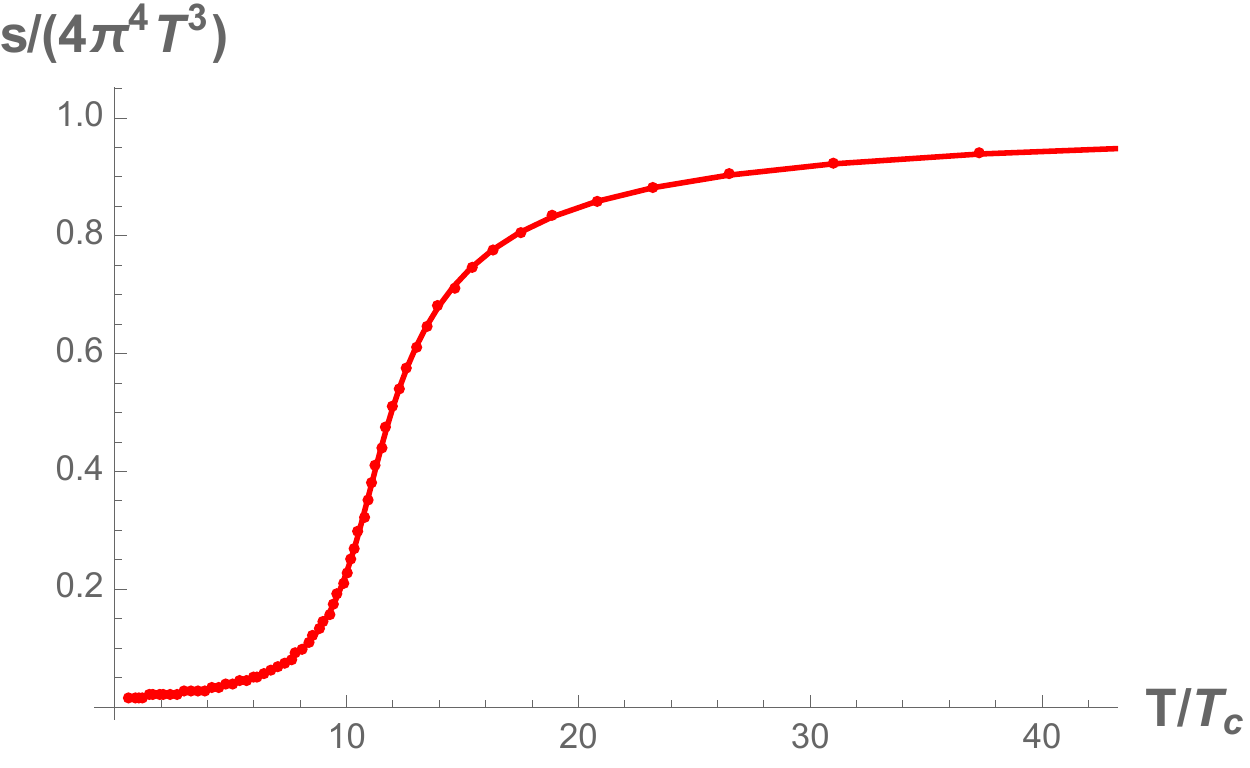}
            \caption[Network2]%
            {{\small $V_{\text{QCD}}$}}    
            \label{VQCDcomplexity2}
        \end{subfigure}
         \centering
        \begin{subfigure}[b]{0.35\textwidth}  
            \centering 
            \includegraphics[width=\textwidth]{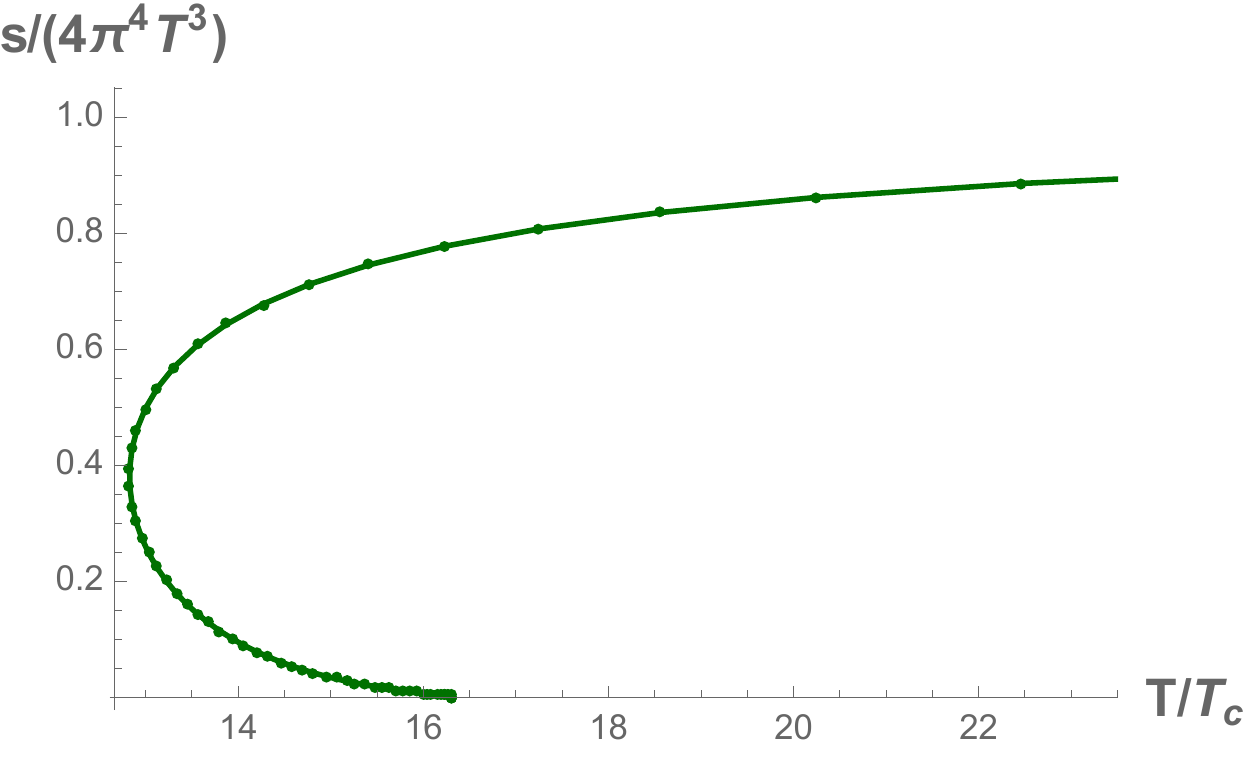}
            \caption[]%
            {{\small $V_{\text{1}}$}}    
            \label{complexityV1}
        \end{subfigure}
         \centering
        \vskip\baselineskip
        \begin{subfigure}[b]{0.35\textwidth}   
            \centering 
            \includegraphics[width=\textwidth]{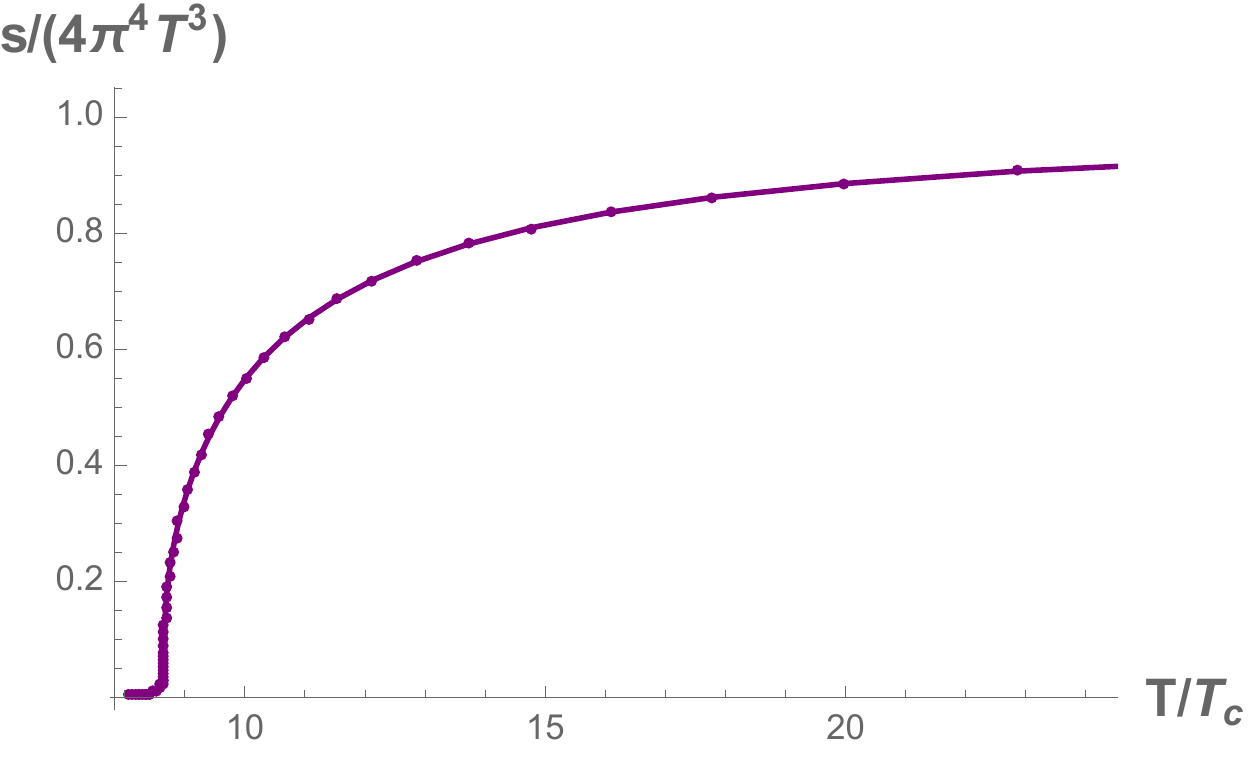}
            \caption[]%
            {{\small $V_2$}}    
            \label{V2complexity}
        \end{subfigure}
        \quad
         \centering
        \begin{subfigure}[b]{0.35\textwidth}   
            \centering 
            \includegraphics[width=\textwidth]{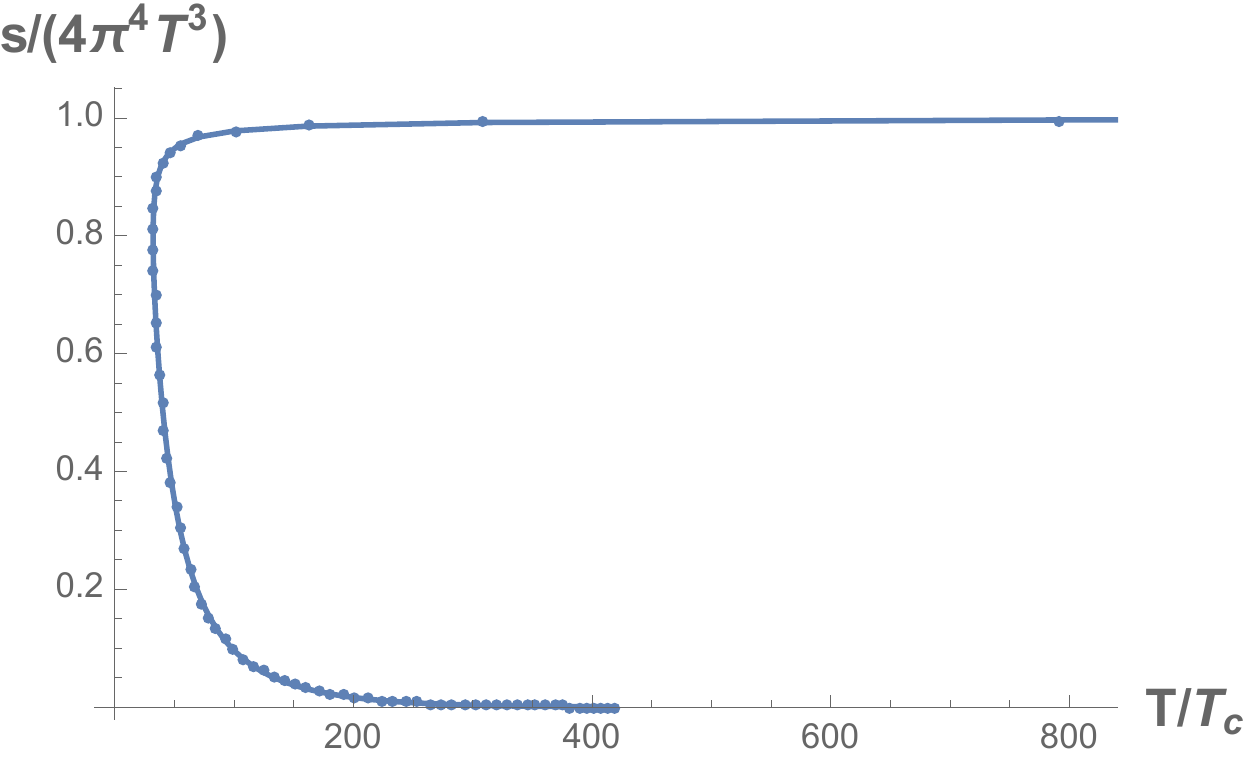}
            \caption[]%
            {{\small $V_{IHQCD}$}}    
            \label{entropyVIHQ}
        \end{subfigure}
        \caption[ Phase transition of entropy versus temperature for four non-conformal cases. ]
        {\small Phase transition of entropy versus temperature for four non-conformal cases.} 
        \label{fig:entropyphases}
    \end{figure*}

\begin{figure*}[ht!]
        \centering
        \begin{subfigure}[b]{0.36\textwidth}
            \centering
            \includegraphics[width=\textwidth]{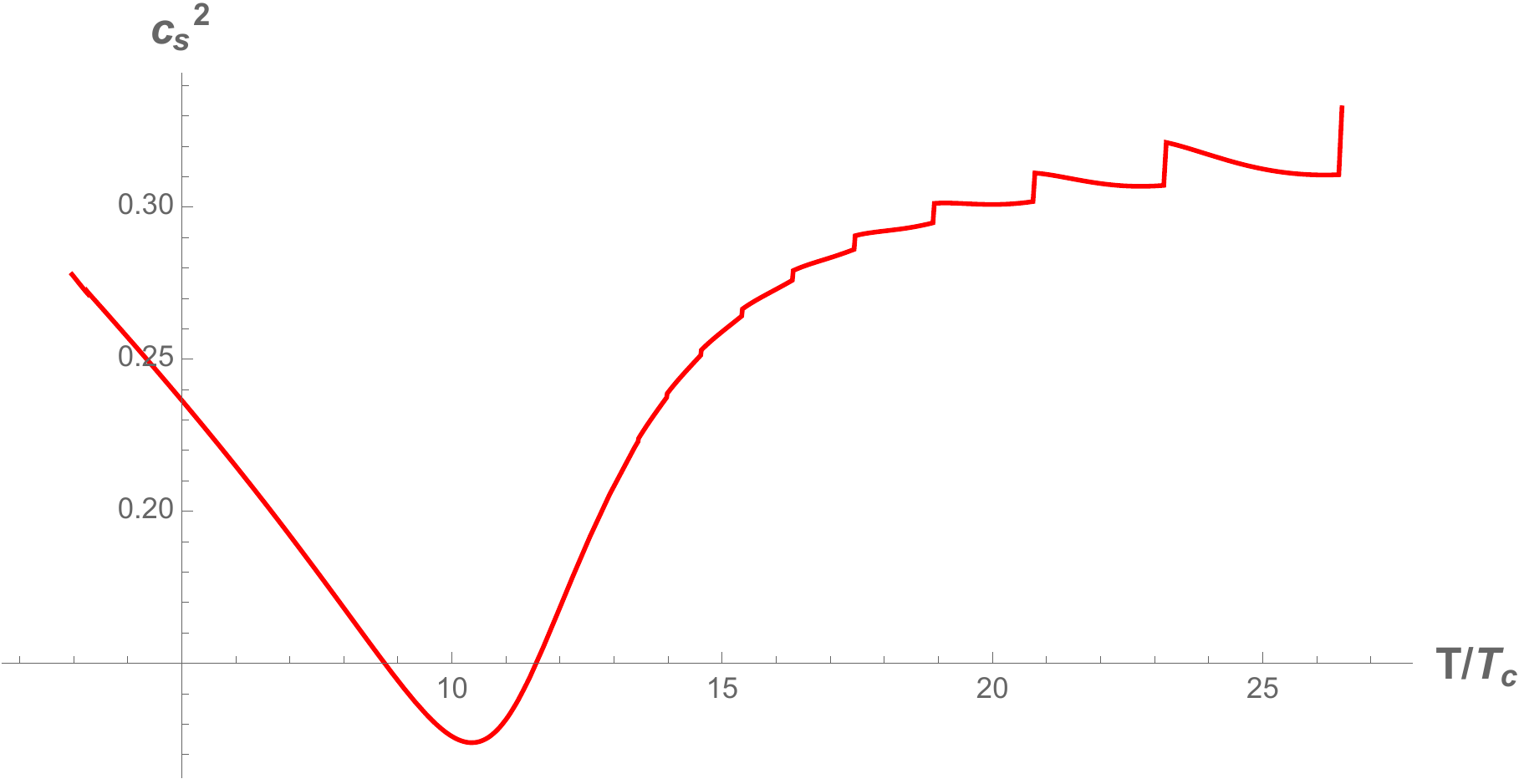}
            \caption[Network2]%
            {{\small $V_{\text{QCD}}$}}    
            \label{soundVQCD}
        \end{subfigure}
         \centering
        \begin{subfigure}[b]{0.36\textwidth}  
            \centering 
            \includegraphics[width=\textwidth]{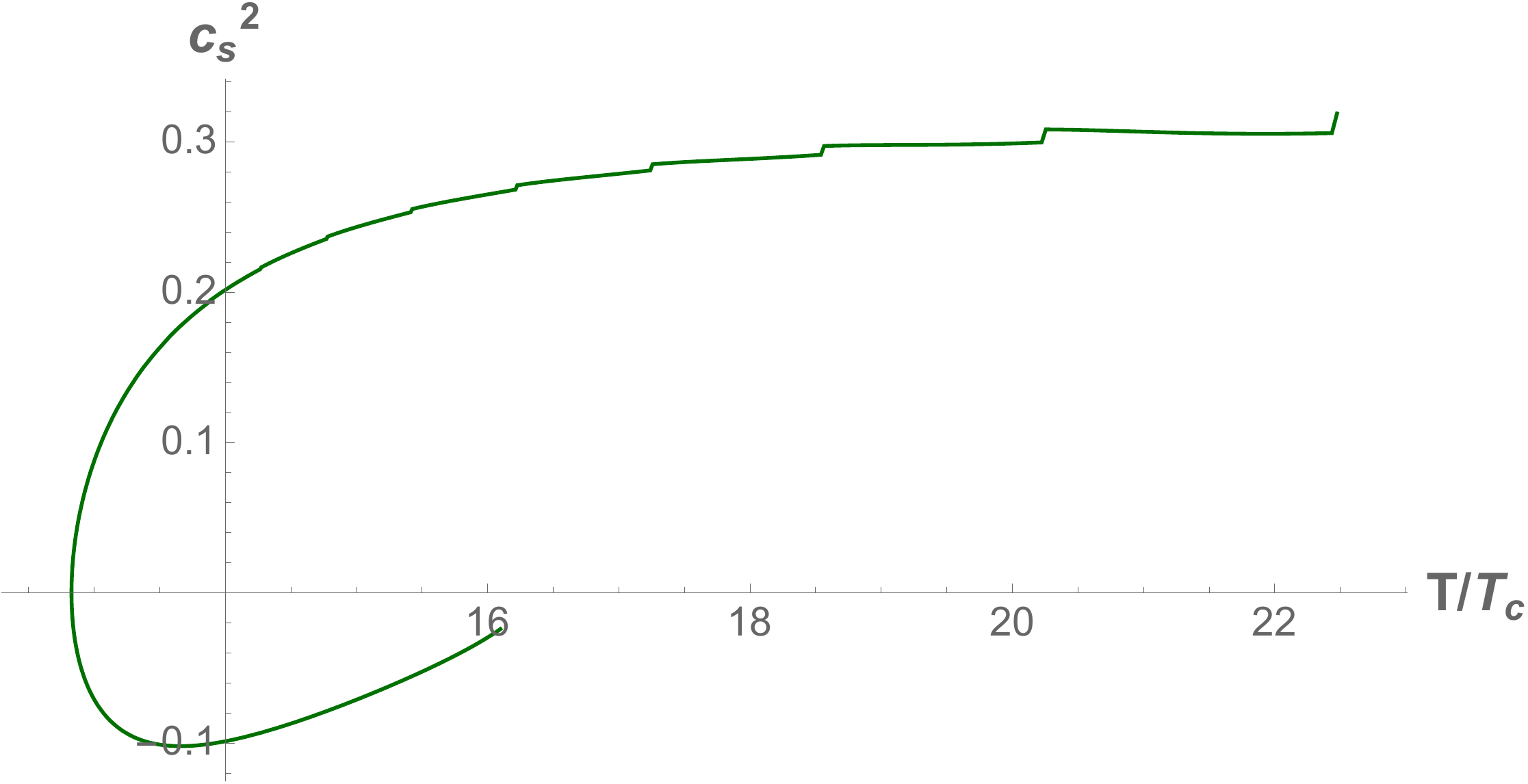}
            \caption[]%
            {{\small $V_{\text{1}}$}}    
            \label{complexityV1}
        \end{subfigure}
         \centering
        \vskip\baselineskip
        \begin{subfigure}[b]{0.36\textwidth}   
            \centering 
            \includegraphics[width=\textwidth]{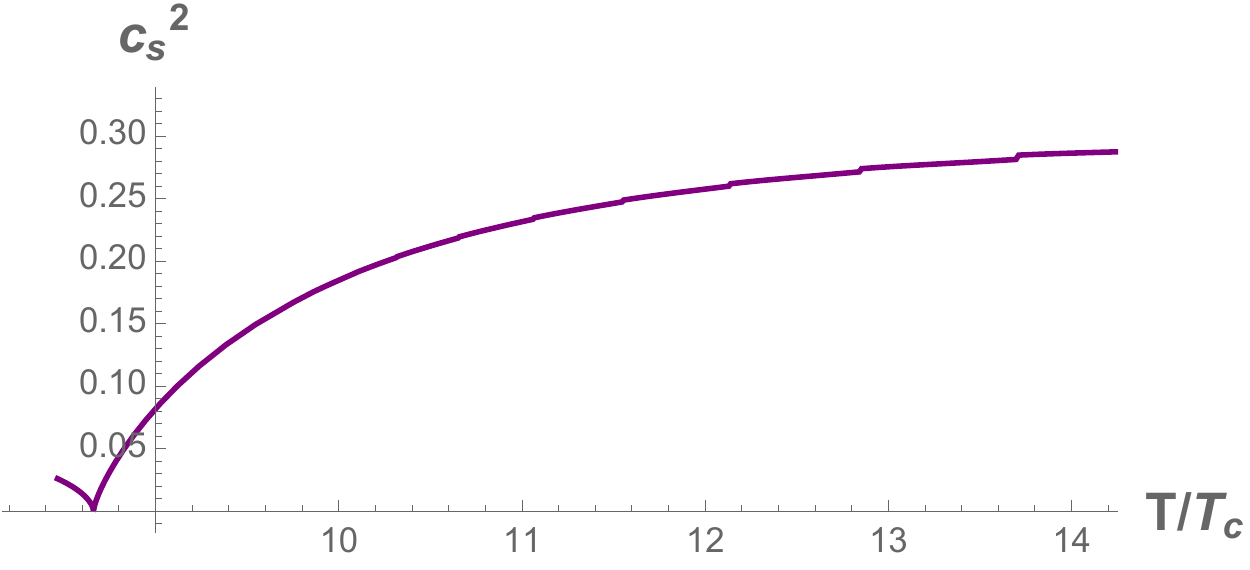}
            \caption[]%
            {{\small $V_2$}}    
            \label{V2complexitysound}
        \end{subfigure}
        \quad
         \centering
        \begin{subfigure}[b]{0.34\textwidth}   
            \centering 
            \includegraphics[width=\textwidth]{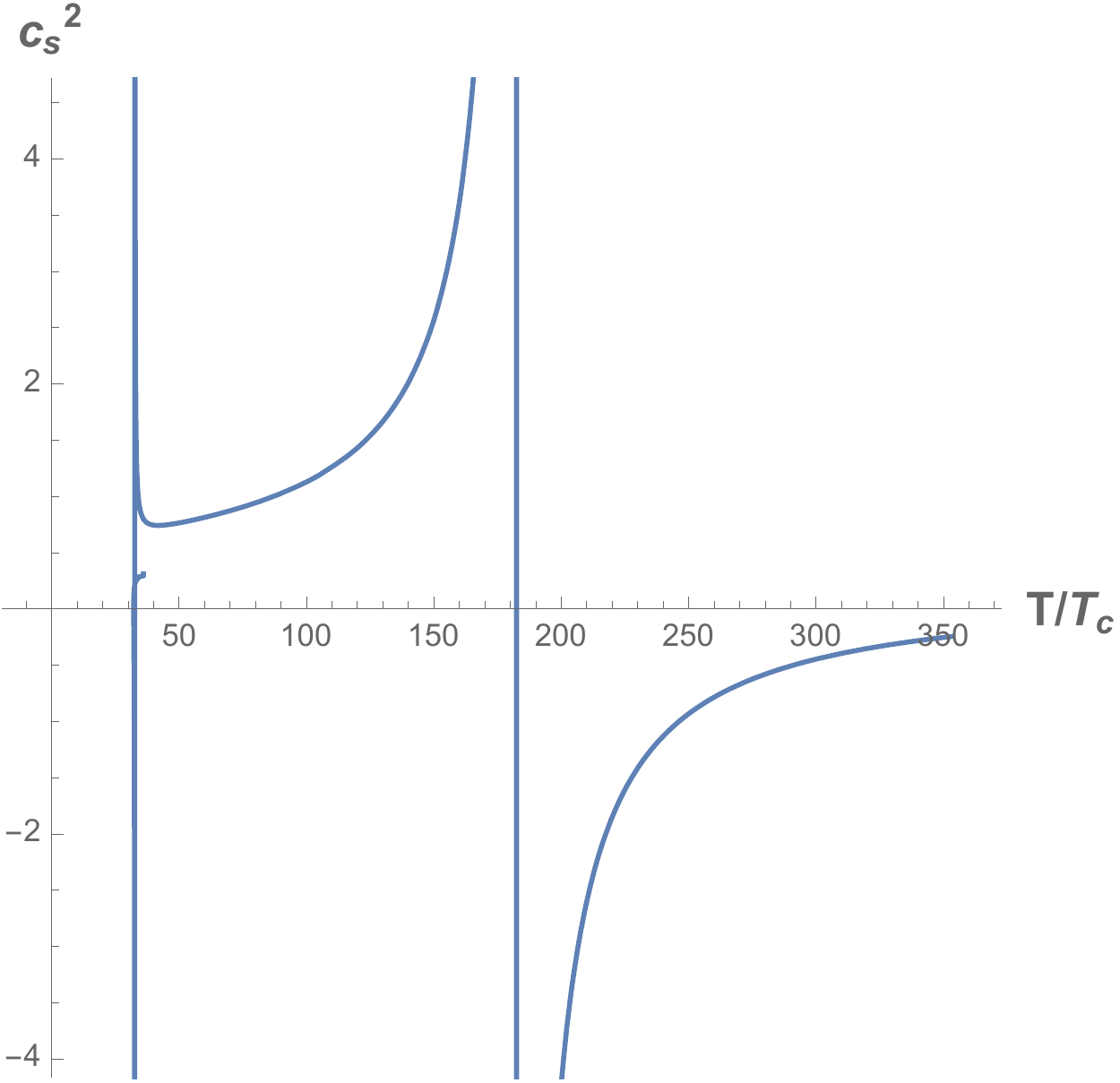}
            \caption[]%
            {{\small $V_{IHQCD}$}}    
            \label{soundVIHQ}
        \end{subfigure}
        \caption[ The diagrams of speed of sound versus $T/T_c$  for four non-conformal cases. ]
        {\small The diagrams of speed of sound versus $T/T_c$  for four non-conformal cases.} 
        \label{fig:speedphases22}
    \end{figure*}

Solving for these parameters numerically, the plots for the entropy and speed of sound versus temperature for the four models of potential could be constructed, which they are shown in Fig. \ref{fig:entropyphases} and Fig. \ref{fig:speedphases22}. Note that in all of these models $ s/T^3$ is actually proportional to the number of degrees of freedom \cite{Zhang:2016rcm}. Therefore, beforehand, one could expect that the behavior of complexity growth rate would be similar to entropy as well.

For the late-time interval $[t, t+ \delta t]$, the bulk term of the on-shell action inside the horizon is
\begin{equation} 
\begin{split}
S_{\text{bulk}} &= \int d^5 x \frac{2}{3 \pi G_N} e^{4A+B} V(r)
 = \frac{2V_3}{3 \pi G_N} \int_t ^{t+\delta t} dt \int_{r_H}^\infty e^{4A+B} V(r) dr,
\end{split}
\end{equation}

and the GHY term would be

\begin{gather}
S_{\text{GHY}} =\delta t V_3 \bigg [ e^{-4A-B} \partial_r ( e^{8A} h) \bigg ] \Bigg |_{r_H}^\infty,
\end{gather}

where $V_3 \equiv \int d^3 \vec{x} $ is the volume of the boundary field system.

So the growth rate of the holographic complexity density $c \equiv \mathcal{C}/V_3$ at late time is
\begin{gather}
\frac{dc}{dt}=\frac{2}{3\pi G_N} \int_{r_H}^\infty e^{4A+B} V(r) dr+\frac{1}{\pi} \Big [ e^{-4A-B} \partial_r (e^{8A} h) \Big ] \Bigg |_{r_H}^{\infty}.
\end{gather}

\begin{figure*}[ht!]
        \centering
        \begin{subfigure}[b]{0.35\textwidth}
            \centering
            \includegraphics[width=\textwidth]{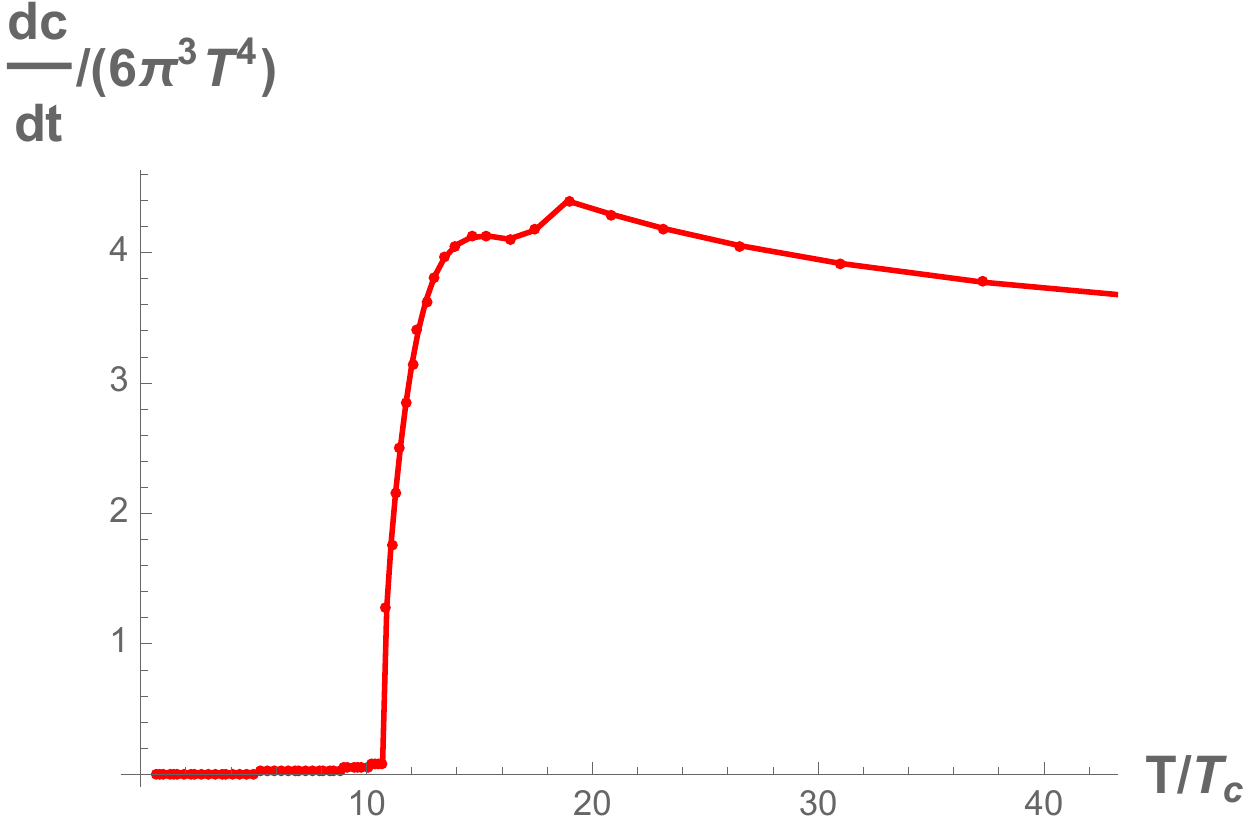}
            \caption[Network2]%
            {{\small $V_{\text{QCD}}$}}    
            \label{VQCDcomplexity2}
        \end{subfigure}
         \centering
        \begin{subfigure}[b]{0.35\textwidth}  
            \centering 
            \includegraphics[width=\textwidth]{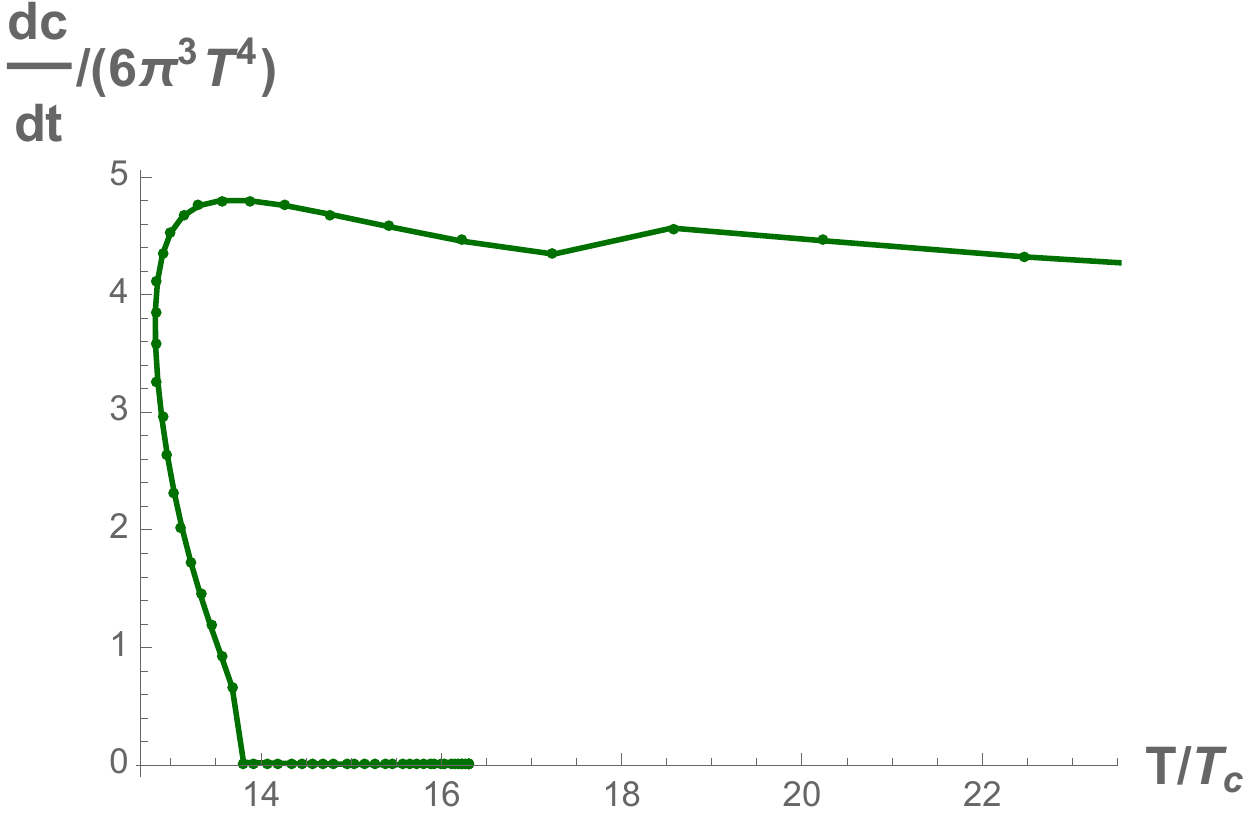}
            \caption[]%
            {{\small $V_{\text{1}}$}}    
            \label{complexityV1}
        \end{subfigure}
         \centering
        \vskip\baselineskip
        \begin{subfigure}[b]{0.35\textwidth}   
            \centering 
            \includegraphics[width=\textwidth]{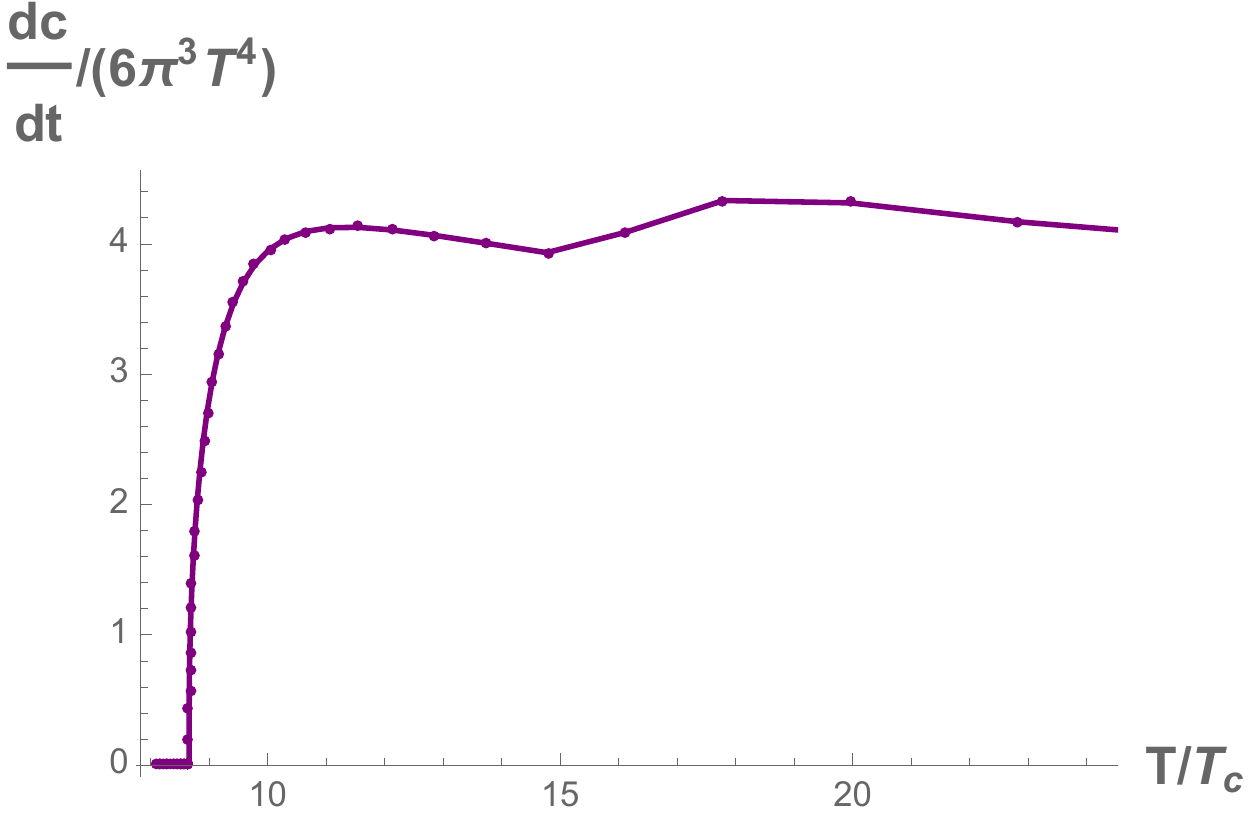}
            \caption[]%
            {{\small $V_2$}}    
            \label{V2complexity}
        \end{subfigure}
        \quad
         \centering
        \begin{subfigure}[b]{0.35\textwidth}   
            \centering 
            \includegraphics[width=\textwidth]{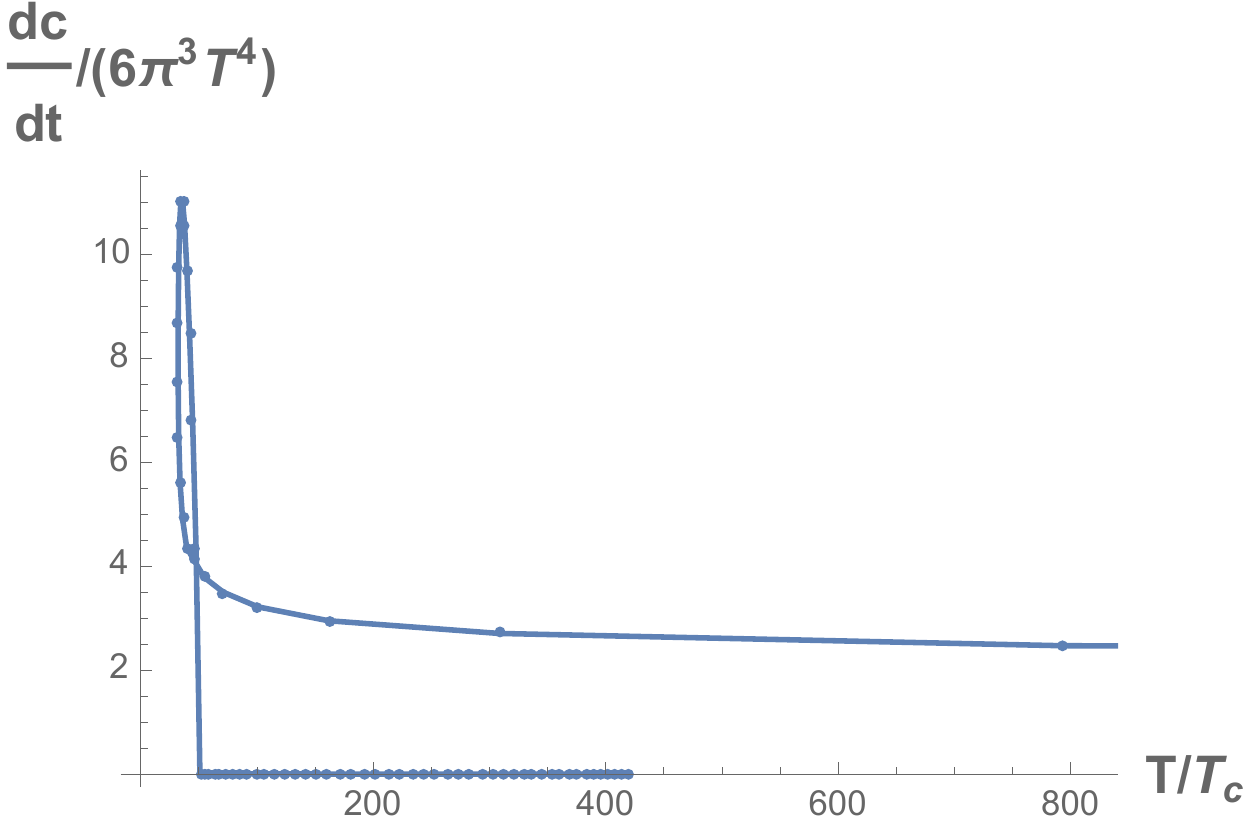}
            \caption[]%
            {{\small $V_{IHQCD}$}}    
            \label{complexityVIHQ}
        \end{subfigure}
        \caption[ Phase transition of complexity growth rate versus temperature for four non-conformal cases. ]
        {\small Phase transition of complexity growth rate versus temperature for four non-conformal cases.} 
        \label{fig:complexityphases22}
    \end{figure*}

The plots for the behavior of late time complexity during each phase transition are presented in Fig. \ref{fig:complexityphases22}.  As one would expect and as found in \cite{Zhang:2017nth} they are very similar to the diagrams of entropy for each potential, demonstrating that complexity could also act as a good probe of confinement and phase transitions. 
 
Interestingly the jump that we see in the complexity for all of these cases during the phase transition is close to the topological jump of $\Delta \mathcal{C}=2\pi$ found in \cite{Abt:2017pmf, Abt:2018ywl}.  So it is worth noting that the results in here coming from the ``complexity=action" conjecture is fairly consistent with the results of \cite{Abt:2018ywl}  which has came from the ``complexity=volume" and the ``subregion complexity" conjectures, \cite{Alishahiha:2015rta, Stanford:2014jda, Susskind:2014moa,Brown:2015bva}.

Comparing these diagrams with those of the speed of sound \cite{Anabalon:2017eri} would be interesting too. Note that for the high temperatures where the first three models become conformal, the speed of sound becomes constant and close to $0.3$ and also the rate of complexity growth would become constant as well. For the low temperature case, note that when the slope of diagrams for the $c_s^2$ versus $T/T_c$ is negative, the slope of $c_s^2$ versus $T/T_c$ would be positive and vice versa. This qualitatively could be explained by the fact that when the speed of sound grows in a region, the information could propagate easier, it would be easier to go from one state to another and therefore, the rate of growth of complexity would decrease.

Now we study the full time behavior of complexity growth rate in this model. Similar to \cite{Lehner:2016vdi, Carmi:2017jqz}, the time evolution of complexity could be found by adding the null boundary and joint terms to the action and Gibbons-Hawking-York (GHY) terms. So,
\begin{gather}
I=\frac{1}{16 \pi G_N} \int_{\mathcal{M}} d^5 x \sqrt{-g} \left[ R-\frac{1}{2} (\partial \phi)^2-V(\phi) \right] \nonumber\\+\frac{1}{8 \pi G_N} \int_{\mathcal{B}} d^4 x \sqrt{|h|} K + \frac{1}{8\pi G_N} \int_\Sigma d^3 x \sqrt{\sigma} \eta \nonumber\\
-\frac{1}{8\pi G_N} \int_{\mathcal{B'}} d\lambda d^3 \theta \sqrt{\gamma} \kappa+\frac{1}{8\pi G_N}\int_{\Sigma'} d^3 x \sqrt{\sigma} a.
\end{gather}

The affine parametrization could be chosen in a way that $\kappa=0$ so to make the contribution of the null boundary vanish.

Now assuming $G_N=1$ and  $t_L=t_R=\frac{t}{2}$, we can divide the evolution of the black hole into two stages of the time before the critical time $t_c$, and after it. In the past critical time,  the past null-boundary intersects the past-singularity and there is the GHY boundary term. After the critical time however, the two null boundaries from the left and right CFTs would intersect with each other and therefore there would be the contribution from the null joint term instead of the GHY term.

The critical time $t_c$ could be found by
\begin{gather}
\frac{t_c}{2}-r^*(\infty)=t-r^*(0),\nonumber\\
-\frac{t_c}{2}+r^*(\infty)=t+r^*(0),
\end{gather} 
so
 \begin{gather}
 t_c=2( r^*(\infty)-r^*(0)).
 \end{gather}

 \begin{figure}[ht!]
 \centering
  \includegraphics[width=8cm] {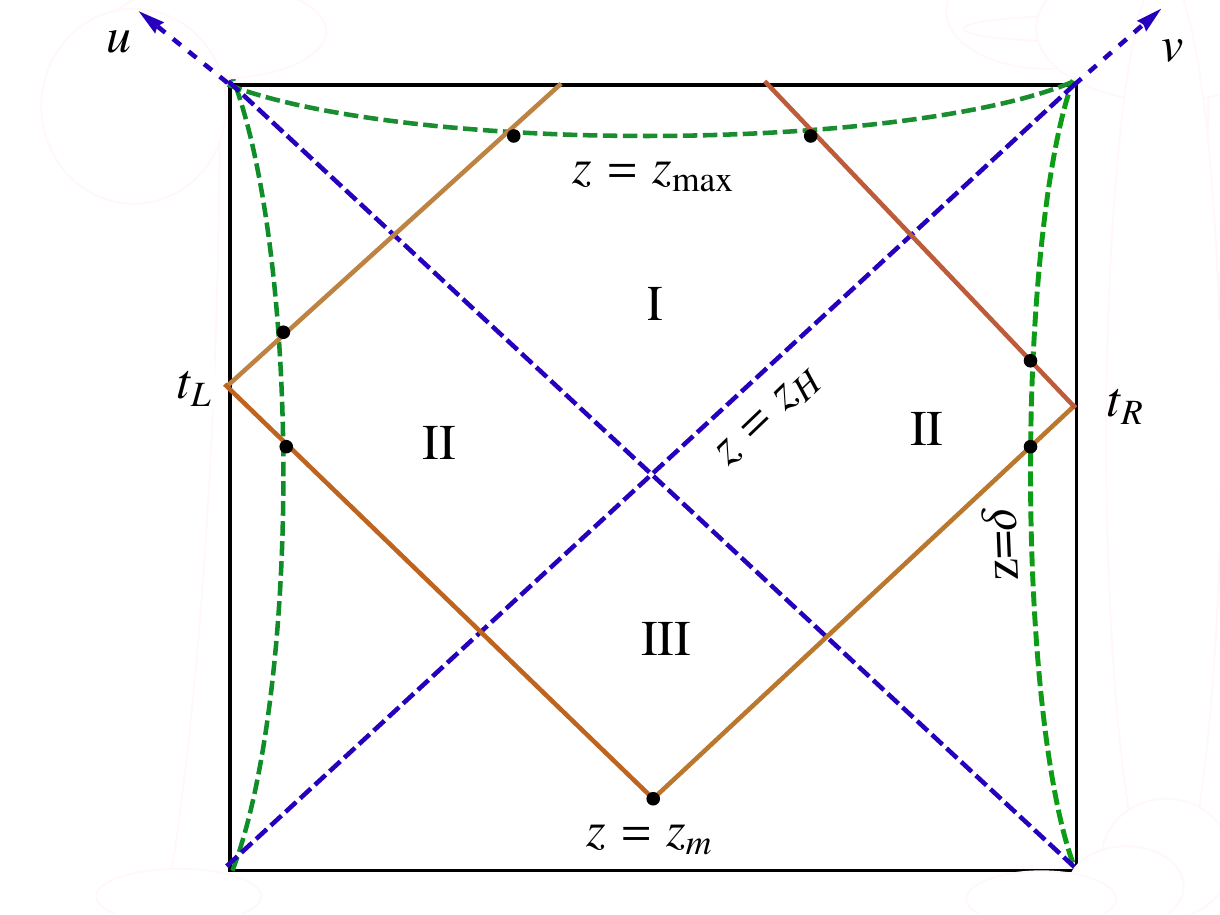}
  \caption{Penrose diagram for the black brane ansatz and the null boudaries. }
 \label{fig:picPenrose}
\end{figure}

From Fig. \ref{fig:picPenrose}, for a state at $r_R=t_L=\frac{t}{2} > \frac{t_c}{2}$, one gets
\begin{gather}
S_{bulk_I} =\frac{2V_3}{3 \pi G_N} \int_{r_H}^{r_{\text{max}}} dr e^{4A+B} V(r) \left( \frac{t}{2} +r^*(0)-r^*(r) \right) , \nonumber\\
S_{bulk_{II}} =\frac{4V_3}{3 \pi G_N} \int_{\delta}^{r_H} dr e^{4A+B} V(r) \left( r^*(0)-r^*(r) \right) , \nonumber\\
S_{bulk_{III}} =\frac{2V_3}{3 \pi G_N} \int_{r_H}^{r_m} dr e^{4A+B} V(r) \left(- \frac{t}{2} +r^*(0)-r^*(r) \right),
\end{gather}

so
\begin{gather}
S_{bulk}=\frac{4V_3}{3 \pi G_N} \int_{\delta}^{r_{\text{max}}} dr e^{4A+B} V(r) \left( r^*(0)-r^*(z) \right)+\frac{2V_3}{3 \pi G_N} \int_{r_m}^{r_{\text{max}}} dr e^{4A+B} V(r) \left( \frac{t}{2} -r^*(0)+r^*(r) \right).
\end{gather}
Note that $r^*(r)=\int \frac{e^{B-A}}{h} dr$. Also the GHY boundary term at $r=r_{\text{max}}$ would be
\begin{gather}
S_{\text{bound}}= \frac{V_3}{8\pi G_N} \left(\frac{t}{2}+r^*(0)-r^*(r_{\text{max}}  ) \right) \left( e^{-4A-B} \partial_r (e^{8A} h) \right)_{r=r_{\text{max}}}.
\end{gather}

Now, for considering the contribution of joint terms which occurs at $r=r_m$, we first take two normal vectors of the null boundaries,
\begin{gather}
k_1^a=\alpha \left( \frac{ e^{-2A}}{h} (\partial_t)^a+e^{-A-B}(\partial_r)^a \right), \ \ \ \ \ \ k_2^a=\beta\left( -\frac{ e^{-2A}}{h} (\partial_t)^a+e^{-A-B}(\partial_r)^a \right).
\end{gather}

Then, the joint action
\begin{gather}
S_{\text{joint}}=\frac{1}{8\pi G_N} \int d^dx \sqrt{\gamma} \log \Big | \frac{k_1\ . \ k_2}{2}  \Big|, 
\end{gather}
would be
\begin{gather}\label{Ijoint}
S_{\text{joint}}=\frac{V_3}{8\pi G_N} e^{3A(r_m)} \left (\log \left |e^{2A(r_m)} \right | -\log \left | h( r_m) \right | \right)+\frac{V_3}{8\pi G_N} e^{3 A(r_m) } \log \alpha \beta.
\end{gather}

In the above action, $\gamma$ is the determinant of the induced metric on the joint point and  $\alpha$ and $\beta$ are two constants that appear because of the ambiguity in normalization of the null boundaries. For removing the ambiguity, one can add a counter term which is in the following form
\begin{gather}
\frac{1}{8\pi G_N} \int d\lambda d^d x \sqrt{\gamma} \log \frac{\Theta}{d},
\end{gather}
where
\begin{gather}
\Theta=\frac{1}{\sqrt{\gamma} } \frac{\partial \sqrt{\gamma} }{\partial \lambda}.
\end{gather}

Here, $\lambda$ is the affine parameter for the null surface. For the null vector $k_1$ it would be
\begin{gather}
\frac{\partial r}{\partial \lambda}=\alpha e^{-A-B}.	
\end{gather}

For the null surface which is associated with $k_1$, we have $\Theta=3\alpha A'  e^{-A-B}$, while for $k_2$ we get $\Theta=3\beta A'  e^{-A-B}$.

Then, the counter term would be
\begin{gather}
S_{ct}=-\frac{V_3}{4\pi G_N} \int_\delta^{r_m} dr 3A' e^{3A} \log(3A' e^{-A-B}) -\frac{V_3}{8\pi G_N}  e^{3A(r_m)} \log \alpha \beta,
\end{gather}
where the second term cancels the contribution from the normalization factors of \ref{Ijoint}.

So summing all the terms and taking the time derivative and also using $\frac{dr_m}{dt}=\frac{h\left(r_m(t)\right)}{2} e^{A(r_m(t))-B(r_m(t))} $, we get
\begin{equation}
\begin{split}
\frac{dS}{dt}/\frac{V_3}{\pi G_N}&= h(r_m) e^{A(r_m)-B(r_m)} \Bigg( \frac{1}{3} e^{4A(r_m)+B(r_m)} V(r_m) r^*(r_m)\nonumber\\&+\frac{3 r_m }{8} e^{9 r_m^2}  \big (2+18 {r_m}^2-3 \log (h(r_m)) -\frac{h'(r_m)}{6 h(r_m) r_m} \big)-\frac{3}{8} A'(r_m)e^{3A(r_m)} \log \big | 3A'(r_m) e^{-A(r_m)-B(r_m)} \big | \Bigg)\nonumber\\&
+\frac{1}{3} \int_{r_m}^{r_{\text{max}}} dz e^{4A+B} V(z) +\frac{1}{16} \Big( e^{-4A-B} \partial_r(e^{8A} h)\Big)\Big |_{r=r_{\text{max}}}.
\end{split}
 \label{eq:totcomplexity}
\end{equation}

\begin{figure*}[ht!]
        \centering
        \begin{subfigure}[b]{0.33\textwidth}
            \centering
            \includegraphics[width=\textwidth]{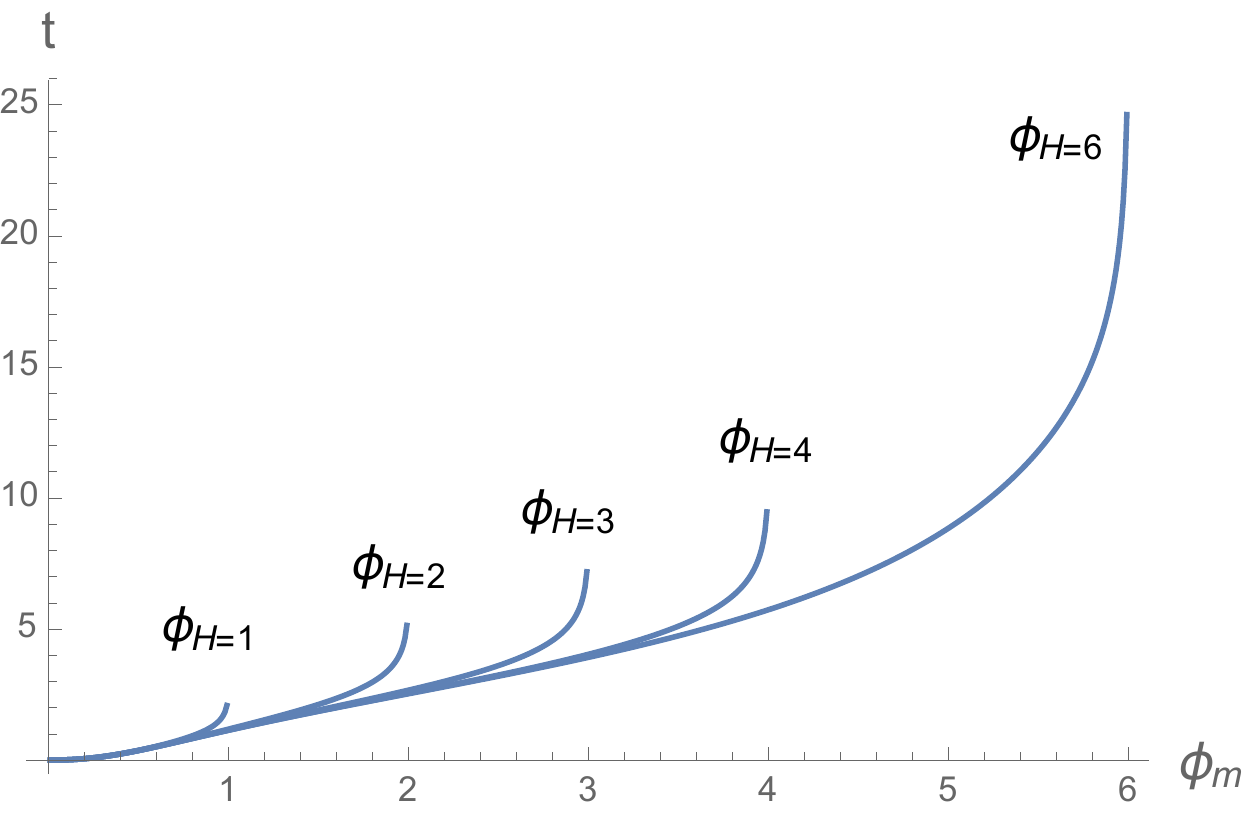}
            \caption[Network2]%
            {$V_{\text{QCD}}$}    
            \label{VQCDcomplexity2}
        \end{subfigure}
         \centering
        \begin{subfigure}[b]{0.33\textwidth}  
            \centering 
            \includegraphics[width=\textwidth]{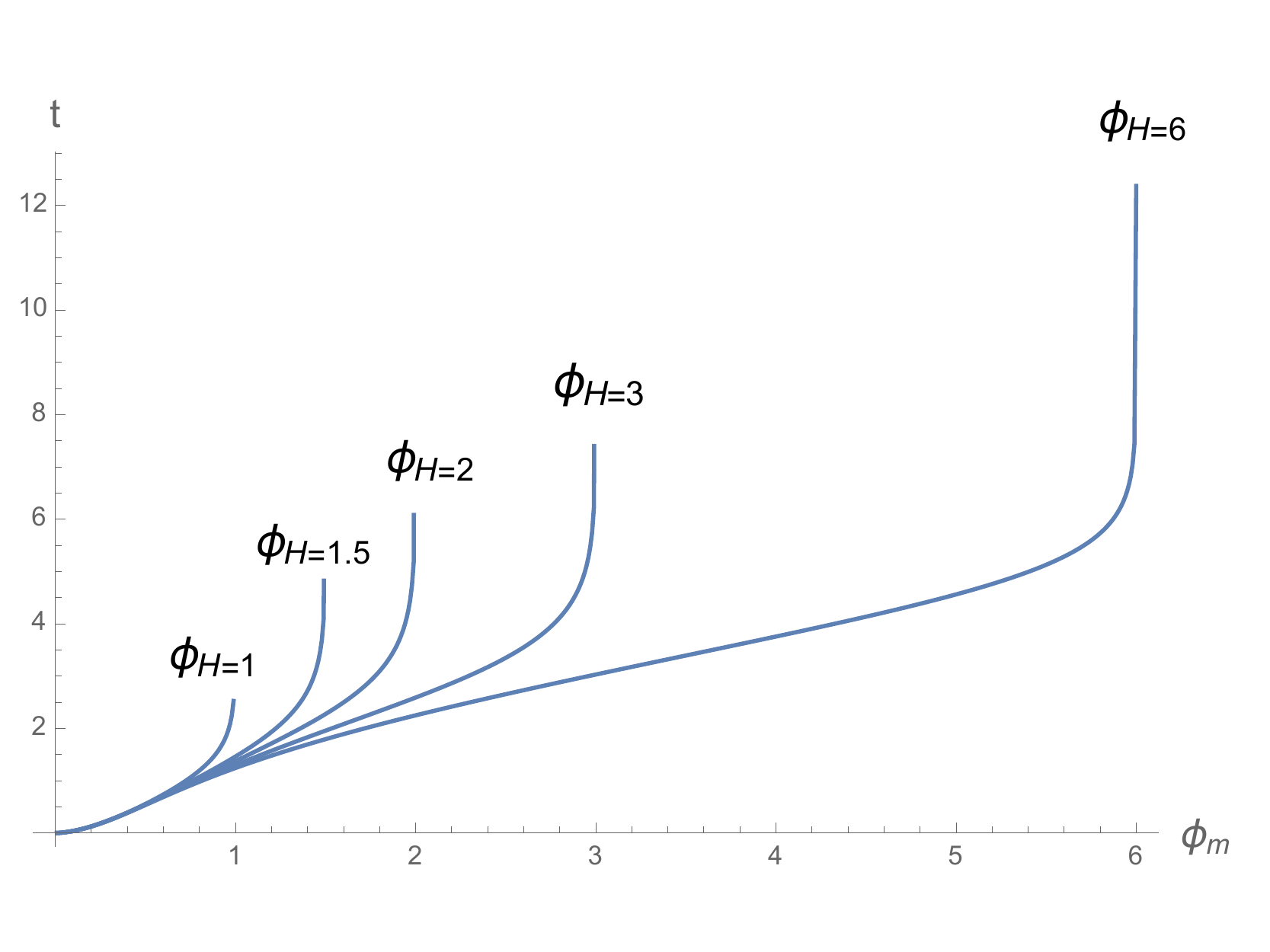}
            \caption[]%
            {$V_1$ }    
            \label{threeComplexity}
        \end{subfigure}
         \centering
        \vskip\baselineskip
        \begin{subfigure}[b]{0.33\textwidth}   
            \centering 
            \includegraphics[width=\textwidth]{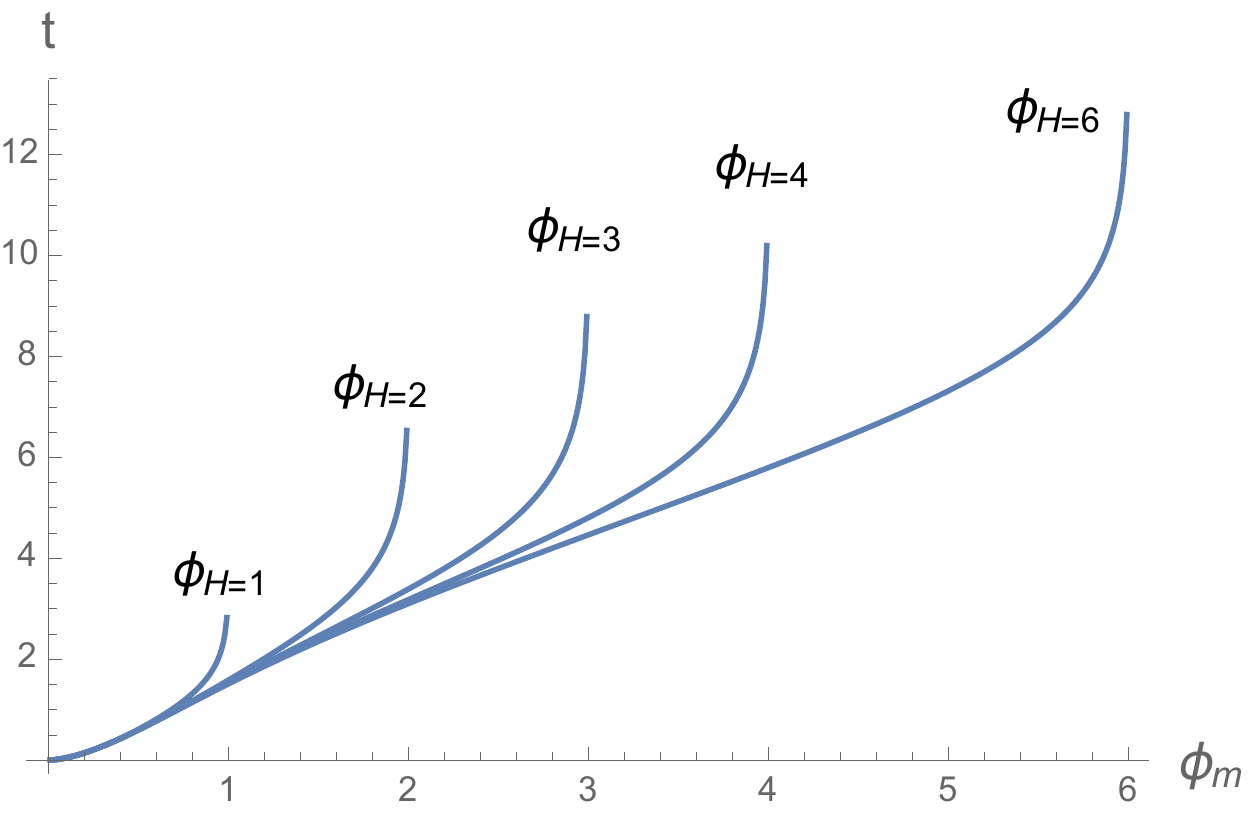}
            \caption[]%
            {$V_2$ }    
            \label{Speed of sound in three models.}
        \end{subfigure}
        \quad
         \centering
        \begin{subfigure}[b]{0.33\textwidth}   
            \centering 
            \includegraphics[width=\textwidth]{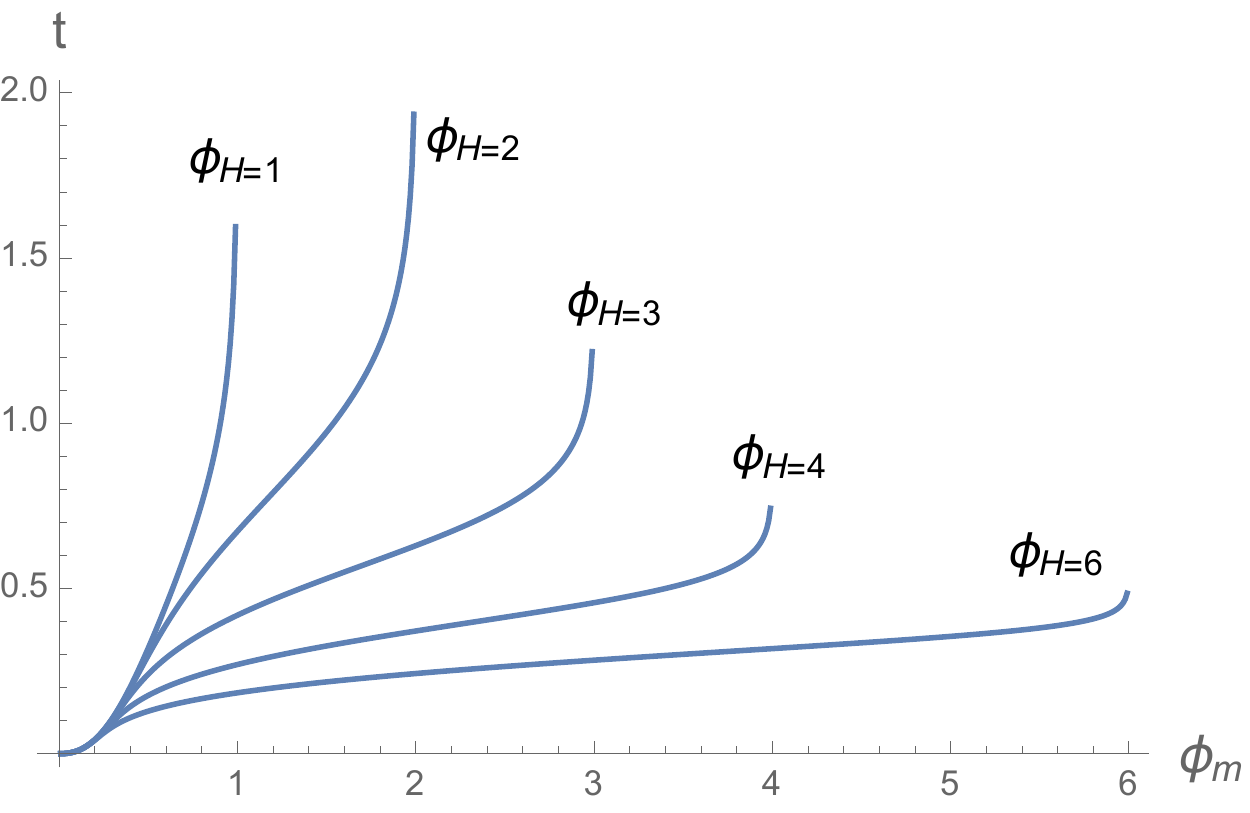}
            \caption[]%
            {$V_{\text{IHQCD}}$ }    
            \label{potentails}
        \end{subfigure}
        \caption[ The relation between $t$ and $\phi_m$. ]
        {\small The relation between $t$ and $\phi_m$.} 
        \label{fig:tphim}
    \end{figure*}

The relationship between $\phi_m$ and $t$ can be read numerically from the relation $ t=2\left ( r^*(0)-r^*(r_m) \right)$ (note $\phi=r$), and the plots for various $\phi_H$ for the different models of potential are shown in Fig. \ref{fig:tphim}.

 Interestingly these plots are very similar to the corresponding plots in \cite{Dudal:2018ztm}. In that work, the confining Einstein-Maxwell-dilaton model has been studied. Authors have found the relationship between the length of a strip $\ell$ of a subsystem which were used for calculating the entanglement entropy, versus $z^*$ which is the turning point of the minimal area surface. Note that $z^*$ actually corresponds to $\phi_m=r_m$ in our model. This actually would make sense  as $t$ and $l$ are directly connected. Note that when $\phi_m$ is smaller than $\phi_H$, the relationship between $t$ and $\phi_m$ is close to linear, but when $\phi_m$ reaches $\phi_H$, then suddenly $t$ blows up.  These feature could be a universal behavior of the confining models.

 \begin{figure}[ht!]
 \centering
  \includegraphics[width=7.5cm] {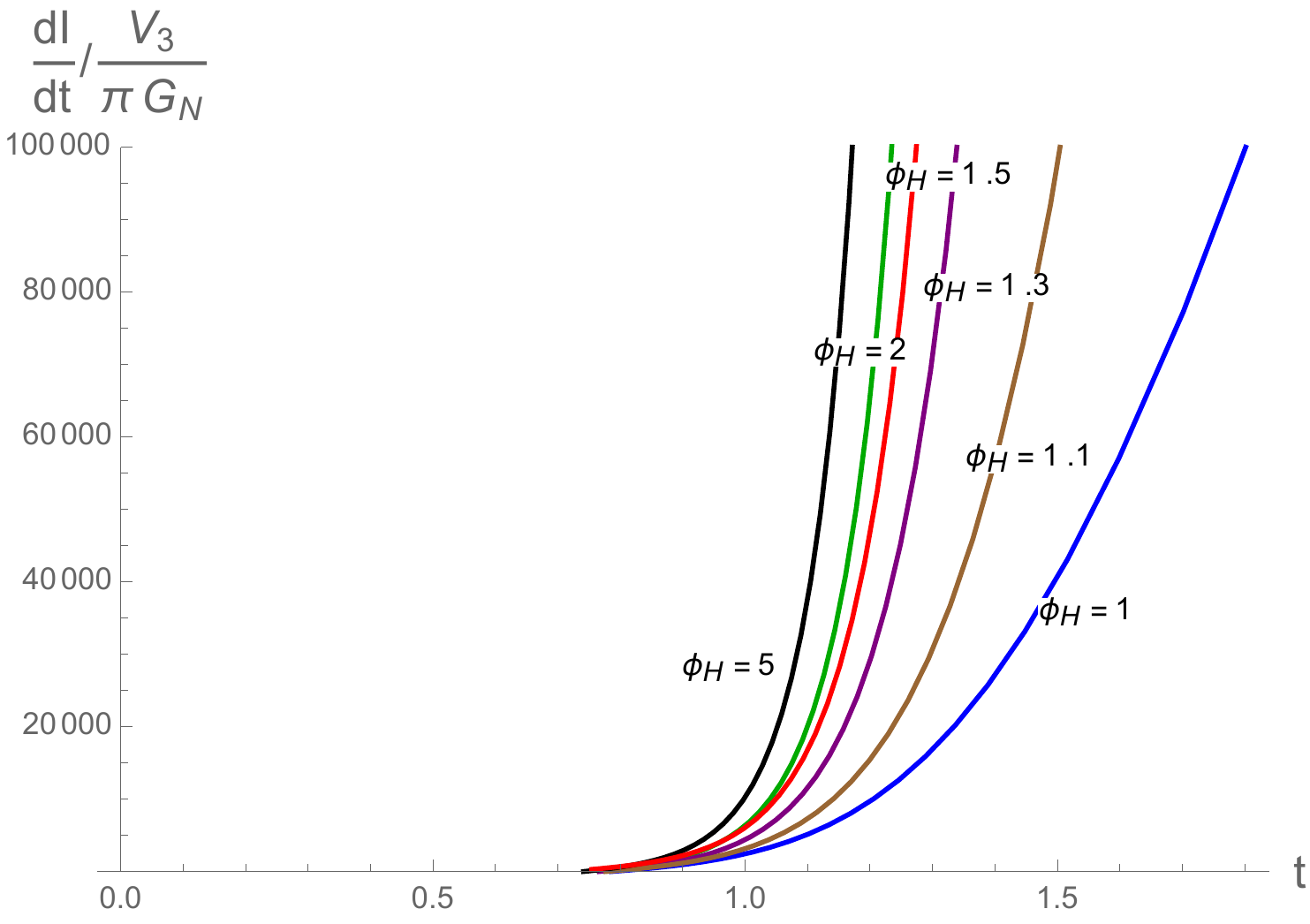}
  \caption{The rate of growth of complexity at early times for various black holes in the $V_{\text{QCD}}$ model, with different $\phi_H$, corresponding to different temperatures.}
 \label{fig:dcdt}
\end{figure}

In Fig. \ref{fig:dcdt}, we present the full, time dependent, behavior of complexity growth rate coming from Eq. \ref{eq:totcomplexity}, for the case of $V_{\text{QCD}}$. For other potential models, the general behavior is qualitatively very similar. From this figure it is obvious that the Lloyd's bound would be violated at early times and it would be violated more strongly for black holes with higher temperatures.

However, one might expect that after this sudden increase, the rate of growth would decrease and then saturate at the Lloyd's bound. However, with only the terms that we considered here we did not observe such behavior. Therefore, considering some other counter terms should be necessary and then adding them to the Eq.  \ref{eq:totcomplexity}, could probably solve this issue and similar to other studies, the Lloyd's bound would be saturated from above. We leave finding such terms for these QCD models to future works.

\begin{figure*}[ht!]
        \centering
        \begin{subfigure}[b]{0.33\textwidth}
            \centering
            \includegraphics[width=\textwidth]{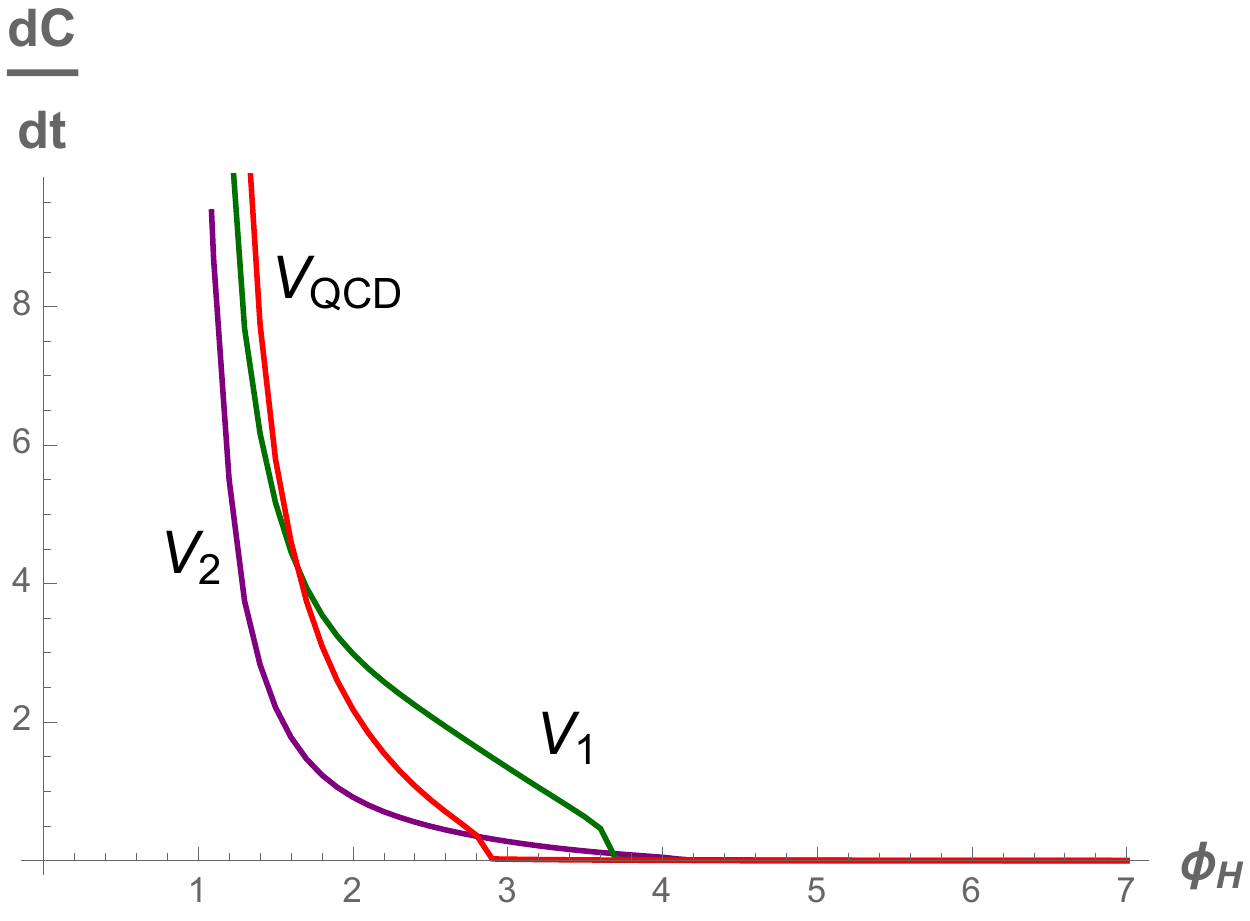}
            \caption[Network2]%
            {$V_{\text{QCD,1,2}}$}    
            \label{VQCDcomplexity2}
        \end{subfigure}
         \centering
        \begin{subfigure}[b]{0.33\textwidth}   
            \centering 
            \includegraphics[width=\textwidth]{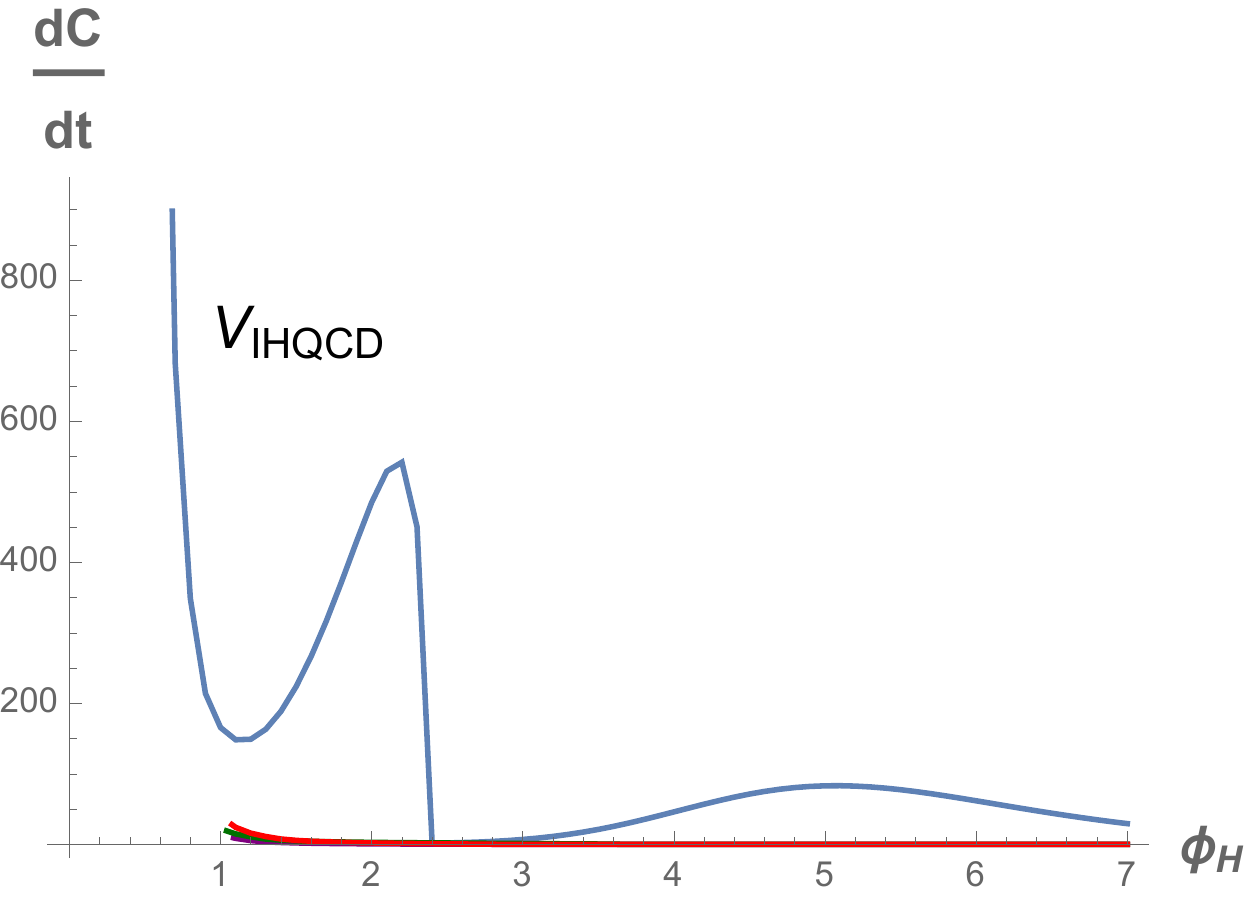}
            \caption[]%
            {$V_{\text{IHQCD}}$ }    
            \label{potentails}
        \end{subfigure}
        \caption[ The relation between $\frac{dC}{dt}$ and $\phi_H$. ]
        {\small The relation between $\frac{dC}{dt}$ and $\phi_H$. } 
        \label{fig:dcdtf}
    \end{figure*}

In Fig. \ref{fig:dcdtf}, we show the behavior of complexity growth rate $\frac{dC}{dt}$ versus $\phi_H$ for late times. One could see that at later times though, increasing $\phi_H$ would decrease the rate of growth of complexity. Also, the range of complexity growth for $V_{\text{IHQCD}}$ is much bigger than the other ones. The phase transitions of this model for higher $\phi_H$ is also evident from the diagram.

By adding gauge fields to the action and using models such as the one in \cite{Gubser:2005ih, Gubser:2008px,Herzog:2009xv} and by doing the same calculations, one can similarly study the full time behavior of complexity growth rate for the \textit{superconducting} phases and specifically during the phase transitions \cite{Momeni:2016ekm}, which could have many practical applications in quantum information.

We are also interested to study the effects of hydrodynamics and non-hydrodynamic modes on the rate of growth of complexity and how these modes affect the different phase transitions during the complexity growth. We will discuss about it briefly in the following parts.

\subsection{$V_{\text{QCD}}$}
The parameters of this confining model have actually been chosen in a way to fit the data of the temperature dependence of the speed of sound which have been obtained from the lattice QCD. 

With this potential we observe a crossover behavior at the zero baryon charge density. In this case, at lower temperatures, near the pseudocritical temperature $T=T_c$, there would be a rapid change in the large momentum dependence of the imaginary part of the hydrodynamic mode. As momentum increases, the imaginary part would flow to the minus infinity. This, as mentioned in \cite{Janik:2016btb}, would point to a novel effect around $T=T_c$ in the sound channel where  there would be a crossing between the hydrodynamic and non-hydrodynamic modes. It could be seen in Fig. \ref{soundVQCD} that actually at this point there is a sharp decrease in the speed of sound. Then from Fig. \ref{VQCDcomplexity2}, one can see that at this point, there would be a sharp increase in the rate of complexity growth.

Also, one could note that in all the figures, at high temperatures, the speed of sound would reach to $1/3$, which is the expected value for the conformal case or plasmas with a very high $T$.

In the higher temperature regions, where the results match with the conformal case, the non-hydrodynamic modes are actually the most effective ones, while in the lower temperature points, the hydrodynamic modes play the most important role. Also, it is worth to notice that the hydrodynamic modes (at lower temperature, and around $T=T_c$) are much more affected by the ``ultra locality property" of the system as it was mentioned in \cite{Janik:2015waa}. This could actually greatly affect the complexity growth rate of the QCD systems as we have observed.

\subsection{$\text{V}_1$}

This model is actually the most interesting one as it has some distinctive features. The parameters for this model have been constructed in a way to generate a first order phase transition. In this case one could observe instability or spinodal region for a certain temperature range where this behavior could also be seen from the phase diagrams of entropy, complexity growth and specially the diagram for the speed of sound.

In fact, the first order phase transition appears in two different scenarios \cite{Janik:2016btb}. One is similar to the Hawking-Page scenario where the transition is between a black hole and a thermal gas. The other one is between two black holes with different sizes. In this case for our $V_1$ model, the first order phase transition is actually between two black holes \cite{Gursoy:2008za}. The fact that both of these cases could happen is due to the functional dependence of the dilaton field potential on the radial distance in the deep infrared region.  Another distinctive feature is the existence of non-propagating sound mode in the very low temperature which also shows its effects on the complexity growth rate phase diagram. 

The most important feature in this model though, is the presence of the \textit{spinodal} region which has been studied in the nuclear physics in the context of spinodal multifragmentation \cite{Chomaz:2003dz}. One can also see from the diagram of the speed of sound (as $c_s^2 < 0$), there exist a dynamical instability during phase transition which are caused from the bubble formation in this region. 

The same kind of behavior has also been observed in gravitational studies of black strings and p-branes in the context of Gregory and Laflamme instability, \cite{Gregory:1993vy}. Note that in this case the hydrodynamical modes are purely imaginary and this shows its effect on the complexity growth rate as well. It could be seen from the fact that, unlike the cross over and second order phase transition where the slopes of $\frac{dC}{dt}$ versus $T/T_c$ are positive, in this case the slope of complexity growth during the phase transition is actually negative. This behavior could be a universal property of complexity for the regions with\textit{``hydrodynamical instabilities"} with purely imaginary mode.

\subsection{$\text{V}_2$}

In this model at a critical temperature $T=T_c$, the speed of sound would vanish which could be checked from Fig. \ref{V2complexitysound}. Also, near the critical temperature the entropy behaves as
\begin{gather}
s(T) \sim s_0 +s_1 \left (T/T_c-1 \right)^{1-\alpha},
\end{gather}
where according to \cite{Gubser:2008ny}, the constant in the power should be $\alpha= 2/3$. As the complexity growth rate phase diagram is very similar to the entropy diagram, we expect that this relation could approximately describe the complexity growth rate behavior near critical temperature as well.

\begin{figure*}[ht!]
        \centering
        \begin{subfigure}[b]{0.43\textwidth}
            \centering
            \includegraphics[width=\textwidth]{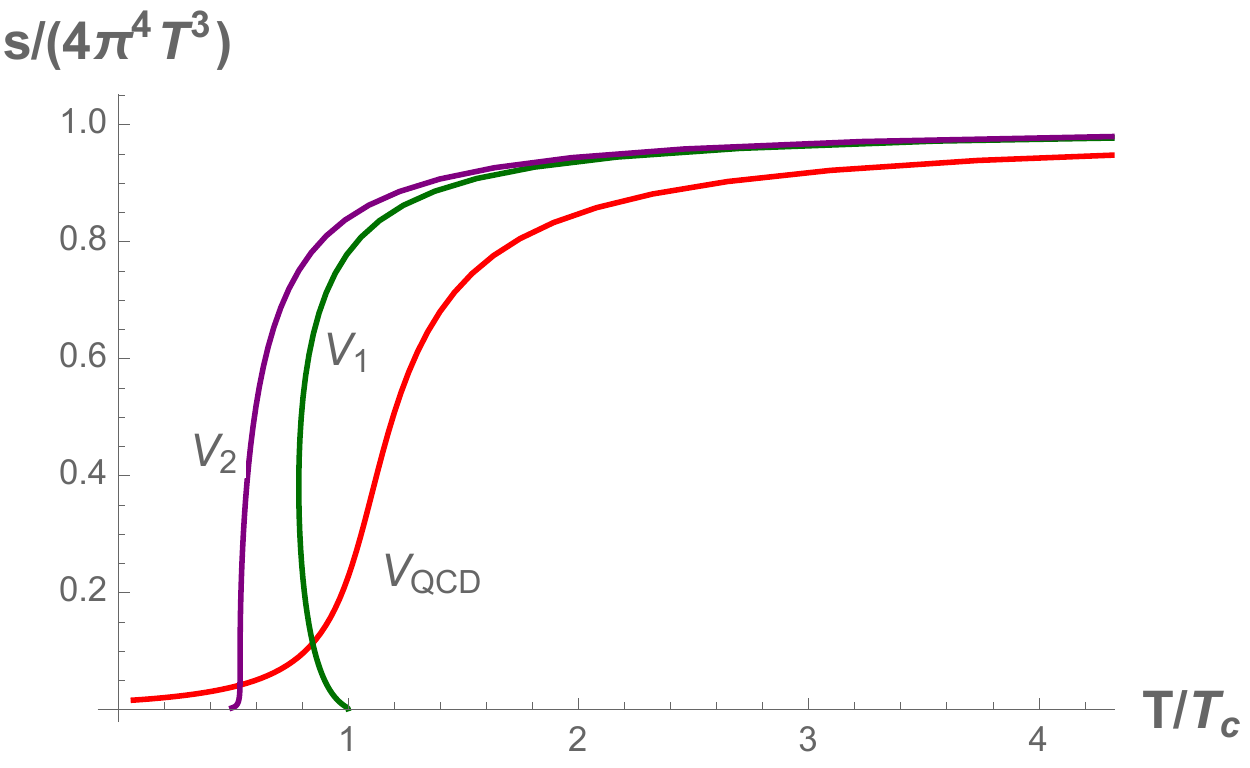}
            \caption[Network2]%
            {Entropy Phase Diagrams.}    
            \label{VQCDcomplexity2}
        \end{subfigure}
         \centering
        \begin{subfigure}[b]{0.43\textwidth}  
            \centering 
            \includegraphics[width=\textwidth]{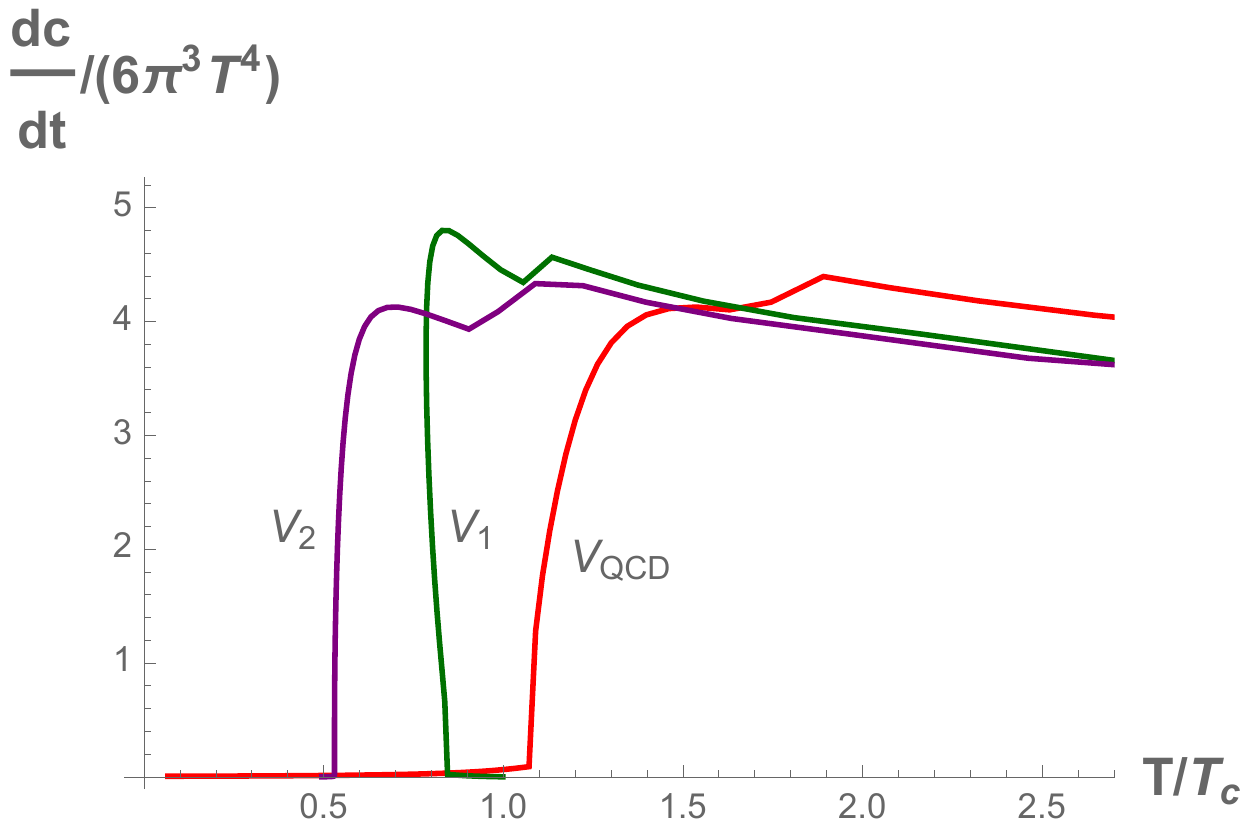}
            \caption[]%
            {{Complexity Phase Diagrams.}}    
            \label{threeComplexity}
        \end{subfigure}
         \centering
        \vskip\baselineskip
        \begin{subfigure}[b]{0.43\textwidth}   
            \centering 
            \includegraphics[width=\textwidth]{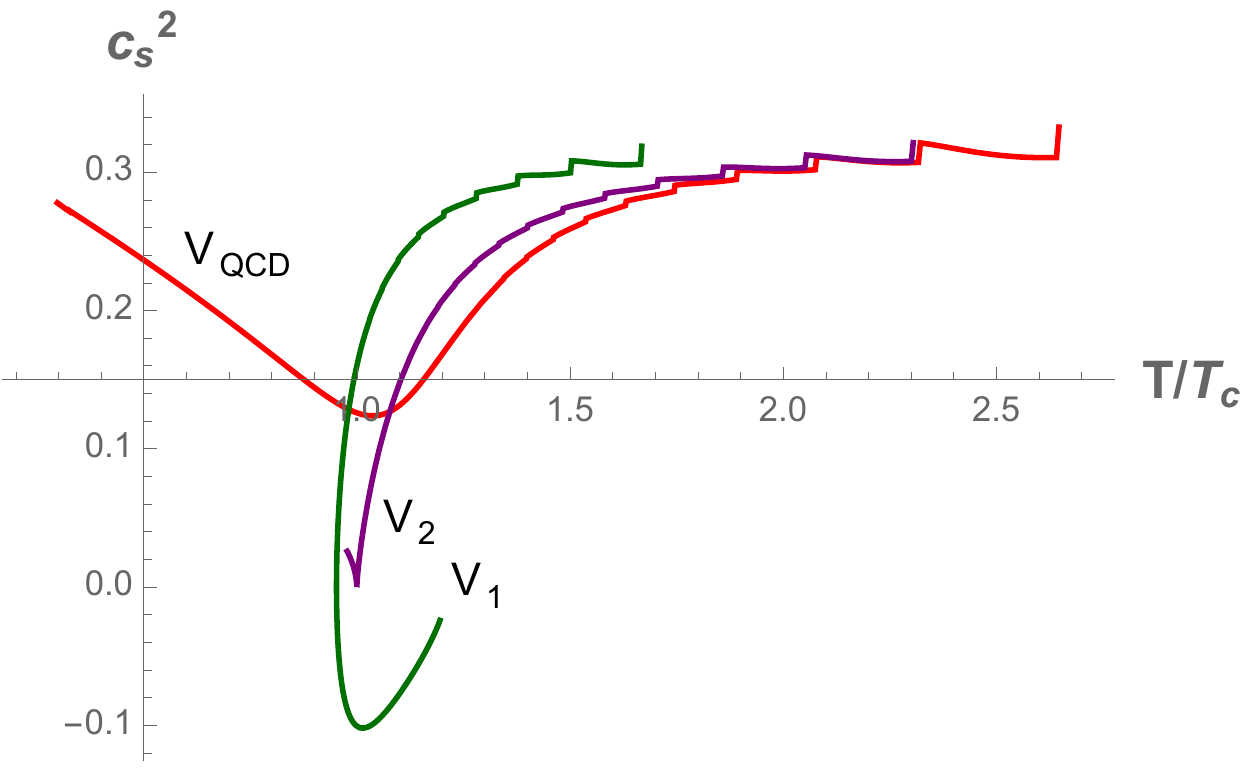}
            \caption[]%
            {Speed of sound in three models.}    
            \label{Speed of sound in three models.}
        \end{subfigure}
        \quad
         \centering
        \begin{subfigure}[b]{0.43\textwidth}   
            \centering 
            \includegraphics[width=\textwidth]{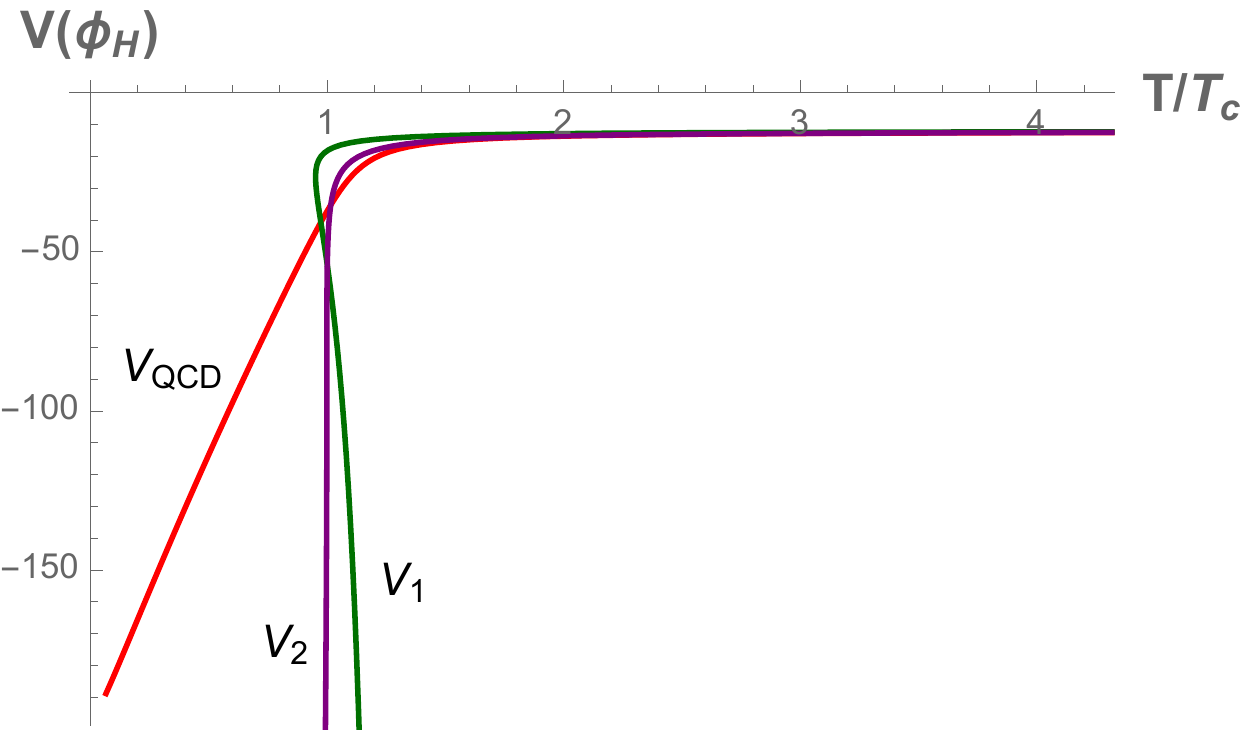}
            \caption[]%
            {Potentials of the three models.}    
            \label{potentails}
        \end{subfigure}
        \caption[Comparing different quantities for three models of QCD.]
        {\small Comparing different quantities for three models of QCD.} 
        \label{fig:allquantities}
    \end{figure*}

Notice that in all the diagrams of entropy, complexity growth rate and speed of sound, (Fig \ref{fig:allquantities}) and also the time dependences of $\phi_m$ (Fig \ref{fig:tphim}), the behavior for the $V_2$ case is very similar to $V_{\text{QCD}}$ case. The main difference is that $V_2$ is actually in the back of $V_{\text{QCD}}$.

As mentioned in \cite{Janik:2016btb}, the generic temperature dependence of the frequencies of the quasi normal modes are also very similar to those of the crossover case. This, therefore, would make all the corresponding quantities of these two models act similar to each other around the critical temperature. 

Actually, the hydrodynamic description of the system would break down at smaller momenta scales, \cite{Janik:2016btb}. This would make the critical temperature of $V_2$ lower than the crossover case. For the higher temperatures, all of these models including $V_2$ would behave the same way and similar to the conformal case.

In order to get some more detailed results about the relationship between the different characteristics of the model and the behavior of the growth of complexity, the numerical data points for the plot of complexity could be increased. This would then increase the calculation time dramatically.

\subsection{$\text{V}_{\text{IHQCD}}$}

The improved holographic QCD model is very unique compare to other models and acts substantially different from the other ones. It has been constructed in order to best model the dynamical properties of QCD, specifically the asymptotic freedom, color confinement and to get a more realistic value for the bulk viscosity. In this model there is a first order phase transition, but between a black hole and a thermal gas or a vacuum confining geometry, similar to Hawking-Page phase transition. This fact could actually explain the sharp drop in the phase diagram of the complexity growth rate of this model, check Fig. \ref{complexityVIHQ}. 

The exact critical temperature were the first order phase transition would occur would be very difficult to determine for this case as there are several solutions with instabilities and also the temperature of the reference geometry which is the thermal gas would be infinite. However, from the diagrams, the phase transition point has a very distinct behavior since at that point the geometry changes dramatically, which is also apparent in the diagrams of speed of sound and specifically complexity growth rate. 

From these diagrams one could actually detect two characteristic temperatures which are very close to each other. As mentioned in \cite{Janik:2016btb}, for the temperatures between these two and for low momenta, the lowest lying excitation modes become purely imaginary which leads to ultralocality violation in this section. This greatly impacts the first phase of the complexity growth rate. Then notice that at $T=T_m$, for some modes, the hydrodynamic mode and the first non-hydrodynamic mode would have the same dispersion relation. At a critical temperature $T_c $, a bit higher than $T_m$, a transition which changes the geometry substantially would take place. This would cause the rapid increase and then the falling down in the plot of complexity growth rate.

The plot shown in Fig. \ref{soundVIHQ} is actually for the unstable smaller black hole. The stable case which models the behavior of the pure glue system \cite{Boyd:1996bx} is not shown here and we leave it for the future studies. The plot of speed of sound shown here still could communicate the existence of spinodal instability and therefore the bubble formation. Specifically for the small black hole region the speed of sound is anomalously large and then become superluminal violating the causality. This actually points out to a \textit{``dynamical"} instability which is different from the ``spinoidal case" seen before. This difference of behavior could also be detected from the behavior of normal modes \cite{Janik:2016btb} and specifically from the complexity growth rate behavior shown in Figs. \ref{entropyVIHQ}, \ref{soundVIHQ}, \ref{complexityVIHQ} and even Fig. \ref{fig:dcdtf}.

By studying the poles of Green's functions, the behavior of hydrodynamic and non-hydrodynamic modes have been discussed in more details in \cite{Janik:2016btb}. It has been shown that the modes would be degenerated close to the minimal black hole temperature and there would exist a region of temperatures where the non-hydrodynamical modes would be unstable.

In \cite{Janik:2016btb}, it has been pointed out that a small gap between the degrees of freedom at low momentums would exist.  The non-hydrodynamical modes show \textit{ultralocal} properties as they have weak dependence on the momentum scale. It would be interesting to study the effects of such non local modes on the behavior of complexity growth rates. Also, a more careful study of the behavior of complexity growth rate close to the mode gap between the two solutions would be very interesting.

 Finally, it is worth to note that as the modes of this system would match the results from the holographic dual of superfluid systems, \cite{Janik:2016btb,Amado:2009ts }, one would expect that the behavior of growth of quantum complexity shown in Fig. \ref{complexityVIHQ}, would actually match the behavior of growth of complexity in superfluid systems which again might be able to be tested experimentally.

\section{Discussion}\label{sec:discuss}

In this work we showed the relationship between different field potentials and complexity growth rate behavior in several models such as charged dilaton, Born-Infeld and dyonic black hole. We conjectured that it would be a universal feature of complexity that, for potential wells, the complexity growth rate would be higher and for parameters of the model which there is a steep potential barrier, the complexity growth rate would decrease to zero. This could indicate that quantum fluctuations such as tunneling would be the main source of complexity growth rate after thermalization point.

Also, we studied the complexity growth rates during different phase transitions. First, for the dyonic black hole, we studied the full time behavior. Similar to other studies, we found that the Lloyd's bound would be violated at early times but then at later times it would be saturated from above. Changing the charges of the black hole could reveal a second order phase transition in the complexity growth rate which is similar to the phase transition of the van der Waals fluid. For very big charges, the complexity growth rate would diverge as we have expected, since at those ranges of charge the black hole would be unstable. Therefore, we found that complexity growth rate could be a very suitable probe to catch different phase transitions and instabilities.

To further explore this point, we considered the complexity growth rate of AdS soliton black holes which for the first time has been studied in \cite{Reynolds:2017jfs}. We compared the behavior of complexity growth rates in AdS soliton and pure AdS case, the potential behavior in the AdS soliton with its tachyonic condensation, and also we compared the Schwinger phase diagrams and again we observed strong correlations.

Finally, for the main part of this work, we studied the Gubser model of QCD \cite{Gubser:2008ny, Gubser:2008yx} which by tuning the parameters of the potential for the $V_{\text{QCD}}$, $V_1$, $V_2$ cases, the crossover, first and second order phase transitions could be generated. We also studied the improved holographic QCD potential model, $V_{\text{IHQCD}}$ which improves modeling the dynamical properties. Solving for the entropies, complexity rates, speed of sounds and potentials numerically, we observed similarities in their behavior near the phase transition points in each model. By considering the crossings between hydrodynamic and non-hydrodynamic modes, we then explained different features of complexity growth rates and the phase diagrams.

For the full, time dependent phase diagrams, however, we showed that additional counter terms in the action for this QCD model would be needed to produce the desired behavior considering the Lloyd's bound.

For the future works, one could make the argument of relationship between complexity growth rate and particle creation and annihilation rates more precise. For doing that, one could imagine Schwinger mechanism as the motion in the space of unitary operators and then using the intuition from \cite{Jefferson:2017sdb} and \cite{Chapman:2017rqy}, calculate how much pair creation or vacuum quantum fluctuations would increase the distance between the initial state at $t_0$ and the state at later times after only one pair creation. Holographic methods of considering an open string, writing the Nambu-Goto action and the induced metric on the string worldsheet,  and then adding proper boundary and joint terms for calculating the action could be used to solve this problem. The first few steps have been taken in the final part of \cite{deBoer:2017xdk}.  The main challenge that one would face is determining the correct Wheeler-DeWitt. Solving these issues and comparing with the results from field theory methods such as the one in \cite{Hackl:2018ptj}, one then could check if actually the creation of these particles would move on the optimal path of increasing the computational complexity which would be a very interesting problem.

As a side note, in \cite{Hashimoto:2014yya, Ghodrati:2015rta} for different backgrounds, the decay rate is found to increase with the magnetic field parallel to the electric field, while it decreases with the magnetic field perpendicular to the electric field which seems to be a universal feature of Schwinger effect. One then could check how in the presence of electric and magnetic fields for two setups of magnetic field parallel and perpendicular to the electric field, the complexity growth rates would be different. Comparing the results could strengthen our conjecture then.

The relationship between ``shear viscosity" of black holes and the rate of complexity growth rate  specially around the critical temperature could also be studied. This specifically would be an interesting problem with practical applications. For instance as it has been suggested in \cite{Karsch:2007jc}, in quark-gluon plasmas, there would be a sharp rise in the bulk viscosity near the deconfinement transition which actually points to the ``soft statistically hadronization" of the plasma.  So, it would be interesting to study the complexity growth rate at this point and depict the relationship between complexity and shear or bulk viscosity of black holes and the corresponding QCD phases.

We hope to study these questions in the future works.

\section*{Acknowledgement}
I thank Hesam Soltanpanahi and Saeedeh Sadeghian for useful discussions.

\bibliography{NewSchwingerComplex}
\bibliographystyle{JHEP}
\end{document}